%% file: main.tex
\documentclass[manuscript,screen,nonacm]{acmart}

\input{macros}

\input{metadata}

\begin{document}
\raggedbottom

% ACM requires the abstract before \maketitle.
\input{sections/abstract}
\maketitle

\input{sections/introduction}
\input{sections/preliminaries}
\input{sections/system-overview}
\input{sections/evaluation}

\input{sections/related-work}
\input{sections/conclusion}
\input{sections/code-and-artifact-availability}

\begin{acks}
\input{sections/acknowledgments}
\end{acks}

\begingroup
\sloppy
\hbadness=10000
\bibliographystyle{ieeetr}
\bibliography{main}
\endgroup

\newpage
\appendix
\section*{Appendices}
\input{appendices/device-construction}
\input{appendices/semantic-profiles}
\input{appendices/execution-model}
\input{appendices/resource-estimation}
\input{appendices/zsz-construction}

\end{document}

%% file: macros.tex
\usepackage{listings}
\usepackage{wrapfig}
\usepackage{multicol}
\usepackage{caption}
\usepackage{booktabs}
\usepackage{tikz}
\usetikzlibrary{quantikz2}     % Draw quantum circuits with TikZ.
\usetikzlibrary{arrows.meta}   % Latex[] arrow tips in block diagrams.
\usetikzlibrary{decorations.pathreplacing}  % Braces in block diagrams.
\usepackage{enumitem}          % Compact description lists (Sec. 3.3).

\makeatletter
\newcommand{\QLXNeedspace}[1]{%
  \par\begingroup
  \dimen@=#1\relax
  \vskip\z@\@plus\dimen@
  \penalty-100
  \vskip\z@\@plus-\dimen@
  \vskip\dimen@
  \penalty9999
  \vskip-\dimen@
  \vskip\z@skip
  \endgroup
}
\makeatother

\definecolor{qlxoutlinered}{HTML}{B00020}
\definecolor{qlxcodebg}{HTML}{F5F8FB}
\definecolor{qlxcodeframe}{HTML}{B8C8D8}
\definecolor{qlxcodekeyword}{HTML}{7254A3}
\definecolor{qlxcodestring}{HTML}{27864A}
\definecolor{qlxcodecomment}{HTML}{627D98}
\definecolor{qlxcodeapi}{HTML}{147D92}

\newcommand{\profile}[1]{\textsc{#1}}

\lstdefinestyle{qlx-python}{
  language=Python,
  basicstyle=\small\ttfamily,
  keywordstyle=\color{qlxcodekeyword}\bfseries,
  stringstyle=\color{qlxcodestring},
  commentstyle=\color{qlxcodecomment}\itshape,
  emph={qlx,pinnacle},
  emphstyle=\color{qlxcodeapi},
  backgroundcolor=\color{qlxcodebg},
  frame=single,
  rulecolor=\color{qlxcodeframe},
  framesep=5pt,
  xleftmargin=0.35em,
  xrightmargin=0.35em,
  aboveskip=0.65\baselineskip,
  belowskip=0.65\baselineskip,
  showstringspaces=false,
  columns=fullflexible,
  keepspaces=true,
  breaklines=true,
  breakatwhitespace=false,
  tabsize=4,
  upquote=true
}

  \definecolor{qlxcodetype}{HTML}{A34F16}
  \definecolor{qlxcodevalue}{HTML}{356FA3}

  \lstdefinelanguage{QLXMLIR}{
    sensitive=true,
    alsoletter={.},
    morecomment=[l]{//},
    morestring=[b]",
    morekeywords=[1]{
      module,attributes,iter
    },
    morekeywords=[2]{
      qlx.decl.action,
      qlx.decl.instrument,
      qlx.program,
      qlx.apply,
      qlx.return,
      qlx.measure,
      qlx.instrument,
      qlx.prepare,
      cflow.repeat,
      cflow.yield,
      event.selection,
      func.func,
      func.return,
      arith.constant,
      cf.br,
      cf.cond_br,
      scf.for,
      scf.while,
      scf.yield,
      lvm.kernel,
      lvm.apply,
      lvm.return,
      lvm.prepare,
      lvm.measure,
      lvm.domain,
      lvm.channel,
      lvm.stream,
      lvm.space
    },
    morekeywords=[3]{
      qlx.logical_qubit,
      lvm.logical_qubit,
      qlx.clifford_action,
      qlx.action,
      i1,i8,i16,i32,i64,
      f16,f32,f64,
      index,none
    },
    morekeywords=[4]{
      clifford_action,
      matrix,
      phases,
      ports,
      kind,
      mode,
      qlx.stage,
      lvm.capability
    }
  }

  \lstdefinestyle{qlx-mlir}{
    language=QLXMLIR,
    basicstyle=\small\ttfamily,
    keywordstyle=[1]\color{qlxcodekeyword}\bfseries,
    keywordstyle=[2]\color{qlxcodeapi}\bfseries,
    keywordstyle=[3]\color{qlxcodetype},
    keywordstyle=[4]\color{qlxcodekeyword},
    stringstyle=\color{qlxcodestring},
    commentstyle=\color{qlxcodecomment},
    literate=
      *{\%}{{{\color{qlxcodevalue}\%}}}1
       {@}{{{\color{qlxcodeapi}@}}}1
       {!}{{{\color{qlxcodetype}!}}}1
       {\#}{{{\color{qlxcodekeyword}\#}}}1
       {^}{{{\color{qlxcodevalue}\textasciicircum}}}1,
    backgroundcolor=\color{qlxcodebg},
    frame=single,
    rulecolor=\color{qlxcodeframe},
    framesep=5pt,
    xleftmargin=0.35em,
    xrightmargin=0.35em,
    aboveskip=0.65\baselineskip,
    belowskip=0.65\baselineskip,
    showstringspaces=false,
    columns=fullflexible,
    keepspaces=true,
    breaklines=true,
    breakatwhitespace=false,
    tabsize=2,
    upquote=true
  }

\lstdefinestyle{qlx-mlir-numbered}{
    style=qlx-mlir,
    numbers=left,
    firstnumber=1,
    stepnumber=1,
    numberblanklines=false,
    numberstyle=\scriptsize\ttfamily\color{qlxcodecomment},
    numbersep=7pt,
    xleftmargin=2.6em,
    framexleftmargin=2.1em,
    framesep=4pt
}

%% file: metadata.tex
\title{CUDA-Q Logical: Retargetable Compilation for Fault-Tolerant Quantum Computing}

\author{Alexander McCaskey}
\author{Justin Lietz}
\author{Adam Holmes}
\author{Kohei Nakaji}
\author{Vadym Kliuchnikov}
\author{Yifan Hong}
\author{Amalee Wilson}
\author{Andres Paz}
\author{Bettina Heim}
\author{Bruno Schmitt}
\author{Krysta M.~Svore}
\affiliation{%
  \institution{NVIDIA}
  \city{Santa Clara}
  \state{California}
  \country{USA}
}

\authorsaddresses{%
Corresponding authors: Alexander McCaskey,
\href{mailto:amccaskey@nvidia.com}{amccaskey@nvidia.com}, 
Justin Lietz, \href{mailto:jlietz@nvidia.com}{jlietz@nvidia.com}%
}
  
\acmConference[QLX Working Draft]{QLX Paper Working Draft}{September 2026}{}
\acmBooktitle{QLX Paper Working Draft}
\acmYear{2026}
\copyrightyear{2026}
\setcopyright{none}
\acmDOI{}
\acmISBN{}
\makeatletter
\ifnum\ACM@format@nr=\z@
  \def\@authorfont{\fontsize{10.5pt}{13pt}\selectfont\sffamily}%
\fi
\def\QLX@typeset@groupedauthors{%
  \begingroup
  \sloppy
  \@tempcnta=\z@
  \def\par{%
    \ifnum\@tempcnta=\z@
      \@tempcnta=\@ne
    \else
      ,\space
    \fi}%
  \@authorfont\fontsize{10.5pt}{13pt}\selectfont\@currentauthors
  \endgroup
  \par}
\patchcmd{\@typeset@author@bx}
  {\@authorfont\@currentauthors\par\@affiliationfont}
  {\QLX@typeset@groupedauthors\@affiliationfont}
  {}
  {\PackageError{qlx-paper}{Could not configure grouped author display}{}}
\makeatother

%% file: sections/abstract.tex
\begin{abstract}
% Fault tolerant quantum program compilation and subsequent execution requires 
% mapping to an error-correcting architecture, scheduling on physical
% resources, and coupling to real-time syndrome processing, decoding, and feedback
% across the quantum register and its classical control hardware. Researchers have developed
% powerful software tools for these tasks, but combining them into one compilation path
% still requires manual translations. Those handoffs can obscure assumptions,
% detach estimates from the artifacts they describe, and make it difficult to
% attribute performance differences when the same application is compiled for
% different architectures. 
%Compiling and executing a fault-tolerant quantum program is a heterogeneous
%systems problem. Fault tolerance adds code-specific logical operations,
%resource preparation, physical scheduling, syndrome transport, decoding, and
%feedback, which must be coordinated across a quantum register and heterogeneous classical
%control and compute hardware. Researchers have developed strong specialized
%software tools for code design, circuit construction, simulation, decoding, and
%resource analysis. Using these tools within a complete compilation and execution workflow requires shared
%interfaces that connect their results to the logical program, carry
%architectural choices between stages, and check the transformations between
%them. 

Realizing fault-tolerant quantum computing requires mapping logical programs to heterogeneous quantum 
error correction (QEC) codes and diverse fault-tolerant execution models, scheduling  physical resources, 
and coupling to real-time classical control and feedback. Specialized tools exist for each step but rely on 
manual composition and translation that discard assumptions and provenance, separating resource estimates 
from the compiler artifacts they describe, and making it difficult to validate correctness, compare architectures, 
or attribute costs to specific design choices. We present CUDA-Q Logical, an extensible compiler infrastructure 
for retargetable fault-tolerant compilation, analysis, and execution. Interoperable with CUDA-Q and 
other mainstream front-ends, CUDA-Q Logical progressively lowers target-independent logical programs 
through a constrained logical virtual machine, QEC microcode, physical gate schedules, and real-time 
control plans, with each layer preserving semantics and provenance, while verifying composition and resource constraints. 
By deriving every resource estimate directly from compiler artifacts, the framework unifies compilation 
and resource analysis, enabling successively refined estimates and principled cross-architecture 
comparison while permitting QEC codes, execution models, decoders, and hardware architectures to be 
introduced as modular extensions. Across workloads ranging from application-architecture studies 
to qLDPC surgery and detector-error-model composition, we show that schedule-derived estimates 
reconcile with established independent models. Crucially, this compiler-visible structure exposes 
cost drivers hidden by aggregate analytical formulas, carries QEC artifacts intact into 
simulation, and demonstrates a complete compilation pipeline for fault-tolerant quantum computing.

\end{abstract}

% ACM frontmatter commands belong after \begin{document} and before \maketitle.
\begin{CCSXML}
<ccs2012>
  <concept>
    <concept_id>10010520.10010553.10010562</concept_id>
    <concept_desc>Computer systems organization~Quantum computing</concept_desc>
    <concept_significance>500</concept_significance>
  </concept>
  <concept>
    <concept_id>10011007.10011006.10011008</concept_id>
    <concept_desc>Software and its engineering~General programming languages</concept_desc>
    <concept_significance>300</concept_significance>
  </concept>
</ccs2012>
\end{CCSXML}

\ccsdesc[500]{Computer systems organization~Quantum computing}
\ccsdesc[300]{Software and its engineering~General programming languages}

\keywords{fault-tolerant quantum computing, quantum error correction,
  compiler intermediate representations, resource estimation,
  quantum architecture, microarchitecture co-design}

% WORDING Attempts
% Compilation, analysis, and execution of a fault-tolerant quantum program requires more than selecting a code
% or estimating logical error rates. A compiler must transform logical application semantics
% into code-specific operations, resource-bound physical schedules, and timed
% control and decoding actions on a distributed classical system. Existing QEC software 
% workflows distribute these transformations across disparate tools with incompatible 
% representations, which forces key steps to be done manually and leaves estimates 
% disconnected from execution. 
% Compiling logical quantum programs into fault-tolerant executables requires 
% a systems-level approach that span multiple layers of abstraction. The typical workflow 
% must take logical intent to code-specific operations, resource preparation, syndrome preparation, and decoding.
% Numerous specialized software tools have been developed that address pertinent aspects of 
% this lowering pipeline, but 
% Researchers have
% developed numerous specialized software tools for studying and enabling aspects of the problem,
% including code design, circuit construction, simulation, decoding, and
% resource analysis. 

%% file: sections/introduction.tex
\section{Introduction}
\label{sec:introduction}

Recent demonstrations of below-threshold surface-code memories with real-time feedback
\cite{google2025belowthreshold, atom2026toric} and rapid progress in high-rate quantum
low-density parity-check (qLDPC) codes suggest multiple viable paths toward fault-tolerant quantum computing
\cite{bravyi2024highthreshold, Xu_2024_constant, Hong_2024_four,
reichardt2024tesseract, webster2026pinnacle, cain2026oratomic, gu2026qgpu,
zhao2026ultra, okada2026pair, hong2026zszlp, bhardwaj2026mitten,
yang2026gala, Gottesman_2014, Fawzi_2020_constant, Tamiya_2026_FT,
Nguyen_2025_FT, Zhang_2026_Accelerating}.
Yet neither a memory experiment nor a code family defines a complete fault-tolerant computing system.
Realizing useful fault-tolerant programs requires coordinating logical operations, encoded storage, non-Clifford resources, concurrency, syndrome extraction, decoding, and real-time feedback under a common accuracy and timing contract. The choices made at each layer can shift application resource estimates by orders of magnitude for a given algorithm 
\cite{gidney2025factor, Zhou_2025, webster2026pinnacle,
cain2026oratomic}, complicating cross-architecture comparison and attribution of cost differences to specific design decisions. 
Compiler infrastructure must therefore preserve program intent 
while exposing architecture-specific trade-offs, enabling researchers and QPU builders to innovate on codes, execution models, and control strategies without forcing developers to rewrite applications.
Holding the application fixed enables principled attribution of architectural costs and performance advantages.

Existing quantum software provides many of the components needed for these workflows.
However, these capabilities are typically developed and used in isolation, with limited support for preserving the relationships among logical workloads, resource estimates, schedules, detector models, and decoder outputs across tool boundaries.
Frameworks and languages such as Qiskit, CUDA-Q, Guppy, and
PennyLane provide program representations and compilation workflows for
quantum and hybrid quantum--classical programs
\cite{javadiabhari2024qiskit,cudaqteam_quake,koch2024guppy,
bergholm2018pennylane}. Specialized fault-tolerant compilers scale
surface-code layout and QEC-circuit synthesis or compose QEC passes with
classical callbacks
\cite{paz2026qstack,watkins2024compiler,yin2025qeccsynth}, while resource estimation methods and frameworks decompose 
workloads and pass logical resource counts to cost models \cite{beverland2022assessing,harrigan2024qualtran}.
However, composing these capabilities into one fault-tolerant compilation workflow from application-level program to physics-level control representation requires
shared, layered abstractions and contracts that keep estimates, schedules, detector models, and decoder
results tied to the logical workload and architectural choices that produced
them. 
This is essential for evaluating codes, compiler components, and machine implementations using the same workloads and metric definitions.
An extensible fault-tolerant compiler infrastructure enables this capability by separating application specification from implementation choices. 

A compiler supporting this workflow must preserve program meaning while progressively resolving architectural choices required by the selected target. It must scale to
large, repeated computations and accept components from independent tool
authors and QPU providers through stable, typed interfaces. Its outputs must support
validation and reproducibility, controlled architecture comparisons, and estimates
whose changes can be traced to decisions made during compilation.
Table~\ref{tab:ft-compiler-requirements} summarizes these requirements in
dependency order.

\input{tables/requirements}

Standard compiler engineering and IR development provide an established model for 
realizing these enumerated requirements. A prototypical example in the field is 
MLIR \cite{lattner2021mlir}, which leverages explicit 
types, operations, and verifiers to define the information available to each compiler
pass, while explicit lowerings connect representations at different levels of
abstraction. Quantum IRs have applied this model to
hybrid control and quantum dataflow
\cite{cross2022openqasm3,mccaskey2021mlir,ittah2022qiro,peduri2022qssa}.
Fault-tolerant compilation requires similar guarantees for logical programs,
QEC gadget and protocol bindings, detector error models, physical schedules, and control plans. These
artifacts make different assumptions and support different operations. A
compiler or external tool consuming one of them must know which choices have
been made, which remain open, and which invariants have been checked. IR abstraction hierarchies 
(e.g., MLIR Dialects and progressive lowering) make this information explicit, 
while verified lowerings check that each transformation preserves the earlier 
program as it adds the next architectural choices.

We present \emph{\textbf{CUDA-Q Logical}}, an extensible compiler infrastructure
for fault-tolerant quantum computing, provided as a standalone, modular component 
within the CUDA-Q platform and built upon MLIR. 
It supports CUDA-Q's end-to-end path from application 
kernels to fault tolerant execution, and its compiler contracts can also be leveraged apart from the full path. 
CUDA-Q Logical extends existing quantum compilation flows with compiler-visible representations of fault-tolerant architectural choices.

\begin{figure}[t]
\centering
\includegraphics[width=\textwidth]{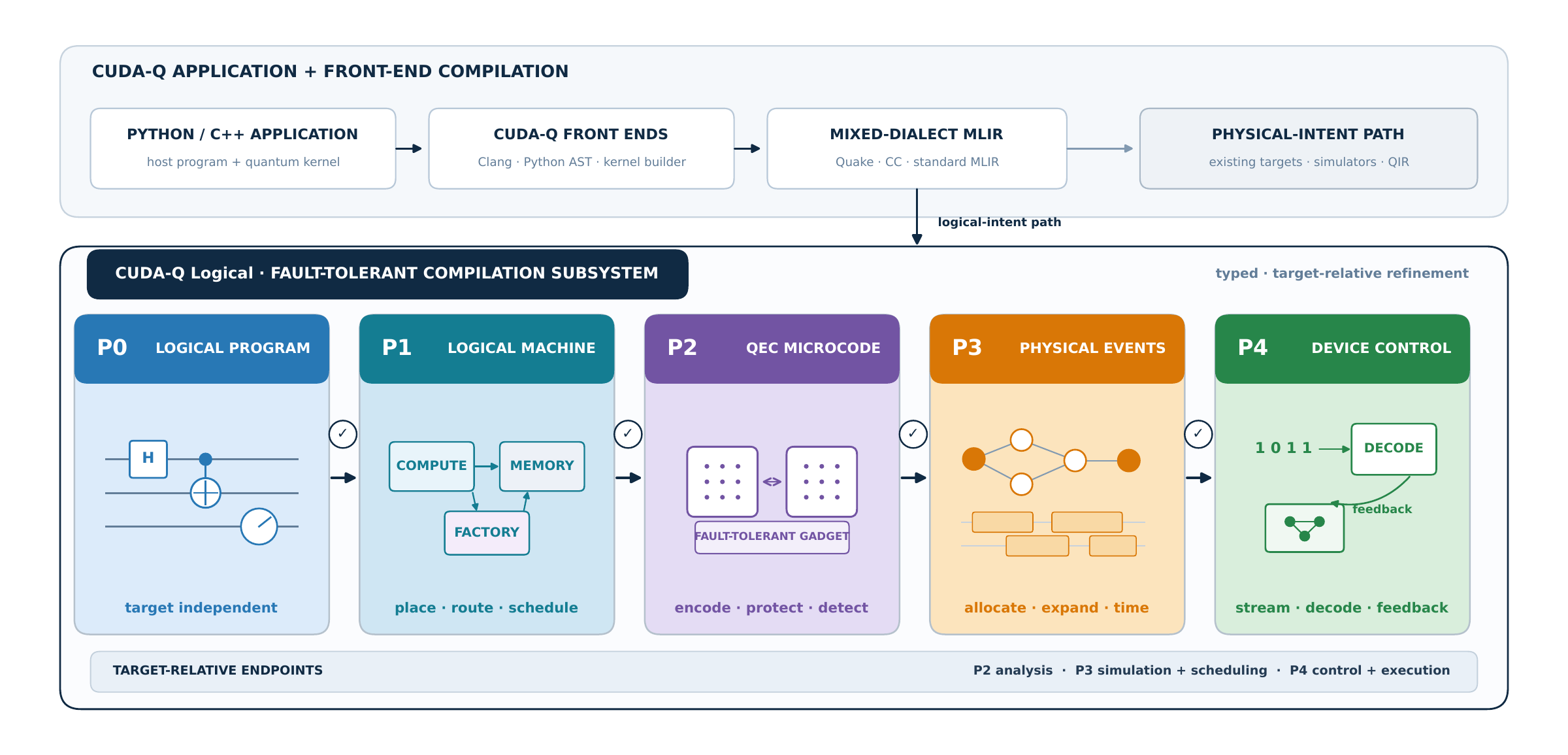}
\caption{CUDA-Q Logical progressively refines one logical program from target-independent intent to real-time control. Each profile adds a new
class of target commitment, while research components attach at typed
boundaries and analyses observe the same evolving build. Compilation may
stop at any verified profile accepted by the selected target.}
\Description{A horizontal pipeline from a logical kernel through the P0
  logical intent, P1 logical architecture, P2 QEC binding, P3 physical
  materialization, and P4 real-time control levels, with research components
  attached at the boundaries between levels and analyses reading the build.}
\label{fig:qlx-compiler-spine}
\end{figure}

The stack exposes typed intermediate representations throughout the compilation flow, allowing researchers to start from a chosen IR abstraction level, analyze artifacts up to any stage, and integrate external tooling at typed boundaries.
These abstractions provide stable integration points for QEC codes, decoders, execution models, and machine-specific implementations developed independently of the compiler.
%Researchers may enter at an intermediate representation, stop at the artifact required by an analysis, or connect external tools at typed boundaries. 
%CUDA-Q already progressively lowers quantum application kernels toward target backends, but its compiler IR did not represent the intervening fault-tolerant logical architecture and QEC binding as explicit, verifiable states.
CUDA-Q already provides a progressive lowering path from quantum application kernels to target backends. CUDA-Q Logical extends that path through explicit, typed representations of fault-tolerant architectural states and QEC bindings, introducing a verifiable refinement pipeline on semantic \emph{profiles} (see Figure \ref{fig:qlx-compiler-spine}) 
beneath the application IR. 
This compiler pipeline is retargetable: the user-selected target determines how far a program must be refined and which architectural choices must be resolved.
We define the lowering workflow for fault tolerant 
compilation as: 
\[
\begin{aligned}
\text{logical kernel}
&\rightarrow \mathrm{P0}\ \text{logical intent}
\rightarrow \mathrm{P1}\ \text{logical architecture}
\rightarrow \mathrm{P2}\ \text{QEC binding} \\
&\rightarrow \mathrm{P3}\ \text{physical materialization}
\rightarrow \mathrm{P4}\ \text{real-time control}.
\end{aligned}
\]
MLIR dialects express these profiles, and verified lowerings connect them.
\textbf{P0} records target-independent logical behavior and success
requirements. \textbf{P1} places that behavior on a code-independent logical
machine, making residency, communication, capacity, and resource-state demand
explicit (Section~\ref{sec:logical-virtual-machine}). \textbf{P2} selects codes
and encodings and realizes logical actions through typed gadgets and
protocols, together with their detector, observable, decoding, and retry
semantics (Section~\ref{sec:qec-microcode}). \textbf{P3} maps that QEC binding
onto physical resources, events, movement, timing, noise, and schedules
(Section~\ref{sec:qec-microcode-physical-machine}). \textbf{P4} binds scheduled
work to instruction delivery, detector transport, decoding, deadlines, and
feedback (Section~\ref{sec:overview-execution}). Appendix~\ref{sec:levels}
defines the complete compiler contract for each profile. Every lowering
preserves the meaning established above it, adds the commitments owned by its
destination profile, and verifies the resulting artifact.

The profile boundaries also define a practical division of labor. Application
developers state target-independent workloads at P0. QEC researchers can supply
new codes, encodings, gadgets, protocols, and extended detector error models at
P2, then exercise them on application workloads without constructing the rest
of the compiler. QPU providers can describe hardware topology, native
operations, resources, noise, and timing at P3, together with instruction
delivery, detector transport, decoding, and feedback capabilities at P4.
Because these contributions enter below P0, the same application can be
analyzed on a new QEC construction or a proprietary machine architecture
without being rewritten around either one.

Our evaluation follows these boundaries. Two architecture studies ask whether
analysis over compiled schedules can reproduce established resource models and
identify the synthesis and scheduling decisions behind their costs. Two QEC
studies ask whether gadget semantics and detector models compose exactly, and
whether a code family developed outside CUDA-Q Logical can enter the compiler
as verified, typed data. A final study joins these capabilities by compiling one
unchanged P0 program through four QEC routes across two codes and two providers.
Together, the studies test schedule-derived attribution, exact semantic
composition, external extensibility, and retargeting without changing the
application.

The remainder of the paper is organized as follows. We first provide some preliminary 
concepts that span this work and place it in the context of compiler engineering, 
QEC research, and computer architecture (Section~\ref{sec:preliminaries}). 
Section~\ref{sec:overview} presents the CUDA-Q Logical architecture and follows a compiler build through
the logical-machine, QEC, physical, analysis, and execution boundaries.
Section~\ref{sec:evaluation} reports the evaluation studies, and
Section~\ref{sec:related-work} positions CUDA-Q Logical relative to quantum
compiler, QEC, and resource-estimation systems. Section~\ref{sec:conclusion}
summarizes the findings, scope, and remaining limitations. The appendices
go into greater detail on device construction (Appendix~\ref{sec:construction}), specification of the
P0--P4 compiler dialect and interfaces and underlying execution model
(Appendices~\ref{sec:levels} and \ref{sec:execution-model}), and develop the
progressive resource-analysis framework (Appendix~\ref{app:resource-analysis}).

%% file: tables/requirements.tex
\begin{table}[b]
  \caption{Priority-ordered requirements for any cross-target fault-tolerant compiler stack.}
  \label{tab:ft-compiler-requirements}
  \small
  \renewcommand{\arraystretch}{1.16}
  \begin{tabular}{@{}p{0.34\linewidth}p{0.61\linewidth}@{}}
    \toprule
    \raggedright\textbf{Requirement} & \raggedright\textbf{Concise definition} \tabularnewline
    \midrule
    \raggedright\textbf{R1. Semantic separation without disconnection} &
    \raggedright Assign every fact to one semantic level and require connected lowerings to preserve program meaning while making new target commitments explicit. \tabularnewline
    \cmidrule(lr){1-2}
    \raggedright\textbf{R2. Compilation toward execution} &
    \raggedright Materialize every resource, event, schedule, detector, and control dependency required by the selected endpoint rather than leaving backend semantics implicit. \tabularnewline
    \cmidrule(lr){1-2}
    \raggedright\textbf{R3. Validation and reproducibility} &
    \raggedright Qualify every claim by a verified boundary and preserve immutable provenance sufficient to diagnose, reproduce, and replay each derived artifact and result. \tabularnewline
    \cmidrule(lr){1-2}
    \raggedright\textbf{R4. Scalable representation, throughput, and bandwidth} &
    \raggedright Keep repeated structure symbolic until expansion is required, then measure representation growth and every quantum and classical rate that can bound feasibility. \tabularnewline
    \cmidrule(lr){1-2}
    \raggedright\textbf{R5. Open contracts and typed extensibility} &
    \raggedright Allow independently developed codes, gadgets, architectures, estimators, decoders, and providers to compose through versioned typed capabilities without changing the semantic core. \tabularnewline
    \cmidrule(lr){1-2}
    \raggedright\textbf{R6. Controlled comparison and iterative co-design} &
    \raggedright Hold workload, success criteria, metrics, and compatible assumptions fixed across targets, and version deliberate co-design changes so that their effects remain attributable. \tabularnewline
    \cmidrule(lr){1-2}
    \raggedright\textbf{R7. Estimates survive refinement} &
    \raggedright Attach every estimate to its semantic level, method, assumptions, uncertainty, and metric scope, and reconcile changes as compilation adds detail. \tabularnewline
    \bottomrule
  \end{tabular}
\end{table}

%% file: sections/preliminaries.tex
\section{Preliminaries}
\label{sec:preliminaries}

CUDA-Q Logical sits at the intersection of quantum error correction, compiler infrastructure, and computer architecture. This section defines the QEC, compiler, and architecture terms used throughout this work. 

\subsection{Encoded Computation and Fault Tolerance}
A \emph{physical qubit} is a hardware degree of freedom subject to noise. A
\emph{logical qubit} is information protected by an error-correcting code. An
$\llbracket n,k,d \rrbracket$ stabilizer code embeds $k$ logical qubits in $n$ physical qubits.
Its commuting Pauli checks define the codespace, and its distance $d$ is the
minimum weight of an undetectable nontrivial logical Pauli. A 
distance-$d$ code detects up to $d-1$ errors and corrects up to
$\lfloor(d-1)/2\rfloor$~\cite{gottesman1997stabilizer}. Measuring the checks
produces a \emph{syndrome}: classical information about errors that reveals no
protected logical state. We use \emph{QEC round} or \emph{code cycle} for one
architecture-defined round of check extraction and associated processing.
Below a threshold defined for a particular circuit, noise model, and decoder,
increasing the code size can suppress logical failure
~\cite{dennis2002topological}.

Fault tolerant execution also requires encodings, live blocks, gadgets, protocols, 
and resource factories. An \emph{encoding} fixes a
compiler-facing interface to the code, while an encoded \emph{block} (or a
geometric \emph{patch}) is a live instance that carries logical data. A
\emph{gadget}, following Ref.~\cite{kliuchnikov2026composing, kliuchnikov2023stabilizer, WuDeq, microsoft2026deq, beverland2024FT}, pairs a
\emph{logical action}---an ideal stabilizer circuit such as preparation, storage,
measurement, or a logical gate---with a physical \emph{realization} that
implements it on zero or more input and output code blocks, together with an
affine \emph{outcome map} from realization outcomes to logical-action outcomes. The
logical action must equal the realization sandwiched between the encoders and
unencoders of its input and output codes, and the unencoders' syndromes must be
determined by the realization outcomes and the input
syndromes~\cite{kliuchnikov2023stabilizer}. Limiting how physical faults spread
is a property one verifies of a gadget, not part of its definition. A
\emph{protocol} composes gadgets, classical decisions, retries, and resource
flow into a larger behavior; when the classical control is fixed, a protocol is
itself a composite gadget. Lattice surgery, for example, implements logical
operations by changing measured checks and jointly measuring patch boundaries
~\cite{horsman2012lattice}. Universal schemes often consume non-Clifford
resource states made by a repeated preparation or distillation protocol, which
we call a \emph{factory}~\cite{litinski2019game}. 

\subsection{Detectors, Observables, and Decoding}
Raw check measurements become useful only after their interpretation is fixed.
A \emph{detector} is a parity of measurement results with a deterministic value
in a fault-free circuit. A \emph{logical observable} is a parity that records
a protected program result. Flipping it constitutes a logical failure. A \emph{detector error model}
(DEM) expresses each modeled elementary fault by its probability and the detectors
and observables it flips. It is therefore a classical interface between a noisy QEC
circuit and a decoder \cite{gidney2021stim, derks2025dem}.

Detectors do not respect gadget boundaries: a detector typically compares
syndromes extracted by different gadgets. Ref.~\cite{kliuchnikov2026composing},
following \texttt{deq}~\cite{microsoft2026deq,WuDeq} and \cite{beverland2024FT}, restores gadget locality by
inserting \emph{virtual} stabilizer measurements at gadget boundaries. A \emph{code presentation} fixes the, possibly over-complete,
stabilizer generators measured at a boundary and thereby the \emph{virtual
syndromes} that cross it. A \emph{gadget detector contract} fixes the code presentations of a
gadget's input and output codes, its \emph{virtual detectors}, which may
involve incoming virtual syndromes, and its \emph{virtual output syndrome map},
an affine map expressing each outgoing virtual syndrome in terms of realization
outcomes and incoming virtual syndromes. The resulting \emph{extended detector
error model} (EDEM) generalizes the DEM to a gadget with open boundaries: a
linear map over $\mathbb{F}_2$ from elementary faults, incoming
virtual-syndrome flips, and incoming boundary Pauli errors to detector and
observable flips and the corresponding outgoing data. EDEMs compose as gadgets
do, provided connected ports carry matching codes and code presentations, and wiring
them together recovers the DEM of the composite circuit. This paper represents
codes, code presentations, gadgets, gadget detector contracts, and EDEMs as typed compiler
artifacts; their mathematics is developed in Ref.~\cite{kliuchnikov2026composing}.

A \emph{decoder} maps detection events to a recovery class or logical-frame update. 
It need not reconstruct the exact physical fault, but it must agree with the circuit's 
detector and observable conventions. Decoder throughput and latency also become system 
constraints when a later operation depends on the result \cite{higgott2025sparseblossom}. Because the decoder depends on measurement order, 
boundary conditions, and record identity, a transformation must preserve them or 
rebuild the decoder interface. 

\subsection{Intermediate Representations and Lowering}
An \emph{intermediate representation} (IR) is the compiler's typed form of a 
program between source and machine-specific executable code. In this work we make use of 
the Multi-Level IR (MLIR) framework for flexible construction, composition, and lowering of domain-specific typed IR structure and semantics \cite{lattner2021mlir}. In MLIR, 
a \emph{dialect} defines a namespace of domain specific operations, types, attributes, interfaces, and verification rules. Multiple dialects may coexist in one module. This supports a family of representations at different abstraction levels rather than one universal instruction set. A \emph{pass} analyzes or transforms IR within a level. A \emph{lowering} translates between levels while preserving established meaning and introducing facts owned by the destination. MLIR often terms these \emph{conversions} and \emph{translations} -- lowering within MLIR and lowering out to other code formats, respectively. We use \emph{progressive lowering} for a sequence in which each step adds target commitments without silently revising earlier ones.

Static single assignment (SSA) associates every value with one definition and
makes dependencies available through def--use chains. Quantum IRs adapt this
idea to the no-cloning and no-deleting constraints. Value-oriented forms expose
quantum dataflow by making an operation consume and produce quantum values,
whereas reference-oriented forms encode state changes through side effects.
Existing IRs make different choices between these models
~\cite{peduri2022qssa,ittah2022qiro,cudaqteam_quake}. In this paper,
\emph{value semantics} refers to explicit dependency and ownership flow, not to
permission to copy a quantum value.

A \emph{target} states which artifact form and capabilities a downstream system can
consume. A \emph{verifier} rejects an IR object whose invariants or required
capabilities are not satisfied. A physical \emph{event graph} makes operations,
resources, and precedence constraints explicit. A \emph{schedule} additionally
assigns events to times and resolves resource conflicts. Finally,
\emph{provenance} records the source program, target, transformations,
parameters, and evidence that produced an artifact. These terms separate
semantic refinement from optimization: a lowering may be correct without being
cost-optimal, and a valid schedule may be complete without accurately modeling
hardware performance.

\subsection{Architectures and Estimates}
A fault-tolerant architecture specifies how a logical computation is encoded,
supplied with resources, moved, executed, decoded, and controlled. It spans the
logical instruction set, the organization of storage and communication,
gadgets and factories, physical connectivity and timing, and the classical
control path. A \emph{resource estimate} is a conditional analysis of a
particular workload and architecture under stated assumptions, not an intrinsic
property of an algorithm. Logical-operation counts, analytical QEC costs, and
schedule-derived occupancy and runtime answer different questions at different
levels of refinement; their assumptions and metric definitions must therefore
remain visible~\cite{beverland2022assessing,gidney2021factoring}.

A \emph{logical architecture} describes the code-independent organization of
logical capacity, services, communication, and resource-state demand. It states
where logical computation and resources may reside and how they may flow.
A QEC binding then selects encodings and fault-tolerant operations, while
physical materialization allocates carriers and determines how communication
and computation are executed.

% We distinguish four classes of evidence, each supporting a different kind of
% claim. Formal evidence, including algebraic checks of code and gadget
% properties and type verification, supports mathematical and structural claims.
% Compilation evidence from verified lowerings and schedules establishes that
% the required work has been represented under a stated machine model. Simulation
% and decoding evidence characterizes behavior under stated circuit, noise, and
% decoder assumptions. Execution evidence records the behavior of a particular
% compiled artifact on a simulator, emulator, or device. No class can substitute
% for another because each answers a different question. Keeping them distinct
% preserves the meaning of each claim; explicit contracts between adjacent
% representations provide continuity by recording what was preserved, what was
% added, and which evidence justified the transition.

% For compiler purposes, QEC records are semantic interfaces rather than
% auxiliary annotations, and an IR level is a contract about meaning rather than
% merely a serialization format. Every reported cost remains conditional on a
% workload, success requirement, operating point, and degree of compiler
% refinement.

%% file: sections/system-overview.tex
\section{CUDA-Q Logical System Overview}
\label{sec:overview}

CUDA-Q Logical compiles one logical program into an \emph{immutable build}. 
As the build moves through P0--P4, it acquires a logical-machine placement, QEC 
binding, physical schedule, and control plan while retaining earlier artifacts. 
It is this semantic layering that enables fault-tolerant retargetability of a logical program. 

\subsection{System Architecture and Context}
As Figure~\ref{fig:qlx-compiler-spine} shows, CUDA-Q Logical sits between existing 
logical application frontend output IR (e.g., CUDA-Q kernel IR) and existing simulation, submission, and real-time
infrastructure. Above it, application kernels, libraries, transformations, and core
IR express and optimize the logical computation. CUDA-Q Logical exposes P0,
P1, and P2 as programmable entry points, so compilation can begin from logical
intent, a logical-machine description, or a concrete QEC binding. Below it,
verified artifacts flow into target lowering, simulation, submission,
and execution.

P0 accepts a CUDA-Q program, represented in the Quake MLIR dialect,
or an imported logical workload such as a \texttt{@cudaq.logical.program} kernel 
or some other frontend IR. It emits a target-independent logical IR instance. 
P1 combines that program with a logical virtual
machine and records code-independent resources, capabilities, residency,
communication, resource-state demand, and a logical schedule. Its output does
not yet select QEC codes, encoded patches, physical carriers, or native timing.
P2 binds the placed operations to codes, encodings, blocks, gadgets,
and resource-state generation protocols. It produces code-specific QEC
microcode together with detector, observable, and retry semantics.
P3 maps that microcode to physical resources, native events, timing,
noise (for simulation or emulation), and an execution schedule. When required by the endpoint, P4
binds the scheduled work to instruction delivery, syndrome transport,
decoding, and feedback. Each lowering checks that the new artifact preserves
the earlier program.

Researchers, developers, and QPU builders can supply their own codes, gadgets, machine descriptions,
schedulers, decoders, and control implementations through the corresponding
typed interfaces. These choices become part of the build while leaving the
application unchanged. 

\subsection{The Logical Virtual Machine}
\label{sec:logical-virtual-machine}
The \emph{logical virtual machine} (LVM) is the code-independent machine model
used by P0-to-P1 lowering. It describes where logical state may reside, which
logical operations are available, and how state and resource demand may move
through the machine. Here, \emph{virtual} denotes a machine contract that
several codes, physical layouts, and controllers can implement.

\begin{wrapfigure}{r}{0.48\textwidth}
  \centering
  \includegraphics[width=\linewidth]{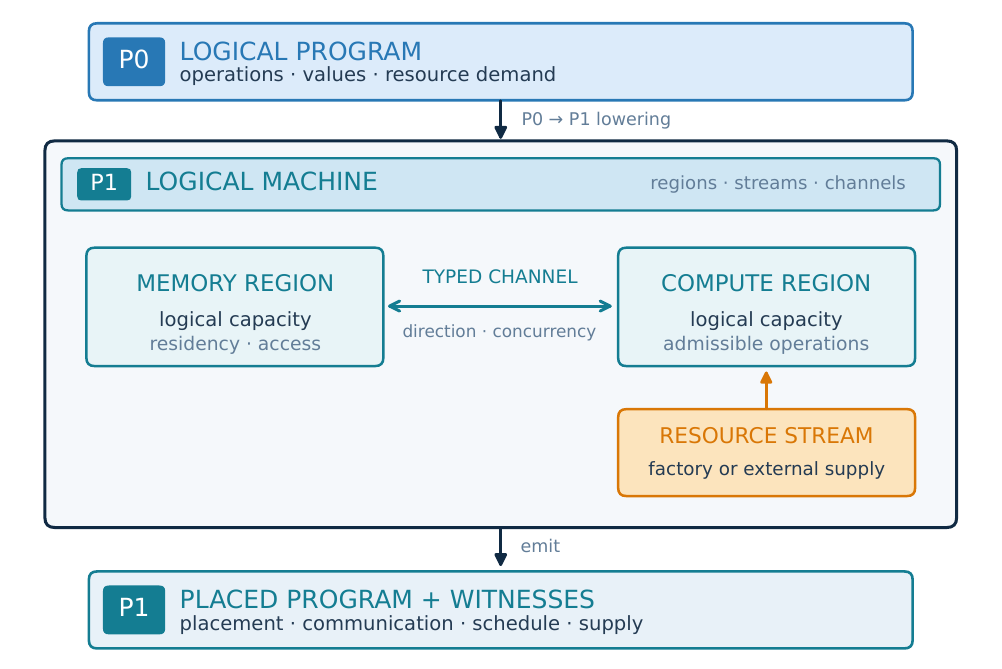}
  \caption{The logical virtual machine for CUDA-Q. Regions expose code-independent
    logical capacity and capabilities, streams expose resource-state supply,
    and channels constrain admissible flow. Encodings and physical routes are
    deliberately absent.}
  \Description{A compact logical-machine diagram places a P0 logical program
    above memory and compute regions connected by a channel. A resource stream
    backed by a factory or external supply feeds the compute region. The output
    is a P1 placed logical program with placement and supply witnesses.}
  \label{fig:logical-virtual-machine}
\end{wrapfigure}
A researcher describes logical capacity with regions, resource supply with
streams, and allowed movement with channels. P2 and P3 later choose the
encodings and physical routes. A \emph{region} is a typed pool of logical
capacity. Compute regions admit active operations, memory regions prioritize
residency, and specialized regions advertise capabilities such as
measurement. Their capacities are expressed in logical units rather than
encoded blocks or physical carriers.

A P1 factory describes an abstract resource supply. Its stream records the
resource kind, availability, and buffering visible to the program. The code,
preparation or distillation protocol, and physical factory that implement the
stream are selected later. An external stream can instead declare that
production remains outside the modeled device.

A \emph{channel} connects regions or streams and specifies the logical
transfers and interactions they support. Direction and concurrency may
constrain placement and logical scheduling, but the channel does not define a
physical route. Physical movement, ancillas, contention, and duration enter
when P2 and P3 implement the communication.

P0-to-P1 lowering maps logical values and operations to region slots, routes
code-independent transfers through channels, and matches resource demand to
streams. The resulting program records logical residency, communication,
scheduling, and resource supply. Researchers can reuse this machine model
across workloads before choosing code-specific gadgets and protocols.

\subsection{QEC Microcode: Codes, Gadgets, and EDEMs}
\label{sec:qec-microcode}

\begin{figure}[t]
  \centering
  \begin{tikzpicture}[
      x=1cm, y=0.8cm,
      box/.style={draw, thick, rounded corners=2pt, minimum width=2.4cm,
        minimum height=1.9cm, fill=blue!4, font=\small\sffamily},
      port/.style={font=\scriptsize\ttfamily},
      wire/.style={-{Latex[length=1.8mm]}, thick},
      fault/.style={-{Latex[length=1.8mm]}, thick, densely dashed},
    ]
    % Single EDEM
    \node[box] (g) at (0,0) {EDEM};
    \draw[fault] (0,2.0) node[above, port] {F} -- (g.north);
    \draw[wire] (-2.4,0.5) node[left, port] {E$_{\mathrm{in}}$} -- (g.west |- 0,0.5);
    \draw[wire] (-2.4,-0.5) node[left, port] {$\Delta$V$_{\mathrm{in}}$} -- (g.west |- 0,-0.5);
    \draw[wire] (g.east |- 0,0.5) -- (2.4,0.5) node[right, port] {E$_{\mathrm{out}}$};
    \draw[wire] (g.east |- 0,-0.5) -- (2.4,-0.5) node[right, port] {$\Delta$V$_{\mathrm{out}}$};
    \draw[wire] (g.south -| -0.5,0) -- (-0.5,-2.0) node[below, port] {$\Delta$D};
    \draw[wire] (g.south -| 0.5,0) -- (0.5,-2.0) node[below, port] {$\Delta$O};
    % Composition
    \begin{scope}[xshift=8.6cm]
      \node[box, minimum width=2.0cm] (a) at (-2.1,0) {sec$_i$};
      \node[box, minimum width=2.0cm] (b) at (2.1,0) {sec$_{i+1}$};
      \draw[fault] (-2.1,2.0) node[above, port] {F$_i$} -- (a.north);
      \draw[fault] (2.1,2.0) node[above, port] {F$_{i+1}$} -- (b.north);
      \draw[wire] (-4.3,0.5) node[left, port] {E} -- (a.west |- 0,0.5);
      \draw[wire] (-4.3,-0.5) node[left, port] {$\Delta$V} -- (a.west |- 0,-0.5);
      \draw[wire] (a.east |- 0,0.5) -- (b.west |- 0,0.5);
      \draw[wire] (a.east |- 0,-0.5) -- (b.west |- 0,-0.5);
      \draw[wire] (b.east |- 0,0.5) -- (4.3,0.5) node[right, port] {E};
      \draw[wire] (b.east |- 0,-0.5) -- (4.3,-0.5) node[right, port] {$\Delta$V};
      \draw[decorate, decoration={brace, amplitude=3pt}, thin]
        (-1.0,0.85) -- (1.0,0.85) node[midway, above=3pt, port] {boundary\_for};
      \draw[wire] (a.south) -- (-2.1,-2.0) node[below, port] {$\Delta$D$_i$};
      \draw[wire] (b.south) -- (2.1,-2.0) node[below, port] {$\Delta$D$_{i+1}$};
    \end{scope}
  \end{tikzpicture}
  \caption{Left: an EDEM as a linear map over $\mathbb{F}_2$. Inputs are the
    elementary-fault configuration F, the profiled boundary Pauli error
    E$_{\mathrm{in}}$ (syndrome and logical-basis lanes), and virtual-syndrome
    flips $\Delta$V$_{\mathrm{in}}$; outputs are detector flips $\Delta$D,
    observable flips $\Delta$O, and the outgoing boundary data. Right: two
    syndrome-extraction instances wired through \texttt{boundary\_for}; the
    contraction of the wired diagram is the DEM of the composite circuit.}
  \Description{Two block diagrams. The left shows one box labeled EDEM with a
    fault input on top, two boundary inputs on the left, two boundary outputs
    on the right, and detector and observable outputs below. The right shows
    two such boxes in series with their boundary wires connected.}
  \label{fig:edem-ports}
\end{figure}
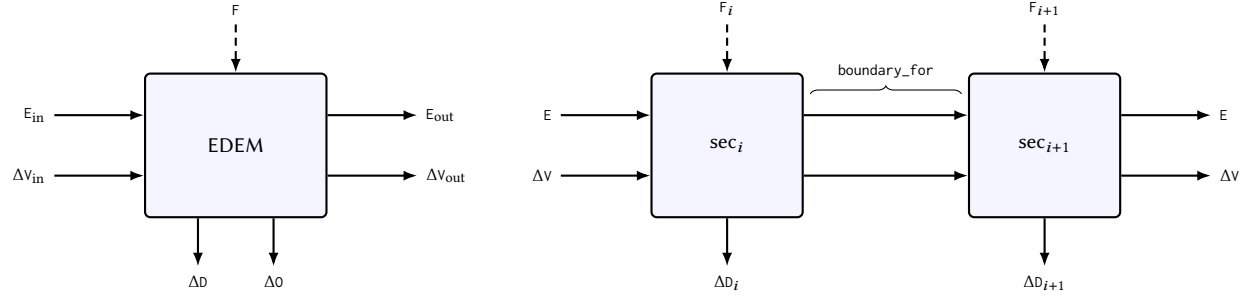

Once P1 has placed a logical program on an LVM, the compiler knows where
logical state resides and which interactions are admissible, but not how logical state 
is encoded, how logical actions are realized, or which physical resources execute
them. P2 binds the placed program to a QEC architecture and produces 
\emph{QEC microcode}. The underlying P2 object model follows the gadget model of
Ref.~\cite{kliuchnikov2026composing}, and summarized as follows:

\begin{description}[leftmargin=1.2em, itemsep=0.15em]
\item[Code and code presentation] (\texttt{Code}, \texttt{CodePresentation}).
  A code defines the stabilizer algebra. A code presentation selects the
  potentially over-complete generators whose signs form the virtual syndrome
  at a block boundary. An \texttt{Encoding} binds a code and presentation to named
  logical ports, while a \texttt{patch} represents one live, linearly owned
  encoded block.

\item[Gadget] (\texttt{@cudaq.logical.gadget}).
  A gadget declares its logical action with \texttt{implements=} and realizes that
  action with a typed body over patches. The compiler infers an outcome
  map from the returned Boolean results. Measurements in the body become
  records with stable identities.

\item[Gadget detector contract] (\texttt{GadgetDetectorContract}).
  A gadget detector contract defines virtual detectors over records and incoming
  virtual syndromes, together with the virtual output syndrome map
  (\texttt{boundary=}). Contracts are separate from realizations, so an
  analysis can select a different contract without changing the gadget body.

\item[EDEM] (\texttt{ExtendedDetectorModel}).
  An extended detector error model (EDEM) is the gadget's linear map over $\mathbb{F}_2$ from elementary
  faults and incoming boundary data to detector flips, observable flips, and
  outgoing boundary data (Figure~\ref{fig:edem-ports}). It contains no fault
  probabilities; operating points are bound after composition.

\item[Protocol] (\texttt{@cudaq.logical.protocol}).
  A protocol composes gadget invocations and manages repetition, retry,
  scratch resources, and resource flow. With fixed control, it can be treated
  as a composite gadget.

\item[Architecture] (\texttt{QECArchitecture}).
  A QEC architecture defines the available codes, presentations, gadgets, detector contracts,
  protocols, and providers. P1-to-P2 lowering selects among them by
  logical action, code, and encoding, using a deterministic tie-break.
\end{description}

During P1-to-P2 lowering, the compiler matches each placed operation to a
compatible gadget or protocol in the selected QEC architecture. It
instantiates the chosen definition with typed patches and measurement records
and records the choice in the build. Providers can contribute their own codes,
presentations, detector contracts, and gadgets through the same interface. Lowering fails if the
architecture supplies no compatible implementation or omits information
required to check it (Section~\ref{sec:ZSZ demos}).

The compiler can check a selected gadget with a builtin
\texttt{verify\_gadget} operation. A single Choi-state probe of the realization tests both
parts of the gadget definition. The detector-contract check confirms that the virtual
detectors are deterministic and that the output syndromes follow from the
measurement records and input syndromes. The logical-action check confirms that
the realization implements the declared logical operation when its results
are interpreted through the outcome map
(Section~\ref{sec:gadget-study}).

\texttt{DetectorModelBuilder} composes gadget EDEMs by connecting their
boundary data and contracting the resulting network. This contraction
produces the exact detector error model for the supplied components and keeps
it linked to the original P0 logical action. Compared with the reference
implementation of Ref.~\cite{kliuchnikov2026composing}, CUDA-Q Logical
preserves record identities through inlining and lowering, represents detector contracts
as independently selectable objects, and allows EDEMs to be used outside the
compiler. A gadget can also provide a realization without a complete boundary
detector contract. The externally supplied surgery gadgets in
Section~\ref{sec:ZSZ demos}, for example, can be compiled and verified against
their logical actions, but they cannot yet participate in EDEM composition.

\subsection{Physical Machine and Execution}
\label{sec:qec-microcode-physical-machine}

The \emph{physical machine} supplies the architectural substrate for P3. It
describes resource classes and capacities, topology and connectivity, native
actions and instruments, and an operating point containing timing, noise,
calibration, and cost assumptions. Like the LVM, this machine description is
reusable across workloads. It does not contain the allocation, route, or
schedule of one compiled program. Those workload-specific commitments belong
to the P3 physical event graph derived from it.

P2-to-P3 lowering projects the selected QEC microcode onto that machine. It
maps formal physical qubit carrier and workspace roles to concrete physical allocations,
expands QEC operations into topology-legal preparation, movement, interaction,
measurement, reset, wait, and release events, and preserves the record and
detector semantics established at P2. Mapping, routing, record-projection, and
provenance witnesses explain which physical subgraph implements each source
operation. When timing is required, scheduling assigns starts and durations
subject to dependencies and resource exclusion.

The resulting P3 artifact lets the compiler reason about physical occupancy,
movement, contention, critical paths, native-event rates, makespan, and
detector-production bandwidth. These quantities are derived from an explicit
physical materialization, but their accuracy remains conditional on the machine
model and operating point supplied to the build. Lowering adds physical detail;
it does not by itself establish calibration accuracy, optimality, or physical
fault tolerance.

Separating P2 microcode from the P3 machine is central to retargetability. The
same placed logical program may be bound to different codes and protocols,
and the same QEC microcode may be materialized on different physical
machines. Each resulting build records the selected definitions and lowering
witnesses, so differences can be attributed to explicit architectural choices
rather than to an unrecorded rewrite of the workload. The object model below
packages these reusable descriptions, derived artifacts, and witnesses into an
inspectable compilation workflow.

% To tie this discussion on LVM, QEC Microcode, and Physical Machines together, 
% consider a simple nondestructive measurement of $X\otimes Y\otimes Z$. This operation 
% reaches P1 as a placed logical action with stable operands and communication intent. P2 selects 
% compatible encodings and a product-measurement protocol, producing patches, typed 
% records, detector semantics, and selection evidence. P3 then assigns physical carriers 
% and expands that protocol into topology-legal events and, when required, a verified 
% schedule. Across these lowerings, the observable, operand order, outcome meaning, 
% and ownership remain fixed; only the commitments needed to execute them are added.

\subsection{Architecture Construction Model}
Listing~\ref{lst:core-workflow} constructs a device description, compiles a 
workload, and estimates the resulting build. The example binds a logical
compute region to the Pinnacle GB30 high-rate qLDPC architecture and compiles
through P2. Builders enforce cross-field invariants and produce immutable descriptions that can be safely shared between compiler passes and external tools. 
Handles preserve the identities and relationships of programs, devices,
codes, architectures, and targets across compilation stages, allowing derived artifacts and analyses to be traced back to the specific inputs and decisions that produced them.
Compiler passes derive new builds rather than mutating existing ones, while reports are explicitly associated with the build and compilation parameters they analyze.
Together, these mechanisms support reproducibility, provenance tracking, and attribution across the compilation workflow.
Appendix~\ref{sec:construction} defines this object
model in detail.

The workload, architecture binding, compilation request, and analysis request
remain separate objects. Changing the architecture produces a new build
without changing \texttt{pbc\_workload}, so the results can be compared and
replayed. The build records the definitions and providers selected during
compilation.
% \begin{wrapfigure}{r}{0.56\textwidth}
% \begin{minipage}{\linewidth}
\begin{lstlisting}[
style=qlx-python,
aboveskip=0pt,
belowskip=0pt,
caption={Compiling and estimating a QEC-aware build.},
label={lst:core-workflow}
]
import cudaq.logical as ql
from ql.architectures import pinnacle

# Define a fault-tolerant architecture programmatically.
builder = ql.devices.DeviceBuilder("PinnacleStudy")
# Construct the P1 logical virtual machine.
compute = builder.logical.add_compute(capacity=8)
# Add P2 QEC semantics.
builder.qec.bind(compute, architecture=pinnacle.gb30)
# Freeze the architecture description.
device = builder.build()

# Compile the logical workload through P2.
build = ql.compile(pbc_workload, pipeline=ql.compiler.pipelines.qec(), device=device)

# Estimate resources from the resulting build.
estimate = ql.analysis.estimate(build, tier=ql.analysis.Tier.STATIC)
\end{lstlisting}
% \end{minipage}
% \end{wrapfigure}

\subsection{Resource Analysis Across the Compiler Spine}
\label{sec:overview-resource-analysis}
Resource estimates in CUDA-Q Logical are reports over compiler artifacts rather than standalone analyses. 
The source artifact determines both the information available to the analysis and
the claims that its result can support. 
An estimate of a \profile{P0} program,
for example, may describe logical demand but cannot claim a physical-qubit
count. Because each report retains the identity of its source build, it also
retains the program, device choices, QEC binding, and other assumptions
already fixed during compilation.

\begin{table*}[b]
  \caption{Resource analyses available at each semantic profile and from
  provenance-linked observations. Some profiles support more than one
  analysis; samples and run measurements provide statistical evidence.}
  \label{tab:estimation-by-profile}
  \small
  \setlength{\tabcolsep}{3pt}
  \begin{tabular}{@{}p{0.065\textwidth}p{0.15\textwidth}p{0.225\textwidth}p{0.255\textwidth}p{0.235\textwidth}@{}}
    \toprule
    \raggedright Source & \raggedright Analysis
         & \raggedright Authoritative input
         & \raggedright Public result or evidence
         & \raggedright Question answered \tabularnewline
    \midrule
    \profile{P0} & \raggedright Logical demand
      & \raggedright Verified target-independent logical program
      & \raggedright \texttt{LogicalProfile}: actions, instruments, liveness,
        depth, synthesis, and resource demand
      & \raggedright What does the algorithm demand before selecting a device
        or QEC binding? \tabularnewline
    \midrule
    \profile{P1} & \raggedright Placement and capacity
      & \raggedright Verified logical-machine binding, placement witness,
        communication intent, and resource supply
      & \raggedright Witness-backed feasibility, capacity pressure,
        communication, and supply metrics retained by the build
      & \raggedright Can the code-independent logical workload inhabit the
        selected machine contract? \tabularnewline
    \midrule
    \profile{P2} & \raggedright Selected-QEC accounting
      & \raggedright Selected call graph and device closure
      & \raggedright \texttt{FabricCounts}: operations, gadgets, protocols,
        detectors, observables, rounds, and logical peak
      & \raggedright What QEC structure was selected and counted through
        folded control? \tabularnewline
    \midrule
    \profile{P2} & \raggedright Analytical projection
      & \raggedright The same selected closure plus physical-error, scaling,
        timing, failure-budget, acceptance, and evidence assumptions
      & \raggedright \texttt{FabricEstimate}: distance and evidence status,
        modeled error, qubits, time, retry demand, bottleneck, and assumptions
      & \raggedright What does the declared analytical model predict, and does
        it meet the stated budget? \tabularnewline
    \midrule
    \profile{P3} & \raggedright Schedule-derived accounting
      & \raggedright Verified physical schedule plus the exact analytical P2
        and device closure
      & \raggedright \texttt{ScheduleEstimate}: event counts, retry-aware time
        and occupancy, provisioned and active resources, utilization, and
        bottleneck
      & \raggedright What does dependency-, capacity-, and resource-checked
        physical execution require? \tabularnewline
    \midrule
    \profile{P3}/\profile{P4}
      & \raggedright Statistical evidence
      & \raggedright Provenance-bound sample, decoded observation, or run
        measurement
      & \raggedright \texttt{SampleResult}, \texttt{RunResult}, and
        \texttt{TwinEstimate}: acceptance-conditioned error, confidence and
        budget bounds, verdict, and decoder/control provenance
      & \raggedright What behavior was observed, with what uncertainty and
        evidence identity? \tabularnewline
    \bottomrule
  \end{tabular}
\end{table*}

Different costs become visible as the program moves through the compiler.
\profile{P0} exposes logical operations, ownership lifetimes, folded control,
and synthesis demand. \profile{P1} adds the placement, capacity, communication,
and supply constraints of the logical machine. Once a QEC binding has been
selected, \profile{P2} exposes the chosen codes, gadgets, protocols, factories,
detectors, and observables. It can also support analytical projections when
the analysis is given an explicit physical-error model, scaling law, timing
model, and failure budget. A verified \profile{P3} schedule adds costs caused
by dependencies, movement, overlap, and contention. When execution requires
realtime decoding or feedback, \profile{P4} makes controller load, detector
transport, queueing, decoder throughput, and deadlines available for analysis.

Listing~\ref{lst:overview-analysis-execution} illustrates this progression.
Each call to \texttt{ql.estimate} observes a compiler artifact rather than the
result of an earlier estimate. The logical, QEC-aware, and scheduled reports
are nevertheless comparable because their source artifacts belong to the same
compilation lineage. A difference between the analytical \profile{P2} cost and
the scheduled \profile{P3} cost can therefore be traced to work introduced or
made explicit during physical compilation, such as protocol expansion,
factory startup, routing, or resource contention.

A later report has a narrower and more concrete scope, but it is not
automatically a better prediction of hardware behavior. Physical error rates,
operation times, scaling laws, and producer models remain assumptions until
they are calibrated or measured. Samples and run measurements provide evidence
about those assumptions while retaining the build that produced them. They do
not form another semantic stage. Table~\ref{tab:estimation-by-profile}
summarizes the quantities supported at each compiler boundary.

\begin{figure}[t]
\begin{minipage}{\columnwidth}
\begin{lstlisting}[
style=qlx-python,
aboveskip=0pt,
belowskip=0pt,
caption={Resource analysis and offline sampling over one compilation lineage.},
label={lst:overview-analysis-execution}
]
import cudaq.logical as ql
budget = ql.estimate.FailureBudget(0.01)

p0 = ql.compile(program, pipeline=ql.compiler.pipelines.logical())
p2 = ql.compile(p0, pipeline=ql.compiler.pipelines.qec(), device=device)
p3 = ql.compile(p2, pipeline=ql.compiler.pipelines.physical(), device=device)
schedule = ql.compiler.schedule(p3)

logical = ql.estimate(p0)
qec_cost = ql.estimate(p2, p_phys=1e-3, failure_budget=budget)
scheduled_cost = ql.estimate(schedule, p_phys=1e-3, failure_budget=budget)

samples = ql.sample(schedule.build, target=ql.targets.stim, shots=10_000)
\end{lstlisting}
\end{minipage}
\Description{Python code that compiles one program through the logical, QEC,
and physical stages, estimates resources from artifacts in that lineage, and
samples the resulting feedback-free physical build.}
\end{figure}

\subsection{Execution Endpoints}
\label{sec:overview-execution}

Execution begins from the same verified artifacts used for resource analysis.
A P3 build fixes the physical operations, records, dependencies,
noise bindings, and schedule that an endpoint must execute. If a decoded or
otherwise computed classical value affects later work in the same shot, a
P4 artifact also fixes the corresponding controller, decoder,
transport, and feedback obligations. The execution endpoint implements these
commitments; it cannot choose a different code, reinterpret detector records,
or construct physical work missing from the build.

The build, target, and launch request have distinct roles. The build states
what must be executed. The target states which observations it can produce
from that class of artifact, such as circuit emission, offline sampling, or
in-shot execution. The launch request selects an observation and supplies
invocation-specific parameters such as the number of shots. This separation
permits several targets to consume the same compiler result without moving
architecture or QEC decisions into backend-specific code.

The final line of Listing~\ref{lst:overview-analysis-execution} shows the
feedback-free case. The Stim target samples the scheduled physical build, so
the samples retain the event ordering, noise model, detector interpretation,
and provenance fixed by that build. This path is sufficient when decoding is
performed after the physical instruction stream has completed. If a decoded
value controls a later operation, offline sampling is no longer equivalent:
the relevant records must reach a decoder, and its correction must be consumed
before dependent physical work proceeds. Such a program requires an execution
target that supports the corresponding P4 control contract.

The strength of an execution claim depends on the endpoint. Local simulation
can show that physical events, detector records, decoding, and feedback occur
in the required causal order. It does not establish hardware latency,
transport jitter, queue capacity, or sustained decoder throughput. Those
claims require measurements or verified bounds from the deployed controller,
transport, and decoder implementations. CUDA-Q Logical records this distinction
by attaching the source build, target identity, launch conditions, and
available evidence to every emitted artifact, sample, and run result.

%% file: sections/evaluation.tex
\section{Evaluation}
\label{sec:evaluation}

The evaluation examines three compiler capabilities. Schedule-derived resource
analysis should recover established results while exposing the synthesis and
scheduling decisions that determine them. Exact composition at the P2 boundary
should allow independently developed gadgets, detector models, and codes to
enter the compiler without losing their declared semantics. Retargeting should
preserve one P0 program while code and provider choices produce different
verified realizations. The studies use algebraic checks, compiler verification,
simulation, and sampling. They do not include hardware execution or establish
circuit distance, thresholds, or physical fault tolerance. The compiler and
supporting artifacts are planned for public release.

\subsection{Schedule-Derived Resource Analysis and Attribution}
\label{sec:evaluation-schedule-attribution}

We evaluate resource analysis on two markedly different fault-tolerant
architectures. The RSA-2048 study tests whether a compiled
schedule can recover an independently published surface-code estimate and
retain the structure needed to identify its limiting regimes. The Pinnacle
study tests whether explicit synthesis and scheduling can account for the gap
between a qLDPC architecture's analytical model and its compiled realization.
In both cases, the estimate remains an observation of the compiler artifact
that produced it.

\input{sections/evals/rsa-gidney-ekera-v2}
\input{sections/evals/pinnacle-FH}

\subsection{Exact QEC Semantics and External Extensibility}
\label{sec:evaluation-qec-composition}

We evaluate the P2 contract from complementary directions. The
gadget study checks a realization against its declared logical action and
detector contract, then asks whether extended detector error models compose to
the same model as the complete circuit. The ZSZ study supplies a qLDPC code and
graph-surgery gadgets constructed outside CUDA-Q Logical and follows them
through compiler selection, verification, scheduling, and circuit emission.
The shared question is whether QEC components can remain independently
developed while entering a common compilation path with their semantics
intact.

\input{sections/evals/gadgets}

\input{sections/evals/zsz-gadgets}
\input{sections/evals/pinnacle-zsz-retargeting}

%% file: sections/evals/rsa-gidney-ekera-v2.tex
\subsubsection{RSA-2048 Gidney--Eker{\aa} selected-point reconstruction}
\label{sec:results-rsa2048}
\begin{lstlisting}[
  style=qlx-python,
  float=!b,
  % basicstyle=\scriptsize\ttfamily,
  aboveskip=0pt,
  belowskip=0pt,
  caption={Abridged P0 RSA-2048 workload; study-local QROM, carry-runway, allocation, and cleanup helpers are elided.},
  label={lst:ge-qlx-p0}
]
import cudaq.logical as ql 

def lookup_addition(_iteration, *live):
    live = qrom_lookup(live, rows=1_023)
    live = carry_runway_add(live, piece_length=1_062)
    return qrom_unlookup(live, accesses=64)

@ql.program
def rsa2048_resource_program() -> None:
    live = allocate_rsa_workspace(register_length=2_124, address_width=10, carry_pieces=2)
    live = ql.ops.repeat(505_965, carries=live, body=lookup_addition)
    measure_and_release(live)
\end{lstlisting}
We compare CUDA-Q Logical against the parallel RSA-2048 operating point
selected by Gidney and Eker{\aa}~\cite{gidney2021factoring}.  This study does
not rerun their parameter search or execute a modulus-specific factoring
instance.  It asks a narrower question: when their selected arithmetic and
architectural parameters are represented as compiler-visible programs,
protocols, resources, and timing constraints, does the resulting P3 schedule
recover the published single-run resource point?  The imported parameters are
$n=2048$, $n_e=3029$, $c_{\mathrm{exp}}=c_{\mathrm{mul}}=5$,
$c_{\mathrm{sep}}=1024$, $\delta_{\mathrm{off}}=4$, level-1 and level-2 code
distances 15 and 27, 28 AutoCCZ factory lanes, a $1\,\mu\mathrm{s}$ surface-code
cycle, and a $10\,\mu\mathrm{s}$ reaction time.  None of these selections is
claimed as a compiler-discovered optimum.  In the CUDA-Q Logical model, one
AutoCCZ factory lane is an independently scheduled producer pipeline that
emits autocorrected CCZ resource states; multiple lanes operate in parallel to
increase the aggregate supply rate.
The released calculator used below is the ancillary file
\texttt{estimate\_costs.py} distributed with the Gidney--Eker{\aa}
publication.

\begin{table*}[b]
  \centering
  \caption{Reconciliation of the selected parallel RSA-2048 point.  The
    published column transcribes the rounded per-run and retry-adjusted values
    of Gidney and Eker{\aa}; the calculator column evaluates their released cost
    script at the same parameter tuple.  The final column reports the CUDA-Q Logical P3
    single-run schedule and, where marked, conditions that cost on the cited GE
    31\% retry-risk estimate.  The calculator is a machine-readable version of
    the published model, not independent experimental evidence.  Mqd denotes
    million-qubit-days.}
  \label{tab:ge-reconciliation}
  \small
  \setlength{\tabcolsep}{5pt}
  \renewcommand{\arraystretch}{1.12}
  \begin{tabular}{@{}lrrr@{}}
    \toprule
    Quantity
      & \shortstack{GE paper\\(rounded)}
      & \shortstack{GE released\\calculator}
      & \shortstack{CUDA-Q Logical P3\\schedule} \\
    \midrule
    Level-1 / level-2 distance
      & $15/27$ & $15/27$ & $15/27$ (claimed) \\
    GE CCZ factories / CUDA-Q Logical AutoCCZ lanes
      & 28 & 28 & 28 \\
    Per-lane factory footprint
      & not tabulated
      & \shortstack{$13\!\times\!7$ level-2 cells\\142,688 qubits}
      & \shortstack{$13\!\times\!7$ cells $+120$ injections\\142,808 qubits} \\
    Steady-state factory cadence
      & not tabulated & $5d_2=135$ cycles & 135 cycles \\
    Toffoli/CCZ count
      & $2.7\times10^9$
      & $2.698296\times10^9$
      & 2,698,311,345 \\
    Single-run time (h)
      & 5.1 & 5.046158 & 5.046228 \\
    Physical qubits
      & $20\times10^6$ & 19,248,768 & 19,252,128 \\
    Single-run volume (Mqd)
      & 4.1 & 4.0472 & 4.0479 \\
    GE retry-risk estimate
      & 31\% & 31\% & 31\% (GE-conditioned) \\
    Expected time with retries (h)
      & $0.31$ days ($7.4$ h) & 7.3133 & 7.3134 \\
    Expected volume with retries (Mqd)
      & 5.9 & 5.8655 & 5.8666 \\
    \bottomrule
  \end{tabular}
\end{table*}

Table~\ref{tab:ge-reconciliation} separates external agreement from internal
determinism.  The official ancillary calculation gives a 5.046158-hour
single-run time and 19,248,768 physical qubits; the compiled schedule gives
5.046228 hours and 19,252,128 qubits.  The time difference is 0.253 seconds
($0.0014\%$).  The 3,360-qubit difference is exactly 120 explicit raw-state
injection qubits in each of 28 CUDA-Q Logical factory lanes.  Gidney and
Eker{\aa}'s board convention counts the $13\times7$ level-2 patch cells of each
lane but not these injection leaves.  We retain the difference rather than
adding it to the reference result.  Their reported 31\% retry-risk estimate
changes the interpretation of the runtime: 5.1 hours is one run, while the
expected time with retries is approximately 7.3 hours, reported as 7.4 hours
in their table and rounded to 8 hours in the paper title.  The emitted P3
schedule still describes exactly one run: it contains no outer retry node.
Conditioning its 5.046228-hour cost on GE's cited risk gives
$E[N]=1/(1-0.31)=1.4493$ attempts and
$E[t]=5.046228/(1-0.31)=7.3134$ hours.  Applying the same factor to the
4.0479-Mqd single-run volume gives 5.8666 Mqd.  These are GE-conditioned CUDA-Q Logical
projections, not a CUDA-Q Logical-derived logical-error claim; as in the GE calculator,
they assume independent equal-cost attempts and omit reset and classical
verification overhead.

We next trace how the compiled quantities arise from the workload and
architecture.  At P0, the program represented by
Listing~\ref{lst:ge-qlx-p0} preserves the
helper-call graph and quantum-resource ownership dependencies that are lost
when the workload is represented only by total operation counts and qubit
requirements.  Within each lookup addition, \texttt{qrom\_lookup} visits 1,023
nonzero QROM rows.  The carry-runway addition then propagates carries within
two pieces using a forward majority (MAJ) sweep and computes the sum while
uncomputing the carry ancillas using a reverse unmajority-and-add (UMA) sweep.
Finally, \texttt{qrom\_unlookup} uses 64 Toffolis for measurement-based
uncomputation.  Each carry piece also includes a coset-padding term
$c_{\mathrm{pad}}$, which is derived from the problem parameters using the
Gidney--Eker{\aa} rule rather than selected independently:
\[
  c_{\mathrm{pad}}
  = \left\lceil \log_2(n^2 n_e) \right\rceil
    + \delta_{\mathrm{off}}
  = 34 + 4 = 38.
\]
Consequently, $c_{\mathrm{sep}}=1024$ produces two pieces of length
$1024+38=1062$.  The explicit helper graph contains
\[
  1023 + 2(1062) + 2(1061) + 64 = 5333
\]
Toffolis per lookup, or 2,698,311,345 over 505,965 folded lookups.  The released
Gidney--Eker{\aa} script uses a continuous uncomputation term and evaluates to
$2.698296\times10^9$; both calculations round to the published
$2.7\times10^9$, while their 0.000586\% difference remains visible in
Table~\ref{tab:ge-reconciliation}.

To derive the P3 schedule from this P0 workload, CUDA-Q Logical requires an
explicit physical architecture, including the production of resource states.
A detailed model of one factory lane contains eight 15-to-1 distillation
protocols whose outputs are used to produce a CCZ state, followed by an
autocorrection stage that emits an AutoCCZ resource.
Compiling this lane to P3 yields a 379-cycle startup latency, a 135-cycle
steady-state production interval, and a 142,808-qubit footprint.

\texttt{DeviceBuilder} instantiates 28 copies of this compiled lane model and
binds them to the application's factory resources.  Thus, factory capacity is
derived from the compiled lane schedule rather than entered as an independent
throughput constant.  At the same operating point, the Gidney--Eker{\aa}
calculator models one CCZ factory as a $13\times7$ array of 91 level-2 cells.
At distance 27, this corresponds to 142,688 physical qubits under the same
$2(d+1)^2$ patch convention, compared with 142,808 qubits in the CUDA-Q
Logical model.  The additional 120 qubits are the raw-state injection leaves
represented explicitly by CUDA-Q Logical.  The two factory boundaries are not
identical: the Gidney--Eker{\aa} calculator places autocorrection fixups and
routing in other board regions, while the CUDA-Q Logical lane includes its
nine-patch AutoCCZ ring.  The level-1 and level-2 distances are inputs from the
selected operating point; this study does not independently validate their
logical failure rates.  Listing~\ref{lst:ge-factory-device} constructs the
physical architecture by defining the compute and factory regions and binding
the P3-characterized factory model to the latter.

\begin{lstlisting}[
  style=qlx-python,
  % basicstyle=\scriptsize\ttfamily,
  aboveskip=0pt,
  belowskip=0pt,
  caption={Abridged layered construction of the RSA-2048 device from a compiler-characterized AutoCCZ factory.},
  label={lst:ge-factory-device}
]
import cudaq.logical as ql 

# Compile the factory schedule into a P3 model that produces AutoCCZ states
factory_model = ql.compiler.factory_model(factory_schedule, produces=ql.standard.AUTO_CCZ_STATE)

# Derive the physical footprint of one compiled factory lane 
FACTORY_LANES = 28
lane = factory_model.characterization
lane_footprint = ql.architecture.PhysicalFootprint(lane.physical_unit_kind, lane.physical_units,
    "compiled AutoCCZ P3 schedule")

# Build the Gidney-Ekera architecture / device 
# Declare logical compute and state factory regions
b = ql.devices.DeviceBuilder("GidneyEkeraArch")
compute = b.logical.add_compute(capacity=COMPUTE_PATCHES, name="compute")
factory = b.logical.add_factory(produces=ql.standard.AUTO_CCZ_STATE,
    via=surface_autoccz_factory, capacity=FACTORY_LANES, name="autoccz_factory")

# Bind QEC architectures and encodings to the logical regions
compute_qec = b.qec.bind(compute, architecture=architecture)
factory_qec = b.qec.bind(factory, encoding=CODE, block_capacity=9 * FACTORY_LANES)

# Define P3 level physical resources
compute_grid = b.physical.add_resources("surface_code_patch", COMPUTE_PATCHES, **_surface_resource_options(SURFACE))
routing_pool = b.physical.add_resources("surface_code_patch", 12, **resource_options)
factory_bank = b.physical.add_resources("autoccz_factory_lane", FACTORY_LANES,footprint=lane_footprint)

# Map each QEC region to its physical resource pool
b.physical.bind(compute_qec, to=compute_grid)
b.physical.bind(compute_qec.auxiliary_regions[0], to=routing_pool)
b.physical.bind(factory_qec, to=factory_bank, factory_model=factory_model)

# Supply timing and calibration data needed by the physical model 
b.physical.set_operating_point(timing=_surface_timing(), calibration=...)

# Build the architecture
device = b.build()
\end{lstlisting}

Using this device, the compiler first combines the 135-cycle production
interval of one compiled lane with the 28 parallel lane instances, giving an
aggregate AutoCCZ supply interval of
$135/28=4.821429\,\mu\mathrm{s}$.  It then schedules each lookup against this
supply and the reaction- and code-depth constraints, which set the one-lookup
period to 35.9045 ms.  Because successive lookup iterations do not overlap,
the scheduler represents all 505,965 iterations as one folded repeat and
computes the one-run makespan from the startup, the repeated lookup periods,
and the final measurement.

\begin{figure*}[t]
  \centering
  \includegraphics[width=0.47\textwidth]{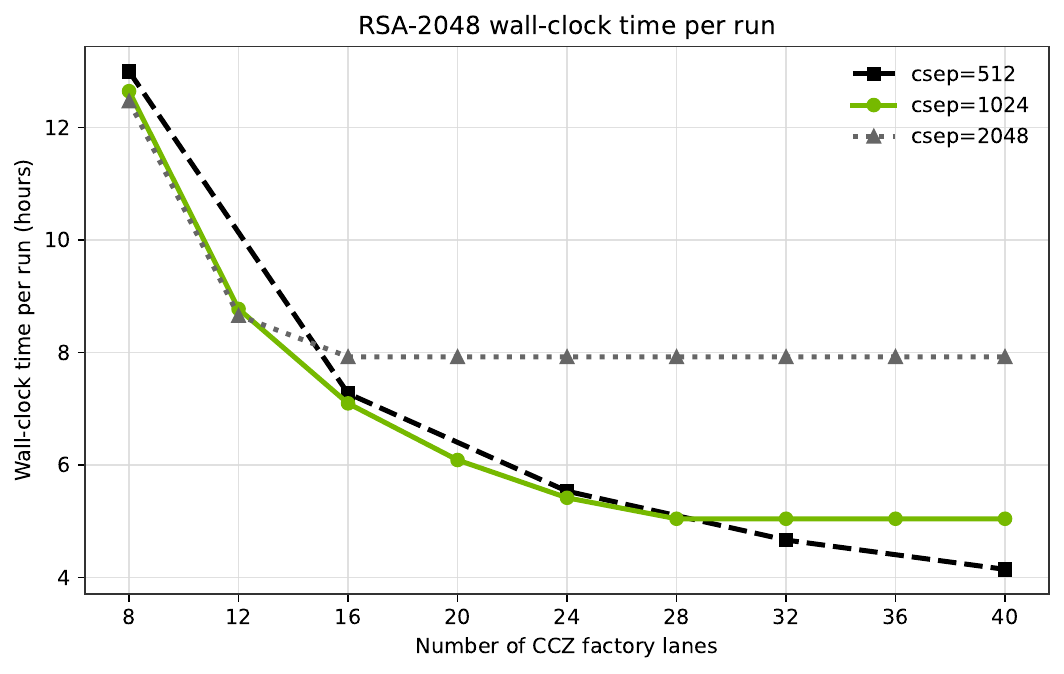}
  \includegraphics[width=0.47\textwidth]{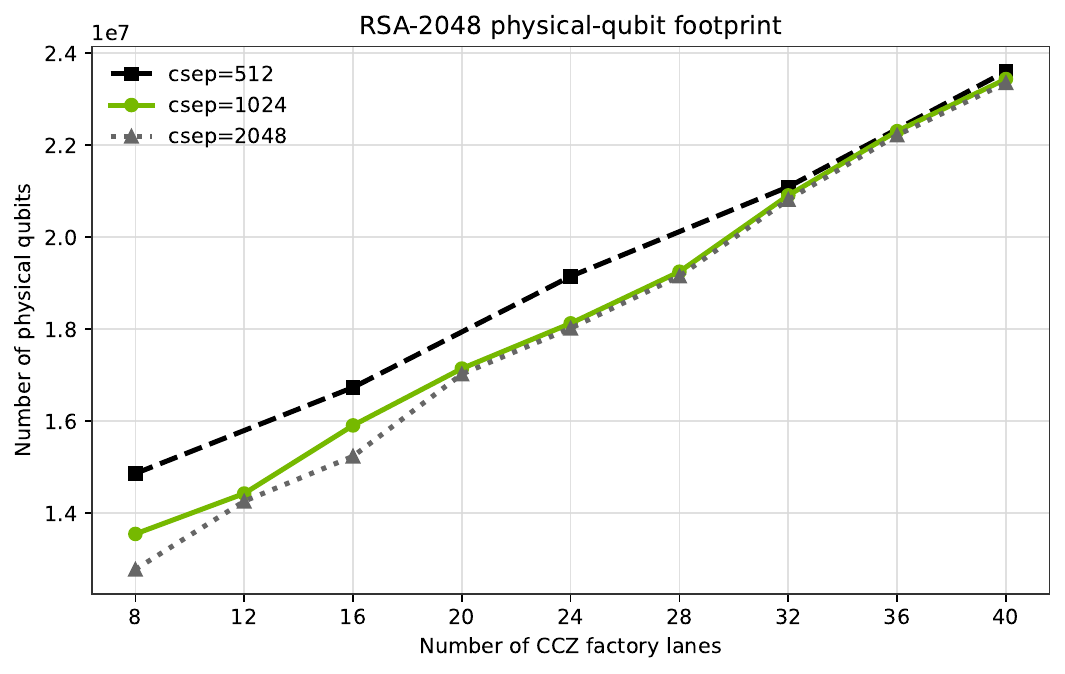}
  \caption{P3 schedule-derived RSA-2048 tradeoffs across AutoCCZ lane counts and
    carry-runway separations.  From left to right: single-run wall-clock time
    and physical-qubit footprint.  Each point is obtained by compiling and
    scheduling the corresponding architecture configuration at P3.}
  \Description{Two line plots. Left: single-run wall-clock time in hours
    versus the number of AutoCCZ lanes for three carry-runway separations;
    each curve falls and then plateaus. Right: physical-qubit footprint versus
    factory lanes for the same separations, rising linearly.}
  \label{fig:ge-factory-csep-sweep}
\end{figure*}

As a reference, Figure~\ref{fig:ge-factory-csep-sweep} reports the P3 sweep over factory-lane
counts for $c_{\mathrm{sep}}\in\{512,1024,2048\}$.  These separations produce
$(\text{pieces},\text{piece length})=(4,550)$, $(2,1062)$, and $(1,2086)$,
respectively.  At $c_{\mathrm{sep}}=512$, four carry pieces require enough
concurrent AutoCCZ supply that every sampled point from 8 through 40 lanes
remains factory-limited; increasing the lane count reduces the one-run
makespan from 12.995 to 4.147 hours.  At $c_{\mathrm{sep}}=1024$, the
two-piece adder crosses from factory-limited to reaction/code-depth limited at
28 lanes, where the makespan saturates at 5.046 hours.  At
$c_{\mathrm{sep}}=2048$, there is no carry-piece parallelism.  Its continuous
supply boundary is 14 lanes, and the sampled sweep first reaches the plateau
at 16 lanes; all points from 16 through 40 remain at 7.925 hours.  Beyond
these supply boundaries, reaction latency and code depth limit the schedule,
so additional lanes increase the physical-qubit footprint without reducing
the runtime.

%% file: sections/evals/pinnacle-FH.tex
\subsubsection{Pinnacle Architecture}
\label{sec:results-pinnacle-cross-target}

We apply CUDA-Q Logical to plaquette-Trotterized Fermi--Hubbard
phase estimation \cite{campbell2022hubbard} using the Pinnacle architecture
\cite{webster2026pinnacle}. We define the workload once as an ordinary CUDA-Q
Logical P0 program and compile it through QEC binding and physical scheduling.
We then compare the schedule-derived result with the published analytical
model and attribute each difference to either synthesis or scheduling.

Via direct compilation with CUDA-Q Logical, we find that for the $L = 16$ case studied by \cite{webster2026pinnacle}, logical demand is $513$ peak logical
qubits ($2L^2+1$; $514$ occupants with an additional ancilla for repeat-until-success rotation synthesis) and, per
Trotter step, $3{,}072$ exact $T/T^\dagger$ actions, $1{,}024$ arbitrary
rotations, $214{,}784$ exact Clifford rotations, and $38{,}336$ fermionic
swaps. These counts arise from analysis on the compiled IR. Synthesis is performed via repeat-until-success (RUS) \cite{Bocharov2015RUS} at per-rotation precision
$5.258\times10^{-7}$, from the optimized error split $x=0.02922$. The declared
requirement is a relative energy error of $0.5\%$ of the total lattice energy
($\varepsilon = 0.005\cdot 1.02\cdot L^2$ hartree, $u/\tau=4$), giving $201$
executable Trotter steps (continuous prefactor $200.232$, assuming that these error bounds are tight). Note that in general these naive estimators for Trotter error may not be tight in practice, and we defer discussion of tightening these calculations to later work. Metric definitions
are shown in the artifact schema: $7{,}419{,}859.28$ paper $T$ states,
$410{,}075.67$ RUS measurements, and the acceptance-corrected expectation
$8{,}654{,}363.76$ logical cycles, recorded separately from the raw
$T{+}\mathrm{RUS}$ count.

\begin{lstlisting}[style=qlx-python, float, label={lst:device},
caption={The complete Pinnacle machine as a CUDA-Q Logical device---processing unit,
memory, and magic engine---with every architecture-declared scratch region
instantiated explicitly. The closing assert is the machine's exact resource
multiset. Table~IV's operating point omits the optional memory module; the
compiled studies above bind this construction without the memory region.}]
import cudaq.logical as ql
from ql.architectures import pinnacle

# The complete Pinnacle machine for the L = 16 workload: one processing unit
# of 33 bridged GB510 blocks, one GB510 memory block, and the paper's
# 4,410-qubit magic engine bound as a typed T-state factory.
engine = pinnacle.magic_engine(p_phys=1.0e-3, target_output_infidelity=1.0e-9)
architecture = pinnacle.gb510
block = pinnacle.processing_block(architecture)

builder = ql.devices.DeviceBuilder("PinnacleFermiHubbardMachine")
compute = builder.logical.add_compute(capacity=514, name="compute")
memory = builder.logical.add_memory(capacity=16, name="memory")
factory = builder.logical.add_factory(produces=ql.standard.T_STATE, via=engine.producer,
    capacity=1, buffer_size=1, name="magic_engine", stream_name="t_states")

compute_qec = builder.qec.bind(compute, architecture=architecture)
memory_qec = builder.qec.bind(memory, architecture=architecture)
factory_qec = builder.qec.bind(factory, encoding=ql.codes.BareQubit)

processors = builder.physical.add_qubits(33 * block.physical_qubits, name="processing_blocks")
memory_qubits = builder.physical.add_qubits(block.physical_qubits, name="memory_block")
engine_qubits = builder.physical.add_qubits(engine.physical_qubits, name="magic_engine_qubits")

# Every architecture-bound region declares its own attempt-local RUS scratch;
# a concrete machine funds each one explicitly.
compute_scratch = builder.physical.add_qubits(block.physical_qubits, name="compute_rus_scratch")
memory_scratch = builder.physical.add_qubits(block.physical_qubits, name="memory_rus_scratch")

builder.physical.bind(compute_qec, to=processors)
builder.physical.bind(memory_qec, to=memory_qubits)
builder.physical.bind(factory_qec, to=engine_qubits)
(compute_aux,) = compute_qec.auxiliary_regions
(memory_aux,) = memory_qec.auxiliary_regions
builder.physical.bind(compute_aux, to=compute_scratch)
builder.physical.bind(memory_aux, to=memory_scratch)

builder.physical.set_operating_point(timing={"cycle_ns": 1000.0, "surface_cycle_ns": 1000.0,
            "condition_ns": 10_000.0, "decode_bit_ns": 10_000.0},
    calibration={"physical_error": 1.0e-3, "magic_engine_output_infidelity": engine.output_infidelity},
    name="paper_operating_point")
device = builder.build()

assert {r.name: r.count for r in device.physical.resource_classes} == {
    "processing_blocks": 53_460, "memory_block": 1_620,
    "magic_engine_qubits": 4_410,
    "compute_rus_scratch": 1_620, "memory_rus_scratch": 1_620}
\end{lstlisting}

\begin{lstlisting}[style=qlx-python, float, label={lst:gadget-author},
caption={Authoring a typed memory gadget. The decorator records the
logical action, name, and metadata, while the function signature declares
the encoded-patch interface.}]
def memory(self, rounds: int | None = None):
    rounds = (self.instance.logical_cycle_rounds if rounds is None else rounds)
    encoding = self.encoding

    @ql.gadget(
        implements=ql.logical.memory,
        name=f"{self.code.name}_memory_r{rounds}",  metadata={"syndrome_rounds": rounds},
    )
    def realization(block: ql.patch[encoding]) -> ql.patch[encoding]:
        for _ in range(rounds):
            block, _ = ops.extract_syndrome(block)
        return block
    return realization
\end{lstlisting}
\begin{table*}[b]
\caption{Comparison of the published Pinnacle estimates
\cite{webster2026pinnacle} with CUDA-Q Logical device footprints, workload
demand, analytical runtime estimates, and schedule-derived P3 results.
Results are reported for $p=10^{-3}$ and $p=10^{-4}$ with
$t_c=1\,\mu$s over the even lattice sizes
$L\in\{4,8,\ldots,32\}$. The $L=4$ point lies below the range reported in
Table~IV of Ref.~\cite{webster2026pinnacle} and is included as an additional
case. The published global runtime bounds of $3.8$\,min for $p=10^{-3}$ and
$1.8$\,min for $p=10^{-4}$ correspond to the maxima of the analytical
columns. Each ratio is the P3 expected runtime divided by the corresponding
analytical expectation.}
\label{tab:pinnacle-summary}
\centering
\small
% Generated from frozen artifacts by make_summary_table.py -- do not edit.
\begin{tabular}{@{}r rr rr rr rr rr rr@{}}
\toprule
 & \multicolumn{2}{c}{Table IV \cite{webster2026pinnacle} (kq)} & \multicolumn{2}{c}{CUDA-Q Logical (qubits)} & \multicolumn{2}{c}{Workload demand} & \multicolumn{2}{c}{Analytical $E[t]$ (s)} & \multicolumn{2}{c}{P3, $p{=}10^{-3}$} & \multicolumn{2}{c}{P3, $p{=}10^{-4}$} \\
\cmidrule(lr){2-3}\cmidrule(lr){4-5}\cmidrule(lr){6-7}\cmidrule(lr){8-9}\cmidrule(lr){10-11}\cmidrule(l){12-13}
$L$ & $10^{-3}$ & $10^{-4}$ & $10^{-3}$ & $10^{-4}$ & $\tau_T$ ($10^6$) & RUS ($10^3$) & $10^{-3}$ & $10^{-4}$ & $E[t]$ (s) & ratio & $E[t]$ (s) & ratio \\
\midrule
4 & --- & --- & 10,890 & 2,400 & 7.200 & 398.5 & 218.4 & 106.6 & 246.2 & 1.127 & 129.0 & 1.210 \\
8 & 19 & 5.6 & 20,610 & 6,016 & 7.396 & 408.8 & 224.3 & 109.5 & 241.7 & 1.078 & 119.3 & 1.090 \\
10 & 25 & 8.3 & 27,090 & 8,728 & 7.410 & 409.5 & 224.7 & 109.7 & 255.6 & 1.138 & 130.0 & 1.186 \\
12 & 35 & 12 & 36,810 & 12,344 & 7.412 & 409.7 & 224.8 & 109.7 & 255.1 & 1.135 & 134.2 & 1.223 \\
14 & 45 & 16 & 46,530 & 15,960 & 7.432 & 410.7 & 225.4 & 110.0 & 199.0 & 0.883 & 102.7 & 0.934 \\
16 & 58 & 20 & 59,490 & 20,480 & 7.420 & 410.1 & 225.0 & 109.8 & 294.1 & 1.307 & 139.9 & 1.274 \\
18 & 71 & 25 & 72,450 & 25,904 & 7.416 & 409.9 & 224.9 & 109.8 & 245.4 & 1.091 & 128.0 & 1.166 \\
20 & 87 & 31 & 88,650 & 31,328 & 7.432 & 410.7 & 225.4 & 110.0 & 261.2 & 1.159 & 125.6 & 1.142 \\
22 & 103 & 37 & 104,850 & 37,656 & 7.431 & 410.7 & 225.4 & 110.0 & 216.4 & 0.960 & 110.3 & 1.002 \\
24 & 123 & 44 & 124,290 & 44,888 & 7.422 & 410.2 & 225.1 & 109.8 & 276.8 & 1.230 & 143.2 & 1.303 \\
26 & 142 & 52 & 143,730 & 52,120 & 7.433 & 410.8 & 225.4 & 110.0 & 257.5 & 1.142 & 131.2 & 1.192 \\
28 & 165 & 60 & 166,410 & 60,256 & 7.439 & 411.1 & 225.6 & 110.1 & 269.4 & 1.194 & 133.2 & 1.210 \\
30 & 187 & 69 & 189,090 & 69,296 & 7.444 & 411.3 & 225.7 & 110.2 & 243.2 & 1.077 & 127.3 & 1.155 \\
32 & 213 & 78 & 215,010 & 78,336 & 7.442 & 411.2 & 225.7 & 110.1 & 310.6 & 1.376 & 148.3 & 1.346 \\
\bottomrule
\end{tabular}

\end{table*}

\begin{lstlisting}[style=qlx-python, float, label={lst:gadgets},
caption={The logical surgery gadget suite selected during compilation, read
back from the architecture the device binds.}]
import cudaq.logical as ql
from ql.architectures import pinnacle

# One line binds the whole Table-I microarchitecture: the [[510,16,24]] code,
# its four seed-operator surgery gadgets and four bridges per block, and the
# typed P2 gadget suite the compiler selects from.
architecture = pinnacle.gb510
assert [d.name for d in architecture.link_roots] == [
    "pinnacle_gb510_prepare_zero",
    "pinnacle_gb510_prepare_one_logical_action",
    "pinnacle_gb510_prepare_plus",
    "pinnacle_gb510_prepare_minus_logical_action",
    "pinnacle_gb510_memory_r1",                       # d+2-round memory cycle
    "pinnacle_gb510_transversal_cx",                  # blockwise logical CX
    "pinnacle_gb510_default_encoding_wsc_mpp",        # WSC generalized surgery
    "pinnacle_gb510_default_encoding_pinnacle_rpp",   # Clifford / T / BRS-RUS
]

# Table I's gadget accounting is architecture data, not user arithmetic:
block = pinnacle.processing_block(architecture)
assert block.physical_qubits == (block.code_block_qubits + 4 * block.gadget_qubits +
                                 4 * block.bridge_qubits) == 1620

# The architecture also declares the scratch it needs for bounded BRS-RUS
# rotation attempts; a built device must fund it (one reserved block).
(scratch,) = architecture.auxiliary_regions
assert scratch.name == "pinnacle_gb510_rus_scratch"
\end{lstlisting}

Table \ref{tab:pinnacle-summary} contains the compiled CUDA-Q Logical results and architectural analysis for this application. There are particular cancellations at play that result in an analytically \emph{constant} running time across distinct lattice sizes. In particular we note that, following \cite{webster2026pinnacle}, we require a 0.5\% relative error on the total lattice energy. As a result, the total energy budget grows as $\varepsilon = 0.005(1.02\text{ Hartree/site })L^2$, and therefore the required number of Trotter steps given by typical Trotter error bounds shrinks by $L^{-2}$. To see this, we follow equations F7--F10 from \cite{campbell2022hubbard} and bound the number of steps as the number of applications (for phase estimation) of the Trotter unitary as $N_{PE}$:

\begin{equation}
N_{PE} \approx \frac{6.203 \sqrt{W}}{((1-x)\varepsilon)^{3/2}} \propto \frac{1}{L^2}
\end{equation}

which, because $W \approx 8.3 L^2$, is proportional to $L^{-2}$. Each Trotter step itself requires work proportional to the lattice size as well, with the number of T gates $N_T = 12L^2$ and the number of arbitrary rotations $N_R = 4L^2$. The T-count derived from rotation synthesis error also works out to be lattice size independent:

\begin{equation}
N_{HT} \approx 1.15 \log_2 \frac{N_R \sqrt{3W}}{x\sqrt{1-x}\,\varepsilon^{3/2}} + 9.2 \propto \frac{L^2 \cdot L}{L^3} = L^0,
\end{equation}

which is a constant factor of 33.19 expected T gates overhead per rotation in this context. Finally, we then have that 

\begin{equation}
\tau_T = N_{PE}\,(12 + 4\cdot 33.19)\,L^2 \approx 144.8\, N_{PE} L^2,
\end{equation}

and thus the lattice dimensions cancel. We see this in the table: the T gate column remains constant throughout the estimates despite growing lattice sizes. 

Through CUDA-Q Logical, we can do a deep architectural analysis to better understand expected execution times as a function of quantum architectural decisions. On the Pinnacle processing unit architecture, we discretize time into logical cycles defined by the time required to perform one joint logical Pauli measurement with full error correction. This can be written as:

\begin{equation}
T(\text{logical cycle}) = \max\{\,T(d+2\text{ syndrome rounds}),\; T(\text{one magic-engine attempt})\,\}
\end{equation}

which is assumed to be $26 \mu s$ at $p=10^{-3}$, and $14 \mu s$ at $p = 10^{-4}$ \cite{webster2026pinnacle}. We calculate runtime by computing how many cycles one Trotter step occupies, multiplied by the number of Trotter steps, multiplied by the logical cycle time. With a compiled CUDA-Q Logical program we can measure the length of the scheduled event graph per Trotter step through the native scheduler walking the real dependency structure that dictates which rotations need which logical qubits, when magic-engines will have states ready, and what can be parallelized. We perform this calculation, and find that it is often the case that three loss mechanisms appear and slow down execution times: dependency stalls (two rotations targeting the same logical qubit must serialize, and the cycle's dependency chains may not always have T-consumers ready to consume states immediately upon preparation), magic cadence and logical cycle ratios (accepted magic states can arrive every 28.9 $\mu s$ on average, whereas the syndrome extraction derived logical cycle time is 26 $\mu s$, so delivery and consumption slots misalign), and step-boundary barriers (a subsequent step cannot begin until the previous Trotter step's tail measurement completes). In these cases, the executed schedule stalls and requires more time than is analytically predicted. 

We also find instances where the compiled schedule executes faster than the
analytical serial prediction: at $L{=}14$ (both rates) and marginally at
$L{=}22$ ($p{=}10^{-3}$) in Table~\ref{tab:pinnacle-summary}. To separate
causes, each ratio factors exactly as $D \times S$, where $D$ uses the compiler's certified synthesis circuits and $S$ is the residual scheduling term. The scheduler dispatches disjoint Pauli product measurements concurrently and hides repeat-until-success latency behind magic-engine state creation, recovering up to 19\% ($S{=}0.81$ at $L{=}24$). The faster execution points are synthesis-demand effects, not scheduling effects. The 33.19 expected T gates per rotation is a fitted average; the certified circuit for a specific
(angle, precision) pair has a per-attempt T-count range (20--42 in this sweep) and success probability (0.55--1.00) do not vary smoothly with the angle, and since the tabulated
Trotter norms perturb $\Delta t$ only in its fourth significant digit, every
lattice re-samples this distribution. At $L{=}14$ the two angles
happen to admit unusually cheap certified circuits (27.5 expected T per
rotation), which by itself gives $D{=}0.83$; the executed schedule actually
runs $7\%$ \emph{above} the analytical expectation of that concrete demand
($S{=}1.07$). Across the sweep $D$ spans $0.83$--$1.52$ while $S$ spans
$0.81$--$1.14$, so the oscillation in the ratio column is driven both by
concrete synthesis demand and scheduling. These effects are
discrete combinatorial properties that do not vary smoothly with lattice
size, and the demand term is also a design choice since $\Delta t$ may be
perturbed within the retained Trotter-error budget to select
synthesis-cheap angles before compilation. Only a compiler that synthesizes
the actual rotation circuits and represents data flow and execution streams
can expose these effects, and then use them to modify system designs for
future system builds.

\paragraph{CUDA-Q Logical definitions used in the compilation.}
Listings~\ref{lst:device}--\ref{lst:gb-code} are executable code snapshots showing the implementations for code families, logical gadgets, and architectural devices. Listing~\ref{lst:gb-code} constructs the
generalized-bicycle code from its published parameters in Table I of \cite{webster2026pinnacle};
Listing~\ref{lst:gadgets} shows the surgery gadget suite including preparation,
memory, transversal CX, the generalized-surgery product measurement \cite{webster2025explicitconstructionlowoverheadgadgets}, and
the three-way product-rotation as typed architecture data, together with Table~I's
gadget accounting and the architecture-declared repeat-until-success scratch ancilla;
Listing~\ref{lst:gadget-author} reproduces from the library how
one such gadget is authored (the $d{+}2$-round memory cycle: an ordinary
typed realization over encoded patches); and Listing~\ref{lst:device} builds the complete machine including the
processing unit, memory block, and magic engine, which is then used for the resource estimates and footprints reported in Table~\ref{tab:pinnacle-summary}.

\begin{lstlisting}[style=qlx-python, float, label={lst:gb-code},
caption={The GB code used in compilation, constructed from its published
Table-I parameters with typed, validated provenance.}]
from cudaq.logical import codes

# Table-I generalized-bicycle instance over the cyclic lift l = 255:
#   A(x) = 1 + x^39 + x^55,   B(x) = 1 + x^70 + x^127   ->   [[510, 16, 24]].
gb510 = codes.pinnacle_gb("gb510")
assert (gb510.n, gb510.k, gb510.d.conservative_value) == (510, 16, 24)

# The published logical-operator seeds fix canonical logical ports; CSS
# commutation, rank, and logical pairing are validated at construction, and
# the claimed distance carries its Table-I citation as typed provenance.
instance = codes.pinnacle_gb_instance("gb510")
assert instance.a == (0, 39, 55) and instance.b == (0, 70, 127)
assert gb510.d.status == "claimed"
\end{lstlisting}

%% file: sections/evals/gadgets.tex
\subsubsection{Gadget Verification and EDEM Composition}
\label{sec:gadget-study}

We use the gadget construction of Ref.~\cite{kliuchnikov2026composing} to
evaluate CUDA-Q Logical's support for fault-tolerant gadgets. Each gadget is
represented as a typed compiler artifact with a declared definition and an
extended detector error model. We verify the implementation against its
definition, compose the models from several gadgets, and compare the result
with the model generated from the complete circuit. This evaluation addresses
three claims:
\begin{enumerate}
  \item \emph{Gadget verification is exact and fails closed.}  For a typed
    Clifford realization, CUDA-Q Logical certifies both conditions of the gadget
    definition: the gadget detector contract (virtual detectors deterministic, virtual
    output syndrome map determined by realization outcomes and virtual input
    syndromes) and the logical-action--realization equivalence under the outcome
    map.  False claims are named, not sampled.
  \item \emph{EDEM composition reproduces the monolithic DEM.}  Wiring
    per-gadget EDEMs together and contracting yields, mechanism by mechanism
    and probability by probability, the detector error model that Stim derives
    from the flat circuit.
  \item \emph{Analysis is per gadget type and the artifacts are portable.}
    Each gadget is analyzed once; a memory experiment of any length is a
    composition of the stored EDEMs; probabilities bind after composition; and
    the artifacts are consumed without the compiler, circuit, or code that
    produced them.
\end{enumerate}
% \newpage
% \begin{wrapfigure}{r}{0.55\textwidth}
%   \begin{minipage}{\linewidth}
\begin{lstlisting}[
    style=qlx-python,
    aboveskip=0pt,
  belowskip=0pt,
    caption={Steane transversal $H$: the detector contract swaps the check families; the receipt certifies both the swap and the logical action.},
    label={lst:qec-steane-contract}
  ]
import cudaq.logical as ql
SteanePatch = ql.patch[ql.codes.Steane]

@ql.gadget(implements=ql.logical.h)
def steane_h_round(block: SteanePatch) -> SteanePatch:
  block = ql.h(block.data)
  block, _ = ql.extract_syndrome(block)
  return block

h_in = steane_h_round.inputs.only()
h_out = steane_h_round.outputs.only()
h_checks = steane_h_round.records.syndrome(0).checks

# Reorder the virtual input syndrome into the
# check convention after H.
swapped = tuple(h_in.syndrome[(i + 3) % 6] for i in range(6))

h_detector_contract = ql.gadgets.GadgetDetectorContract(steane_h_round,
  detectors=tuple( ql.gadgets.Detector(swapped[i] ^ h_checks[i]) for i in range(6)),
  boundary={h_out.syndrome: swapped},
)
receipt = ql.gadgets.verify_gadget(h_detector_contract)
assert (receipt.detector_contract.outcome_code_rank, receipt.detector_contract.detector_rank) == (12, 12)
assert receipt.logical_action.action == "ql.logical.h"
\end{lstlisting}
% \end{minipage}
% \end{wrapfigure}
Speed is not a claim of this study.  The prototype's composition runs in pure
Python and is slower in wall-clock time than Stim's native monolithic
construction; the point is that the same model is obtained from modular,
independently verified pieces without a global circuit.

\paragraph{Setup.}
A gadget is authored as a typed function body over encoded patches
(Listing~\ref{lst:qec-repetition-contract}); its logical action is declared with
\texttt{implements=}.  A detached \texttt{GadgetDetectorContract} names the virtual
detectors and the virtual output syndrome map over typed endpoints and stable
records.  \texttt{verify\_gadget} runs both checks from one Choi probe of the
realization: every logical input is Bell-paired with a reference qubit, the
encoded input is placed in an arbitrary stabilizer coset, and the realization's
Clifford projection is run through an outcome-complete symbolic stabilizer
simulation over $\mathbb{F}_2$.  This is the encoder--unencoder sandwich of
Ref.~\cite{kliuchnikov2026composing}.  The resulting outcome code certifies
the detector contract; the post-realization Choi state, compared with the Choi state of
the ideal logical action on bare logical qubits, certifies the action.  Fault
analysis reuses the same operation-local Clifford projection to propagate each
typed elementary fault into an EDEM.  Neither path constructs a \profile{P3}
placement or a timed schedule.

\paragraph{Claim 1: verification.}
The smallest complete example is one syndrome round of the three-qubit
repetition code, whose checks are $Z_0Z_1$ and $Z_1Z_2$
(Listing~\ref{lst:qec-repetition-contract}).  The Choi probe yields a
$6\times2$ affine outcome relation: the two virtual input syndromes, two
extraction records, and two virtual output syndromes are three copies of the
same two free coset bits.  Its left kernel has dimension four, spanned by the
two virtual detectors and the two rows of the virtual output syndrome map,
giving the detector-contract receipt $4=4+0$.  The logical-action receipt records that the
realization's Choi state is stabilized by exactly the two Bell stabilizers of
the identity, $X_{\mathrm{ref}}\bar X$ and $Z_{\mathrm{ref}}\bar Z$, with the
signs the logical action predicts.

% \begin{wrapfigure}{r}{0.54\textwidth}
%   \begin{minipage}{\linewidth}
\begin{lstlisting}[
    style=qlx-python,
    float=!t,
      aboveskip=0pt,
  belowskip=0pt,
    caption={One repetition-code syndrome round, its gadget detector contract, and the receipt certifying both gadget conditions.},
    label={lst:qec-repetition-contract}
  ]
  import cudaq.logical as ql 
  RepetitionPatch = ql.patch[ql.codes.Repetition]

  @ql.gadget(implements=ql.logical.idle)
  def syndrome_round(block: RepetitionPatch) -> RepetitionPatch:
      block, _ = ql.extract_syndrome(block)
      return block

  block_in = syndrome_round.inputs.only()
  block_out = syndrome_round.outputs.only()
  checks = syndrome_round.records.syndrome(0).checks

  # Gadget detector contract: virtual detectors and the virtual
  # output syndrome map. Newly measured checks become
  # the next round's virtual input syndrome.
  detector_contract = ql.gadgets.GadgetDetectorContract(syndrome_round, detectors=(
           ql.gadgets.Detector(block_in.syndrome ^ checks)
           ), boundary={block_out.syndrome: checks},
  )

  receipt = ql.gadgets.verify_gadget(detector_contract)
  assert (receipt.detector_contract.outcome_code_rank, receipt.detector_contract.detector_rank,
          receipt.detector_contract.deterministic_logical_outcome_rank) == (4, 4, 0)
  assert receipt.logical_action.action == "ql.logical.idle"
  assert receipt.logical_action.logical_rank == 2
\end{lstlisting}
% \end{minipage}
% \end{wrapfigure}
The Steane transversal-$H$ round (Listing~\ref{lst:qec-steane-contract})
exercises a nonidentity logical action and a nonidentity boundary frame.  Its three
$X$-check supports equal its three $Z$-check supports, so transversal
Hadamards exchange the two families; the detector contract states this six-coordinate
half-swap as the virtual output syndrome map.  The complete relation is
$18\times6$; six virtual detectors and the six-row map span the
$12$-dimensional left kernel, giving $12=12+0$.  The zero logical rank is
expected, since a Hadamard round has no deterministic logical outcome and the
outcome map is empty.  The logical-action check certifies logical $H$: the Choi
stabilizers $X_{\mathrm{ref}}\bar Z$ and $Z_{\mathrm{ref}}\bar X$ hold on the
realization with positive sign.  A transversal CX between two repetition
blocks certifies logical CX with the four-stabilizer Choi state of a two-qubit
Clifford and the virtual output syndrome map
$s'_t=s_c\oplus s_t$ on the target block.  Logical-$Z$ measurement gadgets
certify \texttt{measure\_z} in both the destructive form, where the reference
qubit collapses onto the reported outcome, and the block-preserving form, where
the outgoing patch and the reference both carry it.

The controls establish that verification checks completeness and the
logical action, not only the equations an author happened to supply.  Swapping the
two record references in the repetition detector contract makes both detectors nonzero.
Removing one detector leaves the remaining rows valid but reduces the declared
rank to three, so the verifier rejects $4\ne3+0$.  Retaining the unswapped
adjacent-round detector contract for the Steane $H$ round makes all six detector rows
nonzero.  A syndrome round that declares \texttt{implements=ql.logical.x}
passes detector-contract verification unchanged---the contract is blind to the
logical action---and is rejected by the logical-action check with a sign mismatch on
$Z_{\mathrm{ref}}\bar Z$.  The same round declaring \texttt{ql.logical.h} is
rejected because $X_{\mathrm{ref}}\bar Z$ does not stabilize its Choi state.
A logical-$Z$ readout declaring \texttt{measure\_x} is rejected for the same
reason.  Non-Clifford and parameterized logical actions are not inferred from the
realization body; they fail as unsupported.
Table~\ref{tab:gadget-verification} is the claim ledger.

\begin{lstlisting}[
    style=qlx-python,
      aboveskip=0pt,
  belowskip=0pt,
    caption={Composing stored surface-memory EDEMs into an $N$-round memory DEM. The consumer uses only the stored EDEM artifacts and the detector-model API; no gadget, circuit, code, or compiler build is needed.},
    label={lst:qec-open-analysis}
  ]
import cudaq.logical as ql

prep = ql.load_extended_detector_model("prep_logical_z.odem.json")
sec  = ql.load_extended_detector_model("sec.odem.json")
mz   = ql.load_extended_detector_model("mz.odem.json")

with ql.DetectorModelBuilder(f"surface_memory_{rounds}") as d:
  prepared = d.instance(prep, name="prep")
  d.harvest(prepared.detector, label="round0")
  carried = prepared.boundary_for(sec)   # virtual output syndrome -> input
  for i in range(1, rounds):
      emitted = d.instance(sec, *carried, name=f"sec{i}")
      d.harvest(emitted.detector, label=f"round{i}")
      carried = emitted.boundary_for(sec)
  measured = d.instance(mz, *carried, name="mz")
  d.harvest(measured.detector, label="terminal")
  d.harvest(measured.observable, label="logical_z")
  graph = d.finish()

# Probabilities bind after composition; structure is probability free.
point = ql.ProbabilityBatch(graph.fault_schema, probabilities)
canonical = ql.canonicalize_detector_model(graph, point)
dem_text = ql.to_dem(canonical.model, canonical.probabilities)
# Mechanisms and probabilities equal stim's DEM of the flat circuit.
\end{lstlisting}
\paragraph{Claim 2: composition.}
The surface-memory library consists of three gadgets of the distance-three
rotated surface code: logical-$|0\rangle$ preparation, one
syndrome-extraction cycle, and terminal $Z$ readout.  Each is analyzed once
under Stim's standard circuit-level depolarizing noise at $p=10^{-3}$ and
stored as an EDEM open at its virtual-syndrome ports; the source models carry
436, 427, and 9 fault columns. Uniform depolarizing channels are converted
into equivalent independent Pauli-component mechanisms before DEM composition.
The categorical branch probabilities $p/3$ and $p/15$ for one- and two-qubit
depolarization are not the independent-mechanism probabilities. Listing~\ref{lst:qec-open-analysis} composes
them: one preparation instance, $N-1$ instances of the same extraction EDEM
wired through \texttt{boundary\_for}, and one readout instance, followed by
canonical merging of equal-effect faults and emission as Stim DEM text.

The independent oracle is \texttt{stim.Circuit.detector\_error\_model()} on
the flat circuit, which CUDA-Q Logical also emits and which byte-matches
\texttt{stim.Circuit.generated("surface\_code:rotated\_memory\_z")} under the
same noise parameters.  Table~\ref{tab:edem-composition} reports the
comparison for $N\in\{3,5,10\}$ rounds.  In every case the composed DEM equals
Stim's DEM exactly: the same set of mechanisms, each with the same detectors,
observables, and probability, and with detector indices already in circuit
order, so no relabeling is needed.  The negative control composes three rounds
and compares against the four-round circuit; the comparison fails.

\begin{table}[b]
  \caption{Composed EDEMs versus Stim's monolithic DEM for distance-three
    rotated surface-code memory at $p=10^{-3}$.  Source fault columns count
    the per-gadget elementary faults before canonical merging; mechanisms
    count the merged DEM.  Equality is exact in mechanisms and probabilities
    with the identity detector relabeling.}
  \label{tab:edem-composition}
  \small
  \begin{tabular}{@{}rrrrrl@{}}
    \toprule
    Rounds & EDEM instances & Detectors & Source fault columns & Mechanisms & Composed vs.\ Stim \\
    \midrule
    3  & 1 + 2 + 1 & 24 & 1,299 & 219   & equal \\
    5  & 1 + 4 + 1 & 40 & 2,153 & 443   & equal \\
    10 & 1 + 9 + 1 & 80 & 4,288 & 1,003 & equal \\
    \midrule
    3 vs.\ 4-round circuit & --- & 24 vs.\ 32 & --- & --- & rejected \\
    \bottomrule
  \end{tabular}
\end{table}

\paragraph{Interoperable circuit-fault analysis.}
CUDA-Q Logical's Stim export connects compiled QEC experiments to Stim's
circuit-analysis tools. In this study, we use that interface to search for
undetectable logical errors and trace the returned witnesses to physical fault
locations in the emitted circuits. We demonstrate this workflow on nine
rotated surface-code $Z$-memory circuits,
with $d\in\{3,5,7\}$ and $N\in\{3,5,10\}$ rounds, using Stim 1.16.0.
All four circuit-generator noise parameters are $10^{-3}$:
\nolinkurl{after_clifford_depolarization},
\nolinkurl{after_reset_flip_probability},
\nolinkurl{before_measure_flip_probability}, and
\nolinkurl{before_round_data_depolarization}.
We run Stim's hyperedge-aware
\nolinkurl{search_for_undetectable_logical_errors}~\cite{gidney2021stim}.
Following the \nolinkurl{UndetectableErrorStim} strategy implemented by the
\texttt{codedistance} package~\cite{webster2026codedistance}, the search
increases the maximum intermediate detection-event-set size from two and stops
after returning the same best weight twice. Edges of every detector degree are
included, symptom degree may increase, and circuit errors are canonicalized.
Witness weight counts elementary circuit faults, not affected qubits.
Limits two and three both return weights 3, 5, and 7 for the corresponding
code distances at every tested round count.

These searches produce heuristic circuit-fault-distance estimates because
Stim truncates the explored detection-event sets. We separately check each
returned witness to have zero detector syndrome, flip logical observable
zero, and map every constituent error to a physical circuit location. The
workflow connects QEC compilation to fault analysis: Stim identifies fault
combinations in the emitted experiment that evade detection and change its
logical outcome. The all-nine result includes undecomposed hyperedges and
confirms a physical weight-seven undetectable logical error for each $d=7$
memory circuit. These witnesses establish upper bounds on circuit-fault
distance; repeating the same weight does not establish optimality. None of this supplies
circuit-distance evidence for the ZSZ or Pinnacle surgery circuits.

A second oracle, internal to CUDA-Q Logical, checks composition on the repetition code
against CUDA-Q Logical's own monolithic propagation of the whole $N$-round gadget for
$N\in\{2,3\}$, again with exact agreement and with a dropped-round control.
For a transversal CX between two patches, composing the CX relay
$s'_t=s_c\oplus s_t$ between pre- and post-CX extraction EDEMs reproduces the
monolithic DEM and creates detector mechanisms spanning both patches; replacing
the relay by $s'_t=s_t$ removes every cross-patch mechanism.

\paragraph{Claim 3: analyze once, reuse everywhere.}
The library is produced by the compiler; the composition in
Listing~\ref{lst:qec-open-analysis} is not.  The three artifacts are
self-contained, probability-independent JSON (or a binary container with a
memory-mapped C++ reader) whose semantic hash is independent of storage codec.
A manifest binds each artifact's exact hash to the ordered probability vector
of one operating point, so fault-column order and probabilities cannot drift
independently; loading a library whose probabilities disagree with its
artifact hash is rejected.  Because structure and probability are separate,
one composed graph is rebound at several physical error rates and sampled
without repeating fault propagation, and the same stored extraction EDEM
serves every round of every memory length.  The decoder-tasking consumer of
these artifacts places a finite gadget call sequence and constructs bounded
decoding windows from the stored EDEMs alone, without a global DEM; that
workflow is reported separately.

\paragraph{Scope.}
Gadget verification certifies both conditions of the gadget definition for
Clifford realizations and stabilizer-instrument logical actions; EDEM composition is
exact for the tested fault schemas and agrees with an independent oracle on
the surface code; typed boundaries preserve nonlocal detector structure across
a two-patch operation; and composed graphs are rebound and sampled without
repropagation.  None of these results establishes fault tolerance, circuit
distance, a threshold, non-Clifford semantics, a \profile{P3} schedule, or
hardware execution.

\input{tables/gadget-verification}

%% file: tables/gadget-verification.tex
\begin{table*}[t]
  \caption{Exact gadget-verification evidence. The detector-contract receipt is
    \(\dim\ker(M^\mathsf{T})=\rho_{\mathrm{contract}}+
    \rho_{\mathrm{logical}}\) over the Choi-probe outcome relation
    \(M\); it certifies the virtual detectors and virtual output syndrome map.
    The logical-action receipt reports the rank of the logical action's Choi stabilizer
    group reproduced, with matching signs, by the realization. Neither proves
    fault tolerance.}
  \label{tab:gadget-verification}
  \small
  \begin{tabular}{@{}p{0.14\textwidth}p{0.25\textwidth}p{0.15\textwidth}p{0.13\textwidth}p{0.23\textwidth}@{}}
    \toprule
    \raggedright Case & \raggedright Verified detector-contract claim &
      \raggedright Contract receipt & \raggedright Logical-action receipt &
      \raggedright Rejected control \tabularnewline
    \midrule
    \raggedright Repetition syndrome round &
      \raggedright
      Two extraction records repeat the two virtual input syndromes; the two
      virtual output syndromes are determined by the records. &
      \raggedright
      \(M\in\mathbb{F}_2^{6\times2}\);
      \(4=4+0\); 2 virtual detectors, 2 output-syndrome rows. &
      \raggedright \texttt{idle}; rank 2 &
      \raggedright
      Swapping both record references makes 2 detectors nonzero. Omitting one
      detector gives the incomplete decomposition \(4\ne3+0\). Declaring
      \texttt{x}: sign mismatch on \(Z_{\mathrm{ref}}\bar Z\). Declaring
      \texttt{h}: \(X_{\mathrm{ref}}\bar Z\) not a stabilizer. \tabularnewline
    \addlinespace
    \raggedright Steane transversal-\(H\) round &
      \raggedright
      Six extraction records and the six-row virtual output syndrome map
      account for the swap between the three \(X\)- and three \(Z\)-check
      families. &
      \raggedright
      \(M\in\mathbb{F}_2^{18\times6}\);
      \(12=12+0\); 6 virtual detectors, 6 output-syndrome rows. &
      \raggedright \texttt{h}; rank 2 &
      \raggedright
      Retaining the unswapped adjacent-round detector contract makes all 6 detector rows
      nonzero. \tabularnewline
    \addlinespace
    \raggedright Repetition transversal CX &
      \raggedright
      Control checks repeat the control's virtual input syndrome; target checks
      equal the parity of both incoming syndromes, \(s'_t=s_c\oplus s_t\). &
      \raggedright complete; two blocks in, two out. &
      \raggedright \texttt{cx}; rank 4 &
      \raggedright
      Relay \(s'_t=s_t\) removes every cross-patch mechanism from the composed
      EDEM. \tabularnewline
    \addlinespace
    \raggedright Repetition logical-\(Z\) readout &
      \raggedright
      Outcome map: the product record is the logical result (destructive and
      block-preserving variants). &
      \raggedright complete; 1 result row. &
      \raggedright \texttt{measure\_z}; rank 1 (destructive), 2 (preserving) &
      \raggedright
      Declaring \texttt{measure\_x}: \(X_{\mathrm{ref}}\) not a stabilizer.
      Non-Clifford logical actions (\texttt{t}, \texttt{ccz}) fail closed as
      unsupported. \tabularnewline
    \bottomrule
  \end{tabular}
  \par\smallskip
  \footnotesize Evidence class: exact symbolic Clifford verification over
  \(\mathbb{F}_2\); sampling uncertainty: not applicable.  The executable
  source and acceptance tests are listed in Section~\ref{sec:gadget-study}.
\end{table*}

%% file: sections/evals/zsz-gadgets.tex
\subsubsection{An External Code Family and Its Surgery Gadgets}
\label{sec:ZSZ demos}

This study tests requirement R5 from Table~\ref{tab:ft-compiler-requirements} -- that a code family and its gadgets constructed outside the compiler can enter it as typed data, with the compiler owning selection, verification, scheduling, and emission. Nothing about the code below is built into CUDA-Q Logical. The code, its logical Pauli basis, and its surgery gadgets were constructed and certified elsewhere and are supplied to P2 through a provider; the compiler treats them exactly as it treats its built-in surface-code and Pinnacle gadgets. Two P0 programs exercise the path: a logical CNOT between two of the block's twenty logical qubits, and a Bell bank that measures ten disjoint $\bar X\bar X$ products and applies their sign corrections; the Bell bank is also the workload of the retargeting study in Section~\ref{sec:results-pinnacle-zsz-retargeting}. Both are single-round, noiseless functional validations.

\paragraph{What the researcher supplies.}
% TO DO (Yifan): this paragraph condenses your code and surgery construction to four equations and the merged-code form; the regular representations, seeds, opposite-check basis, expansion/soundness certification, and versatile-graph details moved verbatim in substance to Appendix~\ref{app:zsz-construction}. Please check nothing essential was lost. Delete this comment to sign off.
The code is a $\llbracket 100,20,8\rrbracket$ instance of the ZSZ lifted-product family~\cite{hong2026zszlp, Guo_2026_ZSZ}, a lifted product~\cite{Panteleev_2022_lifted, Breuckmann_2021_balanced} whose parity-check matrices have the $2\times5$ block form
\begin{equation}
 H_X=\begin{pmatrix}
 A&0&B&0&C^{\mathsf T}\\
 0&A&0&B&D^{\mathsf T}
 \end{pmatrix}\;,\qquad
 H_Z=\begin{pmatrix}
 C&D&0&0&A^{\mathsf T}\\
 0&0&C&D&B^{\mathsf T}
 \end{pmatrix} \, ,
 \label{eq:zsz100-five-block}
\end{equation}
with $A,B$ left-regular and $C,D$ right-regular representations of the non-abelian group
\begin{equation}
  G=\mathbb Z_5\rtimes_4\mathbb Z_4
   :=\langle\, x,y\mid x^5=y^4=1,\;yxy^{-1}=x^4\,\rangle
  \label{eq:zsz100-group}
\end{equation}
applied to the trinomials
\begin{equation}\label{eq:zsz100-trinomials}
    a=1+x+y \;,\qquad b=1+x^2+y^2 \;,\qquad c=1+x^2+y \;,\qquad d=1+x+y^2 \, .
\end{equation}
Each check has weight 9 and each qubit sits in at most 6 checks; the non-abelian twist is what avoids the low-distance obstruction of \eqref{eq:zsz100-five-block} for abelian groups~\cite{hong2026zszlp, bhardwaj2026mitten}. The logical basis is symplectic and \emph{equivariant}: writing the five data blocks in the order of \eqref{eq:zsz100-five-block},
\begin{equation}
 L_Z=\bigl(\,I\;\;0\;\;L[v_L]\;\;0\;\;0\,\bigr)\;,\qquad
 L_X=\bigl(\,I\;\;R[v_R]\;\;0\;\;0\;\;0\,\bigr) \, ,
 \label{eq:zsz100-ogs-basis}
\end{equation}
so that every logical $\bar X_i$ ($\bar Z_i$) is a group translate of one seed operator, each of weight 10, with $L_ZL_X^{\mathsf T}=I$. Equivariance is what makes the gadget set small: a surgery ancilla built for the seed is reused for every logical index by the same group action. The regular representations and seeds $v_L, v_R$ are given in Appendix~\ref{app:zsz-construction}.

A logical Pauli is measured by code surgery~\cite{Cohen_2022, Williamson_2026_gauging, cross2025improved, zheng2025high}: the code is deformed into a \emph{merged} code in which the target logical operator has become a stabilizer, one or more rounds of the merged checks are measured, and the deformation is reversed. For an $X$-type measurement the merged code is CSS with
\begin{equation}
 \tilde{H}_X=
 \begin{pmatrix}H_X&0\\ \Gamma_X &H'_X\end{pmatrix}\;,\qquad
 \tilde{H}_Z=
 \begin{pmatrix}\widehat{H}_Z&\Gamma_Z\\0&H'_Z\end{pmatrix} \, ,
 \label{eq:X-meas merged code}
\end{equation}
where $H'_X,H'_Z$ define the surgery ancilla and $\Gamma_X,\Gamma_Z$ attach it to the data code; $Z$-type measurements exchange the roles. The supplied ancillas are of two kinds. A \emph{versatile graph} measures one logical operator or one pair product and, by equivariance, one $X$ seed and one $Z$ seed cover all 40 single-logical measurements while one $XX$ and one $ZZ$ template cover all 190 pairs. A \emph{hypergraph} measures many commuting products in a single deformation but is specific to its pairing. Table~\ref{tab:zsz100-merged-resources} lists the six supplied gadgets; every merged code preserves the static distance $d=8$, certified exactly by exhaustive enumeration as described in Appendix~\ref{app:zsz-construction}. These are static properties of the merged codes and say nothing about circuit distance.

% TO DO (Yifan): table unchanged except the caption; the adjacent-ZZ row was added from the artifact bundle. Delete this comment to sign off.
\begin{table*}[t]
  \centering
  \begin{tabular}{ccccc}
    \toprule
    \textbf{Measurement} & \textbf{Surgery ancilla} &
    \textbf{Merged code} & \textbf{$\tilde{H}_X$: shape; $w_r$/$w_c$} &
    \textbf{$\tilde{H}_Z$: shape; $w_r$/$w_c$.} \\
    \midrule
    $M_X(19)$ & versatile graph & $\llbracket110,19,8\rrbracket$ &
      $50\times110;\ 10/6$ & $41\times110;\ 11/6$ \\
    $M_Z(19)$ & versatile graph & $\llbracket110,19,8\rrbracket$ &
      $41\times110;\ 11/6$ & $50\times110;\ 10/6$ \\
    $M_{ZZ}(0,19)$ & versatile graph & $\llbracket140,19,8\rrbracket$ &
      $61\times140;\ 12/6$ & $60\times140;\ 9/6$ \\
    $M_{XX}(19,1)$ & versatile graph & $\llbracket138,19,8\rrbracket$ &
      $60\times138;\ 9/6$ & $59\times138;\ 12/6$ \\
    ten disjoint $M_{XX}$ & hypergraph & $\llbracket150,10,8\rrbracket$ &
      $80\times150;\ 9/6$ & $60\times150;\ 10/8$ \\
    nineteen adjacent $M_{ZZ}$ & hypergraph & $\llbracket131,1,8\rrbracket$ &
      $50\times131;\ 11/7$ & $80\times131;\ 10/8$ \\
    \bottomrule
  \end{tabular}
  \caption{The six externally supplied surgery gadgets for the $\llbracket 100,20,8 \rrbracket$ ZSZ-LP code. $\tilde{H}_X$ and $\tilde{H}_Z$ are the merged parity-check matrices \eqref{eq:X-meas merged code}; $w_r/w_c$ are maximum row/column weights. The last row is used by no program here but is part of the verified set.}
  \label{tab:zsz100-merged-resources}
\end{table*}

\paragraph{How it enters the compiler.}
% TO DO (Yifan): the CNOT/PPM and Bell-bank descriptions and the selection checks are yours, compressed; the pipeline figure was dropped as redundant with Figure~\ref{fig:qlx-compiler-spine}. Delete this comment to sign off.
Listing~\ref{lst:zsz-gadgets} reads back what the provider binds to the validation device: the code and its parameters, the CNOT protocol, the parallel $XX$ gadget, and the gadgets' own claim discipline (static merged distance 8, circuit distance not certified). The provider supplies these as data.

\begin{lstlisting}[style=qlx-python, float, label={lst:zsz-gadgets},
caption={The executable ZSZ-LP-100 gadget definitions bound to the validation device. The 290 carriers form an all-to-all Stim validation target, not a hardware footprint.}]
from cudaq.logical.qec import zsz_css as zcss
from cudaq.logical.qec.zsz_css_artifacts import load_zsz100_parallel_xx

code = zcss.zsz_lp_100_css
device, _, provider = zcss.zsz100_css_stim_device(carrier_count=290)
mpp_lowering, cx_lowering, *_ = provider.qec_lowerings
assert (code.n, code.k, code.d.conservative_value) == (100, 20, 8)
assert cx_lowering.metadata["service_port"] == 19
assert cx_lowering.metadata["protocol"] == (
    "M_X(19),M_ZZ(c,19),M_XX(19,t),M_Z(19)"
)
assert mpp_lowering.metadata["parallel_outcomes"] == 10
parallel_xx = load_zsz100_parallel_xx()
assert (
    parallel_xx.ancilla_qubits,
    parallel_xx.static_merged_distance,
    parallel_xx.circuit_distance_certified,
) == (50, 8, False)
\end{lstlisting}

\begin{figure*}[t]
  \centering
  \begin{minipage}[t]{0.48\textwidth}
    \centering
    \textbf{(a)\, Intrablock CNOT}\par\vspace{1.5ex}
    \begin{tikzpicture}[x=0.78cm,y=0.65cm]
      \foreach \r in {0,...,3}{
        \foreach \c in {0,...,4}{
          \pgfmathtruncatemacro{\q}{4*\c+\r}
          \node[circle,fill=green!55!black,inner sep=2.8pt] (ca\q)
            at (\c,-\r) {};
          \node[font=\scriptsize,anchor=west] at (\c+0.11,-\r) {$\q$};
        }
      }
      \draw[green!20!black,line width=1.4pt] (ca0.center) -- (ca1.center);
      \fill[green!20!black] (ca0.center) circle[radius=2.8pt];
      \node[circle,draw=green!20!black,fill=green!55!black,inner sep=2.8pt,line width=1.2pt]
        (catarget) at (ca1.center) {};
      \draw[black,line width=1.2pt] (catarget.west) -- (catarget.east);
      \draw[black,line width=1.2pt] (catarget.north) -- (catarget.south);
    \end{tikzpicture}

    \par\vspace{1.5ex}
    \resizebox{0.95\linewidth}{!}{%
    \begin{quantikz}[row sep=0.18cm,column sep=0.23cm]
      \lstick{$c$} & \qw & \qw & \gate[2]{M_{ZZ}} & \qw
        & \qw & \gate{Z^{m_{XX}}} & \qw \\
      \lstick{$a$} & \gate{M_X} & \gate{Z^{m_X}} & \qw
        & \gate[2]{M_{XX}} & \gate{M_Z} & \qw & \qw \\
      \lstick{$t$} & \qw & \qw & \qw & \qw & \qw
        & \gate{X^{m_{ZZ}+m_Z}} & \qw
    \end{quantikz}}
  \end{minipage}\hfill
  \begin{minipage}[t]{0.48\textwidth}
    \centering
    \textbf{(b)\, Intrablock Bell pairs}\par\vspace{1.5ex}
    \begin{tikzpicture}[x=0.78cm,y=0.65cm]
      \foreach \r in {0,...,3}{
        \foreach \c in {0,...,4}{
          \pgfmathtruncatemacro{\q}{4*\c+\r}
          \node[circle,fill=green!55!black,inner sep=2.8pt] (cb\q)
            at (\c,-\r) {};
          \node[font=\scriptsize,anchor=west] at (\c+0.11,-\r) {$\q$};
        }
      }
      \foreach \c in {0,...,4}{
        \pgfmathtruncatemacro{\qa}{4*\c}
        \pgfmathtruncatemacro{\qb}{4*\c+1}
        \pgfmathtruncatemacro{\qc}{4*\c+2}
        \pgfmathtruncatemacro{\qd}{4*\c+3}
        \draw[green!40!black,line width=2.2pt] (cb\qa) -- (cb\qb);
        \draw[green!40!black,line width=2.2pt] (cb\qc) -- (cb\qd);
      }
    \end{tikzpicture}

    \par\vspace{1.5ex}
    \resizebox{0.496\linewidth}{!}{%
    \begin{quantikz}[row sep=0.18cm,column sep=0.28cm]
      \lstick{$\lvert0\rangle$} & \gate[2]{M_{XX}} & \gate{Z^m} & \qw \\
      \lstick{$\lvert0\rangle$} & \qw & \qw & \qw
    \end{quantikz}}
    \\ \(\text{ten copies in parallel}\)
  \end{minipage}
  \caption{P0 operations and their P2 port assignments. Green dots are the 20 logical ports; the $4\times5$ layout is an index map, not connectivity. (a) A logical CNOT on ports 0 and 1 realized as four surgery measurements through borrowed port 19 with frame corrections. (b) Ten disjoint $XX$ measurements with sign corrections, fused by the provider into one surgery epoch.}
  \Description{Two panels show twenty logical qubits as four rows of five green dots. The first panel connects logical zero to logical one with a CNOT and shows its measurement-based circuit. The second pairs every dot with its vertical neighbor and shows an XX measurement followed by a Z correction, repeated ten times.}
  \label{fig:zsz100-logical-demos}
\end{figure*}
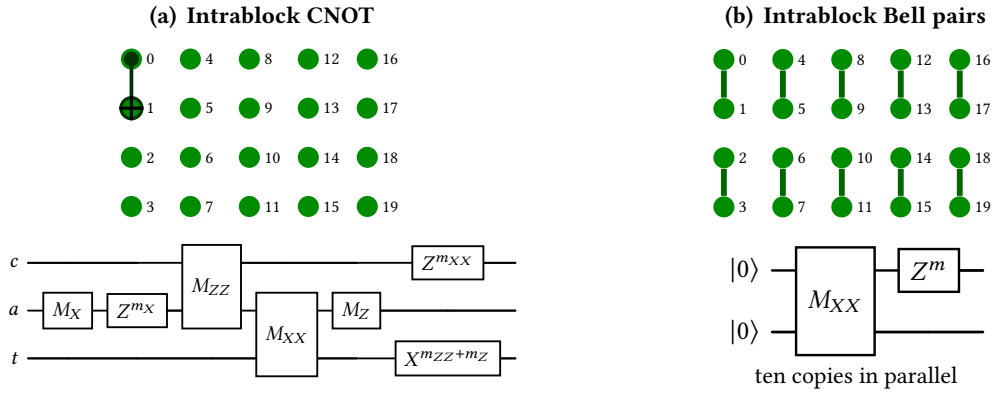

The two programs are ordinary P0 programs and P1 places them without knowing the code: the CNOT program allocates two logical values, prepares a computational-basis input, applies the CNOT, and measures both; the Bell-bank program allocates twenty values in $\lvert0\rangle$, requests $M_{X_{2j}X_{2j+1}}$ for $0\le j<10$, and applies $Z_{2j}^{m_j}$. P1 colocates the values in one compute region. At P2 the provider packs them into logical ports of one block and selects gadgets (Figure~\ref{fig:zsz100-logical-demos}). For the CNOT between control $c$ and target $t$ it selects the measurement-based realization
\begin{equation}
 M_X(a),\quad M_{ZZ}(c,a),\quad M_{XX}(a,t),\quad M_Z(a) \, ,
 \label{eq:zsz100-mbcx}
\end{equation}
on a borrowed service port $a{=}19$, with corrections $Z_a^{m_X}$, $Z_c^{m_{XX}}$, $X_t^{m_{ZZ}+m_Z}$, each measurement a versatile-graph gadget with its own attach and detach; it rejects the realization if port 19 is owned by the program. For the Bell bank it recognizes the canonical disjoint matching and selects the rank-ten hypergraph, one deformation for all ten products, preserving one outcome per product in program order; it does not construct hypergraphs for arbitrary matchings, and with fusion disabled it selects ten pairwise gadgets instead. The compiler lowers the feed-forward the program states; it does not infer corrections.

Selection is gated by checks the compiler runs on the supplied data: the merged matrices have consistent shapes and commute, the attachments agree with the physical representatives of the requested logicals, and the measured stabilizers return the requested outcomes. Beyond these, each of the six gadgets is verified against its declared logical action with the \texttt{verify\_logical\_action} of Section~\ref{sec:gadget-study}: on the $k{=}20$ block each receipt has 20 logical inputs and outputs and Choi rank 40, with the addressed qubits carrying the measured instrument and the rest the identity. The check certifies Clifford logical action without noise using the Choi-state equivalence criterion described in Section~\ref{sec:gadget-study}; negative controls exercise the implementation's rejection of incorrect declarations. Establishing the receipts rejected the provider's declared action for the adjacent-$ZZ$ path on the first run: returned owners were threaded in logical-support order rather than product order, turning the declared path into a star. The realization was correct, the declaration was not, and the fix threads owners by identity. The other half of the gadget definition of Ref.~\cite{kliuchnikov2026composing}, the gadget detector contract, is not claimed: the supplied detectors assume a freshly prepared block rather than an arbitrary incoming virtual syndrome, so these gadgets are verified realizations of their instruments but not yet composable EDEM building blocks.

\paragraph{What the compiler produces.}
% TO DO (Yifan): schedule generation, P3 mapping, and end-to-end validation are yours, compressed into one paragraph. Delete this comment to sign off.
P2 hands P3 a merged code per deformation; P3 reads each nonzero of each check family as a qubit--ancilla interaction and edge-colors the bipartite incidence graph, giving a collision-free schedule of depth equal to the maximum row or column weight, the minimum possible, though without routing or hook-safety guarantees. Each attachment runs one merged-code round and is followed after detachment by one source-code round; preparation is a Clifford encoder derived from the code. P3 assigns data, surgery-ancilla, and check-ancilla roles to carriers on an all-to-all validation target, verifies assignments and record flow, and the \textsf{Stim} backend emits the circuit with detectors and observables exposed. Noiselessly, the CNOT recovers its truth table on all four inputs and every corrected Bell pair satisfies $XX=ZZ=+1$, with no detector events; the terminal Bell-stabilizer probes are appended after emission and are not compiler-selected operations.

The costs of these realizations are the subject of Section~\ref{sec:results-pinnacle-zsz-retargeting}, which compiles the same Bell-bank program with parallel surgery, with serial surgery, and with gadgets from Pinnacle's provider on this encoding, and reports rooted operation counts, serialized depth, and per-route evidence in Table~\ref{tab:pinnacle-zsz-retargeting}. What that comparison holds fixed comes from this section: the encoded-zero preparation, one merged-code round per deformation, and the logical-action-verified gadgets of Table~\ref{tab:zsz100-merged-resources}.

\paragraph{Scope.}
% TO DO (Yifan): this paragraph replaces the caveats previously stated in several places; check the claim discipline matches yours. Delete this comment to sign off.
Everything above is exact algebra, exact static distance, and noiseless functional validation. The circuits carry no noise and produce no nontrivial detector error model or logical-failure estimate; that would need fault-tolerant preparation, repeated rounds, a hook-safe schedule, a noise model, and circuit-distance analysis. The all-to-all target models no routing, motion, connectivity, or timing, and serialized events count operations under unit timing. What the study establishes is that an externally constructed code and its gadgets enter the compiler as data, are selected and verified by the compiler's own checks, and compile through P0--P3 to executable circuits with their outcomes preserved, alongside a second provider's gadgets on the same code from the same program.

%% file: sections/evals/pinnacle-zsz-retargeting.tex
\subsection{Retargeting One P0 Program Across QEC Architectures}
\label{sec:results-pinnacle-zsz-retargeting}

A central goal of CUDA-Q Logical is to separate an application's logical
intent from the choice of error-correcting code and its physical realization.
We test this separation by compiling the same P0 program through four QEC
routes.  The program specifies only a logical measurement-and-feedback
workload; the selected architecture determines the code, block packing,
protected-measurement protocol, detector structure, and physical event graph.
The experiment therefore tests portability above code selection: the source
program remains unchanged even when the chosen code and protected realization
change.

\begin{lstlisting}[
  style=qlx-python,
  label={lst:pinnacle-zsz-portable-p0},
  caption={The portable P0 workload.  It contains no architecture, code,
  block-packing, gadget, or physical-resource choice.}
]
import cudaq.logical as ql

PAIRS = tuple((i, i + 1) for i in range(0, 20, 2))

@ql.program(name="portable_bell_bank")
def portable_bell_bank() -> tuple[
    bool, bool, bool, bool, bool,
    bool, bool, bool, bool, bool,
]:
    data = ql.allocate(20, state=ql.types.zero, name="data")
    outcomes = []
    for left, right in PAIRS:
        data[left], data[right], outcome = ql.mpp(
            ql.types.X(data[left]) @ ql.types.X(data[right])
        )
        outcomes.append(outcome)
    for (left, _), outcome in zip(PAIRS, outcomes):
        data[left] = ql.ops.z_if(outcome, data[left])
    ql.discard(data)
    return tuple(outcomes)
\end{lstlisting}

Listing~\ref{lst:pinnacle-zsz-portable-p0} allocates 20 logical qubits in
$\lvert0\rangle$, measures the ten ordered products
$\overline{X}_{2j}\overline{X}_{2j+1}$, applies a conditional
$\overline{Z}_{2j}$ correction from each result, and returns the ten original
measurement bits.  The same logical program is compiled for every route.  We
also compile a deterministic control that prepares
$\lvert+\rangle^{\otimes20}$, requests the same ordered products, and omits
feedback.  In a noiseless execution, each control observable must be zero.

\paragraph{Experimental design.}
The four routes separate two forms of retargeting.  The first three use the
$\llbracket100,20,8\rrbracket$ ZSZ-LP-100 code and thereby compare protected
realizations on a fixed encoding.  \emph{ZSZ parallel} packs the 20 logical values
into one block and realizes all ten commuting products in a single
hypergraph surgery round.  \emph{ZSZ serial} holds the code, placement,
preparation, and timing model fixed, but instead uses ten graph surgery rounds,
one per product, with the ZSZ provider's own gadgets.  \emph{Pinnacle ZSZ}
keeps the same code, preparation, and ten serial measurements, but the gadgets
come from Pinnacle's provider rather than the ZSZ provider: Pinnacle constructs
its own merged codes for the ZSZ block by the graph-surgery construction it uses
for its GB codes~\cite{webster2025explicitconstructionlowoverheadgadgets, webster2026pinnacle}.
Beyond the different construction, this route differs from ZSZ serial in one
further respect, the extraction schedule of the compiler's current realization
of the Pinnacle gadgets, described with the results below. The fourth route, \emph{Pinnacle
GB510}, exercises the wider portability boundary: it selects Pinnacle's
$\llbracket510,16,24\rrbracket$ GB510 preset, packs the 20 logical values across
two code blocks, and uses a 26-round
protocol~\cite{webster2026pinnacle,hong2026zszlp,Guo_2026_ZSZ}.

The controls are checked on the compiled artifacts.  Within each workload, all
four routes have the same serialized P0 program.  The three ZSZ-code routes
also share the same concrete encoder, preparation operation multiset,
one-round convention, and normalized preparation duration.  For the Bell
workload, each executes an encoded-zero preparation containing 808 CX and 40 H
applications with depth 61.  Consequently, differences among these three
routes arise after the common preparation rather than from an unmatched input
state.  GB510 changes both the code and the protocol, so we report it as a
retargeting result and do not compare it against the matched ZSZ routes.

\paragraph{Resource accounting.}
Table~\ref{tab:pinnacle-zsz-retargeting} counts operations reachable from the
selected P3 \texttt{phys.graph}; inactive module-scope definitions are excluded.
Stored operations measure the representation; two-qubit counts weight each
call template and repeat body by its execution multiplicity.
Physical CX and CZ applications are partitioned into preparation, protected
measurement, and feedback, and these phase counts close to the rooted total.
Detector rows are reported separately because they are P3 sidecar records
rather than executable graph operations.

Every quantum primitive is assigned unit duration, and barriers have zero
duration.  Under this timing, \emph{depth} is the length of the longest
dependency path, and \emph{peak concurrency} is the largest number of physical
qubits active in any one step.  Neither is a hardware quantity.  The
concurrency of 100 for Pinnacle ZSZ, for example, comes from its serial
extraction keeping few qubits active at once, while the construction itself
occupies more qubits than that.

\begin{table*}[t]
  \caption{Executed structural results for the Bell measurement-and-feedback workload. Stored P3 operations count the operations reachable from the selected \texttt{phys.graph} root as stored, so repeated bodies are counted once; they measure the size of the stored representation. Panel (a) partitions physical two-qubit operations by the enclosing top-level call, weighting each call template by its execution multiplicity; GB510 executes its encoded-zero preparation once per code block. Feedback contributes no two-qubit operations on any route. In panel (b), depth is the authenticated makespan under normalized unit timing. Peak concurrency is simultaneous physical-qubit activity. Logical-action evidence counts selected P2 artifacts read and checked for consistency with the requested targets; it is not an independent verification of the realization. Verified is the number of selected gadgets independently checked by the logical-action-equivalence verifier. Control reports the paired deterministic $\lvert+\rangle^{\otimes 20}$ Stim run.}
  \label{tab:pinnacle-zsz-retargeting}
  \centering
  \scriptsize
  \setlength{\tabcolsep}{3.2pt}
  \textit{(a) Rooted executed operations}\\[-0.2em]
  \begin{tabular}{@{}lrrrr@{}}
    \toprule
    Route & \shortstack{Stored P3\\operations} & Prep 2q & Protected 2q & Total 2q \\
    \midrule
    ZSZ parallel & 3,819 & 808 & 2,030 & 2,838 \\
    ZSZ serial & 24,505 & 808 & 17,120 & 17,928 \\
    Pinnacle ZSZ & 71,503 & 808 & 62,087 & 62,895 \\
    Pinnacle GB510 & 8,419,016 & 49,098 & 7,848,896 & 7,897,994 \\
    \bottomrule
  \end{tabular}

  \vspace{0.4em}
  \textit{(b) Schedule, artifact, and execution evidence}\\[-0.2em]
  \begin{tabular}{@{}lrrrrrrr@{}}
    \toprule
    Route & Events & Depth & \shortstack{Peak\\concurrency} & Detectors & \shortstack{Logical-action\\evidence} & Verified & Control \\
    \midrule
    ZSZ parallel & 3,784 & 135 & 215 & 180 & 10/10 & 1/1 & pass \\
    ZSZ serial & 24,452 & 660 & 165 & 1,790 & 10/10 & 10/10 & pass \\
    Pinnacle ZSZ & 71,450 & 67,432 & 100 & 2,774 & 10/10 & 10/10 & pass \\
    Pinnacle GB510 & 8,418,965 & 6,569,447 & 1,210 & 190,504 & 10/10 & 10/10 & pass \\
    \bottomrule
  \end{tabular}
\end{table*}

\paragraph{Evidence for selected logical actions and execution.}
We first confirm that every selected P2 construction corresponds to the
requested logical targets.  For the two ZSZ routes, the authenticated network
plan contains the ten ordered pairs
$(0,1),\ldots,(18,19)$ with operator $XX$.  For the Pinnacle routes, the selected
\texttt{fabric.gadget\_detector\_contract} records are bound back to the requested P0
pairs, including the block-local-to-global mapping required by GB510.  The
detector contracts record commuting checks, inclusion of the requested product but not
either proper factor in the stabilizer span, removal of one logical degree of
freedom, and derivation of the detector basis as the exact Choi left kernel
over input syndromes and measurement records.  All four routes provide such
evidence for all ten requested logical actions.

The \emph{Logical-action evidence} column reports consistency checks on these
provider-emitted P2 records.  The \emph{Verified} column reports a separate
semantic check: every gadget each route selects is checked with the
\texttt{verify\_logical\_action} of Section~\ref{sec:ZSZ demos}, which compares the
realization's Choi state with that of the declared Pauli-product instrument on
the logical algebra.  This verifier is part of CUDA-Q Logical; its check is
independent of the provider's evidence records, not an external software oracle.
All 31 selected gadgets pass, including the ten GB510
gadgets whose logical actions span two code blocks; the receipts certify Clifford logical action in the absence of noise.  We
separately execute 256 noiseless shots as an end-to-end check of P3-to-Stim
lowering.  Every Bell route emits ten observables with a closed detector
boundary, and the $\lvert+\rangle^{\otimes20}$ control produces ten zero
observables and no detector event on every route.  The harness adds no terminal
measurements or observables, so these outputs come from the compiled graph.
The Bell program returns measurement bits obtained before feedback and then
discards the data, while the deterministic control omits feedback.  These
sampling checks therefore do not validate the corrected final state across
the four routes.  The terminal Bell-stabilizer probes reported for the separate
ZSZ demonstration in Section~\ref{sec:ZSZ demos} are not part of this experiment.

\paragraph{Structural results.}
The matched ZSZ comparison isolates the parallel-versus-serial trade-off.
Replacing ten serial deformations with one parallel deformation reduces
protected two-qubit work from 17,120 to 2,030 operations and normalized depth
from 660 to 135, while peak concurrency increases from 165 to 215.  The parallel
route is available only because a hypergraph artifact matching this exact
pairing was constructed offline; the serial gadgets are reusable across all 190
pairs of the block.  Pinnacle ZSZ executes the same 808-two-qubit preparation,
but its protected phase contains 62,087 two-qubit operations and has depth
67,432.  The compiler's current realization of the Pinnacle gadgets measures
each merged check in turn through a single check ancilla and runs three check
rounds per deformation, whereas the ZSZ realization measures all merged checks
of a round in parallel through an edge-colored schedule and runs two.  The
depth comparison is therefore confounded by the extraction schedule and round
count; parallel scheduling alone would not reduce the executed two-qubit count.
Without a matched extraction comparison, these values characterize the current
implementations and do not isolate an intrinsic cost ratio between the two
surgery constructions.

The GB510 route changes more than the protected-measurement construction.  Its
two larger data blocks and 26-round protocol produce 7,848,896 protected
two-qubit operations, normalized depth 6,569,447, and peak concurrency 1,210.
These numbers quantify the resources selected when the unchanged P0 program is
retargeted to GB510.  They are not normalized against the one-round ZSZ routes.

\paragraph{Compilation cost.}
Table~\ref{tab:pinnacle-zsz-compiler-diagnostics} reports one isolated process
per workload--route pair.  P0--P3 time and compiler peak RSS are measured at
the P3 checkpoint.  The remaining columns cover schedule construction and
resource estimation, the independent \texttt{verify\_logical\_action} receipts, and Stim emission and
execution.  Verification is under five seconds for every ZSZ-code route and
about 960 seconds for GB510, whose ten receipts each compare Choi states on
two $\llbracket510,16,24\rrbracket$ blocks.  For GB510, a memory-bounded estimator avoids materializing millions of
individual schedule rows.  These are single-run reproduction diagnostics; they neither benchmark
CUDA-Q Logical throughput nor rank the target architectures.

\begin{table*}[t]
  \caption{Single-run compiler and reproduction diagnostics. P0--P3 is the complete compilation path and peak RSS is measured through P3. Schedule/estimate includes public schedule construction plus authenticated estimation for the three smaller routes and the memory-bounded fused estimator for GB510. Verify is the wall-clock of the independent \texttt{verify\_logical\_action} receipts for the route's selected gadgets. Total covers the complete worker process. Times and RSS are environment-sensitive diagnostics, not provider rankings.}
  \label{tab:pinnacle-zsz-compiler-diagnostics}
  \centering
  \scriptsize
  \setlength{\tabcolsep}{3.5pt}
  \textit{Compiler stages}\\[-0.2em]
  \begin{tabular}{@{}llrrrrrr@{}}
    \toprule
    Workload & Route & P0 (s) & P1 (s) & P2 (s) & P3 (s) & P0--P3 (s) & Peak RSS (MiB) \\
    \midrule
    Bell & ZSZ parallel & 0.031 & 0.009 & 3.394 & 1.236 & 4.669 & 157.203 \\
    Bell & ZSZ serial & 0.009 & 0.006 & 7.174 & 10.107 & 17.297 & 269.516 \\
    Bell & Pinnacle ZSZ & 0.012 & 0.006 & 2.835 & 0.298 & 3.152 & 204.391 \\
    Bell & Pinnacle GB510 & 0.011 & 0.007 & 509.213 & 57.559 & 566.791 & 7807.172 \\
    $XX$ control & ZSZ parallel & 0.033 & 0.016 & 3.916 & 1.156 & 5.121 & 155.312 \\
    $XX$ control & ZSZ serial & 0.011 & 0.006 & 7.122 & 9.754 & 16.893 & 264.047 \\
    $XX$ control & Pinnacle ZSZ & 0.010 & 0.006 & 2.975 & 0.293 & 3.283 & 203.734 \\
    $XX$ control & Pinnacle GB510 & 0.011 & 0.007 & 511.353 & 55.635 & 567.005 & 8780.609 \\
    \bottomrule
  \end{tabular}

  \vspace{0.5em}
  \textit{Downstream reproduction pipeline}\\[-0.2em]
  \begin{tabular}{@{}llrrrrr@{}}
    \toprule
    Workload & Route & Schedule/estimate (s) & Verify (s) & Target emit/execute (s) & Total (s) & Process peak RSS (MiB) \\
    \midrule
    Bell & ZSZ parallel & 0.161 & 0.515 & 0.087 & 5.638 & 171.953 \\
    Bell & ZSZ serial & 1.976 & 4.088 & 0.456 & 24.522 & 408.438 \\
    Bell & Pinnacle ZSZ & 7.371 & 3.698 & 1.061 & 15.780 & 646.188 \\
    Bell & Pinnacle GB510 & 136.094 & 961.001 & 140.389 & 1847.807 & 13374.734 \\
    $XX$ control & ZSZ parallel & 0.147 & 0.582 & 0.090 & 6.140 & 169.609 \\
    $XX$ control & ZSZ serial & 1.945 & 4.093 & 0.462 & 24.085 & 402.797 \\
    $XX$ control & Pinnacle ZSZ & 7.158 & 3.810 & 1.052 & 15.818 & 646.438 \\
    $XX$ control & Pinnacle GB510 & 129.179 & 959.857 & 136.782 & 1836.332 & 13744.344 \\
    \bottomrule
  \end{tabular}
\end{table*}

The three ZSZ-code routes complete P0--P3 compilation in 3.2--17.3 seconds and
the full pipeline in 5.6--24.5 seconds.  Both GB510 workloads also complete
end to end, requiring about 567 seconds through P3, about 960 seconds for gadget
verification, and 1,836--1,848 seconds for the full pipeline, with process
peak RSS near 13.4 GiB.  These measurements
show that the larger construction is compilable and quantify its present
reproduction cost; as single runs on one system, they are not throughput
benchmarks.

\paragraph{Interpretation and scope.}
The four routes demonstrate retargeting above code selection.  One P0 program is preserved while the target selects the encoding,
packing, and protected protocol, yielding different P2 constructions, rooted
P3 graphs, detector sets, schedules, and resource estimates that all lower to
executable Stim.  The matched ZSZ routes isolate realization choices on a fixed
code, while GB510 demonstrates that the same source can cross a code and block
boundary without exposing either choice in P0.

These are compiler and resource-model results.  Timing is normalized, and the validation target omits
connectivity, transport, and physical noise.  Preparation and extraction are
not claimed to be fault tolerant.  Both ZSZ routes' merged codes carry exact static
distance 8 (Table~\ref{tab:zsz100-merged-resources}); the Pinnacle gadgets' detector contracts report
unknown merged-code distance.  No route carries a
circuit-distance certificate, and this Clifford workload does not exercise
Pinnacle's magic-state path.
Section~\ref{sec:ZSZ demos} describes the ZSZ code and gadget constructions
without repeating this retargeting comparison.

%% file: sections/related-work.tex
\section{Related Work}
\label{sec:related-work}
CUDA-Q Logical builds on established work in quantum programming, fault-tolerant compilation, QEC simulation and decoding, and architecture modeling. These systems provide the algorithms, representations, and analysis engines used throughout a fault-tolerant workflow. CUDA-Q Logical supplies compiler
interfaces through which those components can contribute to or analyze the same
application build.

Fault-tolerant QPUs will differ in their codes, logical operations, resource
factories, communication, decoding, and control as well as their physical
hardware. A QPU provider can expose these choices through CUDA-Q Logical while
retaining its own compilation methods and implementations. Developers can then
hold an application fixed, compile it against different provider-supplied
architectures, and identify which choices produce the observed performance
differences. Table~\ref{tab:related-comparison} summarizes the complementary
roles of representative systems discussed in this section.

\begin{table*}[t]
    \centering
    \small
    \caption{Representative software and research components relevant to a
      CUDA-Q Logical workflow. The rows describe complementary roles and possible
      integration points. Only the CUDA-Q, Stim, EDEM, and architecture-study
      paths evaluated in this paper are implemented here; other adapters are
      future integration opportunities.}
    \label{tab:related-comparison}
    \begin{tabular}{@{}p{0.16\textwidth}p{0.22\textwidth}p{0.26\textwidth}p{0.28\textwidth}@{}}
      \toprule
      \raggedright Area &
      \raggedright Representative systems &
      \raggedright Established contribution &
      \raggedright Relationship to CUDA-Q Logical \tabularnewline
      \midrule

      \raggedright Quantum programming and transpilation &
      \raggedright Qiskit~\cite{javadiabhari2024qiskit},
        CUDA-Q and Quake~\cite{cudaqteam_quake},
        OpenQASM~3~\cite{cross2022openqasm3}, and
        MLIR-based quantum IRs~\cite{mccaskey2021mlir,peduri2022qssa,ittah2022qiro} &
      \raggedright Program construction, hybrid control, compiler passes, and
        target-aware circuit lowering &
      \raggedright CUDA-Q programs provide the application path evaluated here.
        Adapters could exchange logical programs or lowered artifacts with Qiskit
        and other frameworks. \tabularnewline
      \midrule

      \raggedright Algorithm decomposition and analytical estimation &
      \raggedright Qualtran~\cite{harrigan2024qualtran} and layered resource
        models~\cite{beverland2022assessing} &
      \raggedright Structured workloads, logical resource demand, and analytical
        cost models &
      \raggedright These outputs can supply logical workloads or independent
        baselines for estimates derived from later compiler artifacts. \tabularnewline
      \midrule

      \raggedright Fault-tolerant synthesis, layout, and routing &
      \raggedright \texttt{qstack}~\cite{paz2026qstack}, lattice-surgery
        compilers~\cite{watkins2024compiler,molavi2025dependency,tan2024lassynth},
        and QECC-Synth~\cite{yin2025qeccsynth} &
      \raggedright QEC and ISA passes, surgery synthesis, physical layout, and
        connectivity-aware routing &
      \raggedright Their algorithms can implement code- or device-specific
        lowerings behind QEC and physical provider interfaces. \tabularnewline
      \midrule

      \raggedright QEC modeling, simulation, and decoding &
      \raggedright Stim~\cite{gidney2021stim},
        Sparse Blossom~\cite{higgott2025sparseblossom},
        \texttt{deq}~\cite{microsoft2026deq}, and composable
        EDEMs~\cite{kliuchnikov2026composing} &
      \raggedright Circuit simulation, detector models, decoding, and modular
        fault analysis &
      \raggedright The evaluated workflow emits Stim circuits and composes EDEMs.
        The same compiler interfaces associate simulator and decoder results with
        the build that produced their inputs. \tabularnewline
      \midrule

      \raggedright Architecture modeling and co-design &
      \raggedright Yoked surface codes~\cite{gidney2023yoked},
        Pinnacle~\cite{webster2026pinnacle},
        Oratomic~\cite{cain2026oratomic},
        AutoQuREO~\cite{oza2026autoqureo}, and cross-layer
        analysis~\cite{su2026resource} &
      \raggedright QPU-specific codes, resource organizations, constraints,
        schedules, and cost models &
      \raggedright Providers can express architecture-specific choices while the
        application and metric definitions remain fixed across compilations.
        \tabularnewline

      \bottomrule
    \end{tabular}
  \end{table*}

\subsection{Quantum Programming Models and Compiler IRs}
Qiskit provides circuit representations, staged transpilation, dynamic-circuit
support, and an extension ecosystem for retargeting programs to new gates and
devices~\cite{javadiabhari2024qiskit}. OpenQASM~3 represents dynamic classical
control, timing, and pulse-level detail at several levels of
specificity~\cite{cross2022openqasm3}. MLIR supplies extensible dialects,
mixed-abstraction modules, verifier interfaces, and conversion infrastructure
for heterogeneous compilation~\cite{lattner2021mlir}. Quantum MLIR systems have
used this infrastructure to lower assembly programs through QIR to executable
hybrid programs~\cite{mccaskey2021mlir}. QSSA, QIRO, and CUDA-Q's Quake dialect
use SSA or value-oriented representations to expose quantum dataflow for
analysis and optimization
~\cite{peduri2022qssa,ittah2022qiro,cudaqteam_quake}.

These frameworks provide the application-level and device-facing compiler
infrastructure on which fault-tolerant workflows can build. Within CUDA-Q,
Quake remains the application IR, and CUDA-Q Logical adds representations for
logical-machine placement, QEC selection, physical event graphs, and control.
This paper evaluates that integration for CUDA-Q programs. Adapters for Qiskit
or other programming frameworks could connect at the logical-program boundary
or consume emitted artifacts, but such integrations are outside the present
evaluation.

Qualtran addresses an earlier part of the workflow by representing algorithms
as hierarchically composable building blocks, supporting simulation and
testing, and deriving architecture-independent resource
counts~\cite{harrigan2024qualtran}. Such structured workloads and counts can
supply application demand for subsequent QEC and architecture studies. They
also provide useful baselines for determining which costs are introduced by
code selection and physical scheduling.

\subsection{Fault-Tolerant and Physical Compilation}
Fault-tolerant compilers provide code- and architecture-specific transformations that QPU builders may wish to expose through a broader software workflow.
Watkins et al.\ introduce a lattice-surgery IR and a streaming pipeline that
compiles large circuits to surface-code operations
~\cite{watkins2024compiler}. Dependency-aware compilation improves
surface-code placement and routing by using operation dependencies
~\cite{molavi2025dependency}. LaSsynth formulates compact lattice-surgery
synthesis as a SAT problem~\cite{tan2024lassynth}, while QECC-Synth maps
syndrome-extraction circuits onto sparse physical connectivity using
code-specific layout flexibility~\cite{yin2025qeccsynth}. CUDA-Q Logical's
provider interfaces are intended to accommodate such specialized algorithms
while allowing their implementations and internal representations to remain in
dedicated libraries.

\texttt{qstack} composes QEC and ISA translations as passes over a quantum IR
and wraps classical callbacks so that static and dynamically generated kernels
follow the same compiler pipeline~\cite{paz2026qstack}. This work and CUDA-Q
Logical share the goal of carrying programs across fault-tolerant abstraction
boundaries without manual rewriting. Their designs emphasize different parts
of that workflow. \texttt{qstack} centers on the composition of transformations
and callbacks, while CUDA-Q Logical records architecture definitions, selected
QEC components, detector semantics, physical schedules, estimates, and control
bindings as build artifacts. A pass pipeline in the style of \texttt{qstack}
could provide transformations within such a build, with the surrounding
artifacts supporting architecture comparison and execution.

\subsection{QEC Modeling, Simulation, and Decoding}
Stim provides scalable stabilizer-circuit simulation, detector sampling, and a
practical circuit-to-DEM boundary~\cite{gidney2021stim}. Derks et al.\ develop
DEMs as a design and verification formalism spanning syndrome extraction,
minimum-weight matching for graphlike detector
models~\cite{higgott2025sparseblossom}. These systems provide the simulation
and decoding evidence needed to test circuit behavior under explicit noise and
decoder assumptions.

Kliuchnikov, Lietz, and Pastawski develop the mathematical model used by the
CUDA-Q Logical code presentation, gadget detector contract, and EDEM
interfaces~\cite{kliuchnikov2026composing}. EDEMs carry virtual syndromes and
boundary Pauli errors between gadgets, allowing gadget-local models to compose
when detector parities cross gadget boundaries. The virtual-stabilizer
measurement used for this composition originates in Microsoft's
\texttt{deq} decoding system~\cite{microsoft2026deq}, which constructs
per-gadget detector descriptions offline and assembles decoding problems,
including sliding windows, online.

CUDA-Q Logical represents these models and conventions at compiler interfaces.
A gadget detector contract records how measurements become detectors and observables; an
emitted DEM remains associated with its circuit or EDEM composition; and a
decoder binding identifies the model and record convention it consumes. This
connection lets simulation and decoding engines participate in a larger
application-to-device study without changing their scientific algorithms.

\subsection{Resource Estimation and Architecture Co-Design} 
Fault-tolerant architecture studies already show how QPUs can differentiate
through software-visible organization. Yoked surface codes separate dense
storage from accessible workspace and account for the hallways that connect
storage to operations~\cite{gidney2023yoked}. Pinnacle combines processing
units, magic engines, and optional memory connected through
ports~\cite{webster2026pinnacle}. Oratomic separates memory, processor, and
factory code blocks and teleports logical state between those
roles~\cite{cain2026oratomic}. Each architecture combines code properties,
logical operations, communication, and physical assumptions in a distinct
system design.

These studies motivate provider-defined compiler components. A QPU builder can
supply its code families, logical operations, factories, resource organization,
physical constraints, and control capabilities. CUDA-Q Logical then compiles a
fixed application against that architecture and records the costs introduced
by its choices. The resulting comparison can expose an architecture's
advantages without reducing it to a generic hardware target or requiring the
application developer to encode provider-specific details.

Beverland et al.\ establish a layered approach to full-stack estimation by
mapping application requirements through logical, error-correction, physical,
and control abstractions~\cite{beverland2022assessing}. The
Gidney--Eker{\aa} analysis shows how factories, code cycles, reaction time,
error budgets, and metric definitions shape application-scale resource
estimates~\cite{gidney2021factoring}. High-rate architecture studies likewise
combine codes, logical instruction sets, hardware connectivity, and application
schedules into distinct system models~\cite{bravyi2024highthreshold}. In this
paper, these models serve as scientific inputs and independent baselines for
compiler-derived estimates.

AutoQuREO combines reusable stack components, surrogate resource models, and
multi-objective optimization for architecture studies
~\cite{oza2026autoqureo}. Su et al.\ expose error-correction abstractions to
programmers and analyze resources across program and hardware
layers~\cite{su2026resource}. CUDA-Q Logical complements these approaches by
keeping estimates attached to the compiler builds they analyze. A researcher
can change a provider-supplied architecture, recompile the same workload, and
attribute the resulting difference to code selection, synthesis, resource
supply, scheduling, or control.

%% file: sections/conclusion.tex
\section{Conclusion}
\label{sec:conclusion}

Fault-tolerant quantum computing becomes a systems problem when an ideal
program must acquire enough architectural, error-correction, physical, and
control detail to run without losing its original logical intent.
This work began from the fragmentation that occurs when
those commitments are made in separate tools: codes become detached from their
gadgets, detector models from their circuits, estimates from their schedules,
and decoder assumptions from the execution they constrain. CUDA-Q Logical addresses that 
problem through a retargetable compilation infrastructure in which one logical workload 
is progressively refined through P0--P4 semantic profiles and every resulting artifact retains the
definitions, assumptions, and evidence that produced it.

Seen in the context of quantum programming IRs, specialized fault-tolerant
compilers, QEC simulation and decoding tools, and layered resource estimators, 
CUDA-Q Logical's distinctive contribution is a compiler-wide semantic contract rather
than a replacement for any one of those systems. Each fact has an owning level:
P0 fixes logical intent, P1 fixes the code-independent logical machine, P2 fixes
the QEC microcode and detector interpretation, P3 fixes physical events and
schedules, and P4 fixes realtime control and feedback. Verified lowerings join
these facts in an immutable build while typed provider boundaries allow the
specialized implementations behind them to remain independently developed.
The central novelty is this combination of semantic separation and continuity:
the stages remain different enough to state and check their own invariants, yet
connected enough that an estimate, simulation, or execution can be traced to
the same source program and to every target commitment added along the way.

The two resource studies show what remains visible when estimates are computed
from compiler artifacts. In the RSA-2048 study, a folded logical workload and a
factory model characterized from a compiled P3 schedule reproduce the
independently reconstructed default point of 5.046228 hours and 19,252,128
physical qubits. The retained schedule also exposes the transition from
factory-limited to reaction- and code-depth-limited execution. In the Pinnacle
study, compiling concrete synthesis circuits and their dependencies separates
discrete synthesis demand from scheduling effects that a smooth average-cost
model cannot reveal.

The QEC studies test the semantic boundary directly. The gadget experiments
certify both conditions of the gadget definition for Clifford realizations and
show that composed extended detector error models reproduce Stim's monolithic
surface-code memory model exactly. The ZSZ demonstrations carry externally
constructed qLDPC surgery gadgets from logical actions to executable Stim
circuits and verify each against its declared action. The retargeting study
then joins these capabilities: one unchanged program compiles through four QEC
routes across two codes and two providers, and the resulting resource
differences remain attributable to the selected realizations. Beyond the
evaluation, the local extended stabilizer path places decoder feedback at the
causal point where it can alter the same shot.
% TO DO: sentence extended at paper freeze to match the scoped ZSZ
% study in Section 5; a coauthor should confirm it reads as intended.
These results do not prove
hardware performance, circuit distance, or physical fault tolerance. They
establish the narrower systems result needed first: heterogeneous QEC and
architecture components can participate in one inspectable compilation lineage
without their evidence claims being conflated.

Together, these capabilities change how fault-tolerant architecture studies
can be conducted. Resource estimates become observations of explicit compiler
state rather than detached answers. Differences between targets or fidelity
levels can be traced to specific choices of code, gadget, factory, route,
timing model, schedule, or control binding. If a compilation path requires a
capability or evidence that is not available, the compiler stops at that
boundary and identifies the missing requirement. These interfaces also provide
a practical division of labor. Researchers can contribute codes, gadgets,
schedulers, simulators, and decoders in the areas they know best. Architecture
studies can combine these contributions without standardizing their internal
algorithms or reducing every system to a universal cost model.

The same interfaces make the origin of a proposal less important than the
evidence that supports it. A new code, gadget, schedule, or provider is
accepted only when it satisfies the relevant compiler checks. This makes
automated search practical. An optimizer or software agent can generate
candidates, but a candidate enters the build only after it passes the
applicable logical-action, detector-contract, and composition checks. If a candidate is
rejected, the compiler identifies both the candidate and the reason. Parts of
the ZSZ integration reported here were developed under this procedure, with
the requirement that no check be weakened to admit a candidate. Constructing
the surgery receipts exposed two defects before any circuit was run. One was
an error in the declared adjacent-$ZZ$ action. The other was an error in
the verifier's handling of MPP signs. This experience motivates future work on
agent-assisted discovery and optimization governed by compiler verification.
% TO DO: paragraph added at paper freeze. A coauthor should confirm the
% description of the AI-assisted ZSZ integration and the two reported defects
% involving the adjacent-ZZ action and MPP sign handling.

The next priority is to strengthen the evidence carried through this
compilation path. Necessary advances include hardware-backed P4 targets and
calibrated noise and timing models. Physical schedules must account for routing
and hook errors. Support must also expand to a broader range of dynamic and
non-Clifford protocols. End-to-end studies should connect composed detector
models to probability sweeps, decoding, and execution. As these capabilities
mature, the central question is no longer simply whether a proposed
fault-tolerant stack can be assembled. An architecture study should instead
identify which explicit decisions determine the stack's behavior, what
evidence supports those decisions, and how the results change when the
workload is retargeted. CUDA-Q Logical provides a common semantic foundation
for asking these questions reproducibly across diverse fault-tolerant
architectures.

%% file: sections/code-and-artifact-availability.tex
\section*{Code and Artifact Availability}

The public CUDA-Q Logical software release as of September 2026  carries CUDA-Q kernels and typed
logical programs through P0--P3. It preserves code-independent placement at
P1, selects QEC realizations at P2, and produces device-bound physical
schedules and schedule-derived resource estimates at P3. An opt-in driver
reproduces the RSA-2048 operating point reported in
Section~\ref{sec:results-rsa2048}.

Other studies use research capabilities beyond that release: compositional
detector error models, gadget verification, Clifford and non-Clifford simulation
backends, the external ZSZ integration, provider-specific retargeting
experiments, and P4 realtime control. The full capability set reported
here is planned to enter supported releases over time as its interfaces, verification,
tests, and documentation mature.

Readers may request the evaluation scripts and
configurations for each study from the corresponding authors at
\href{mailto:amccaskey@nvidia.com}{amccaskey@nvidia.com} and \href{mailto:jlietz@nvidia.com}{jlietz@nvidia.com}.

%% file: sections/acknowledgments.tex
We thank Sam Stanwyck and Fernando Pastawski for their contributions to the
design of this work and for their thoughtful review and feedback. This work
was conducted at NVIDIA. We acknowledge the use of AI-assisted tools in the
design and implementation of this work. The authors reviewed and validated all
resulting technical content and take responsibility for the final manuscript.

%% file: appendices/device-construction.tex
\section{Device Architecture Expression and Construction}
\label{sec:construction}
An architecture description details exactly what a machine can support, and a compiler 
build artifact records how one workload was compiled for that machine. This distinction is 
essential to cross-target compilation. If placement, code selection, routing, 
or scheduling are stored in the architecture, then changing the workload 
changes the machine model. If machine capabilities live only in compiler options, then the 
resulting experiment cannot be inspected or reproduced. CUDA-Q Logical therefore seeks to make architecture 
facts typed and immutable, while leaving workload-specific commitments to compiler passes and lowering. 

The object model introduced in Section \ref{sec:overview} gives this separation a small vocabulary. \emph{Descriptions} 
define reusable codes, gadgets, protocols, machines, and targets. Typed \emph{handles} connect those 
definitions and \emph{Builders} assemble and validate a description graph, and then freeze it. Compilation 
targets a device, a P0 workload, and a target-specific pipeline to produce an immutable \emph{build} 
containing the selected definition closure, artifacts, witnesses, and evidence. Figure \ref{fig:architecture-construction} 
shows the device boundary and the progressive construction sequence. 

\subsection{Devices}
\begin{figure}[b]
  \centering
  % TO DO: figure text contains "realize"; rename to "bind" or "materialize" (realization is reserved for a gadget's circuit)
  \includegraphics[width=\textwidth]{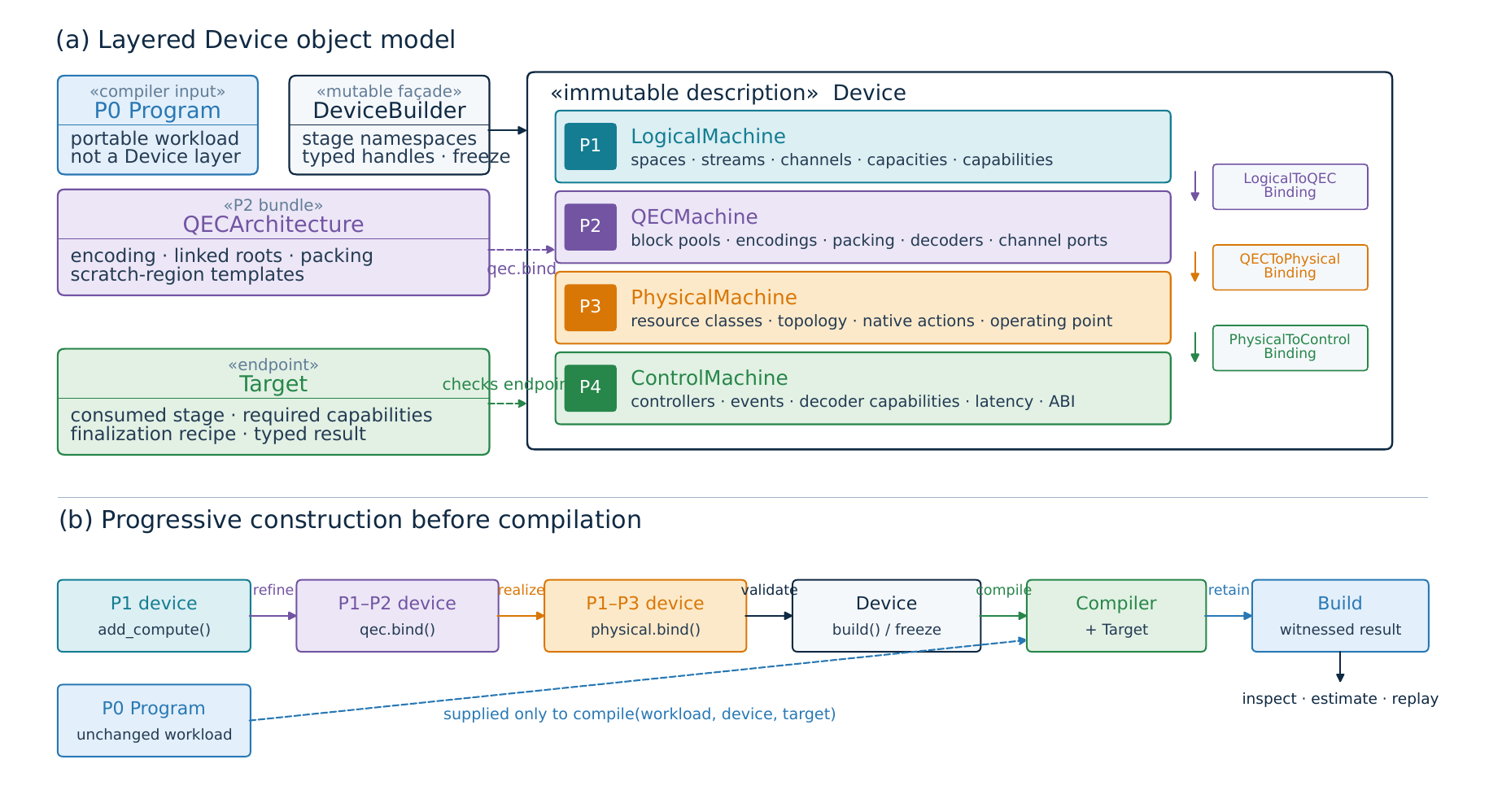}
  \caption{Layered device construction in CUDA-Q Logical. (a) The immutable
    \texttt{Device} composes a contiguous prefix of P1--P4 virtual machines and
    explicit adjacent-stage bindings. P0 workloads, P2 architecture bundles,
    the mutable builder, and targets remain separate inputs. (b) Authoring adds
    P1 logical facts, P2 QEC binding, and P3 physical materialization before
    validation freezes the device. Only then does compilation combine the
    unchanged P0 workload, device, and target into a witnessed build.}
  \Description{A two-panel UML-style diagram. The upper panel places the P0
    program, DeviceBuilder, QECArchitecture, and Target outside an immutable
    Device boundary. Inside the Device are four stacked boxes: a P1
    LogicalMachine with spaces, streams, channels, capacities, and capabilities;
    a P2 QECMachine with block pools, encodings, packing, decoders, and channel
    ports; a P3 PhysicalMachine with resource classes, topology, native actions,
    and an operating point; and a P4 ControlMachine with controllers, events,
    decoder capabilities, latency, and an ABI. Labeled binding objects connect
    each adjacent pair. The lower panel progresses from a P1 device through P2
    and P3 refinements to a Device, then combines it with a separate P0
    program and Target to produce a Build for inspection, estimation, or replay.}
  \label{fig:architecture-construction}
\end{figure}
The device stack begins at the P1 semantic layer. A \texttt{DeviceBuilder} exposes four stage-owned members and permits 
only typed refinements between adjacent layers. A device may stop at any completed layer required by its analysis or 
execution endpoint. 

\begin{wrapfigure}{l}{0.54\textwidth}
\begin{minipage}{\linewidth}
\begin{lstlisting}[
  style=qlx-python,
  basicstyle=\footnotesize\ttfamily,
  aboveskip=0pt,
  belowskip=0pt,
  caption={Constructing a logical machine and its P2 refinement.},
  label={lst:device-construction}]
compute = builder.logical.add_compute(capacity=1)
memory = builder.logical.add_memory(capacity=1)
builder.logical.add_channel(
    memory, compute, direction="bidirectional", concurrency=1)
t_states = builder.logical.add_stream(
    ql.standard.T_STATE, name="t_states", external=True)
builder.logical.add_channel(t_states, compute)
compute_qec = builder.qec.bind(compute, encoding=ql.codes.Steane)
memory_qec = builder.qec.bind(memory, encoding=ql.codes.Steane)
device = builder.build()
\end{lstlisting}
\end{minipage}
\end{wrapfigure}
A CUDA-Q Logical device is the result of this construction. Its P1
\texttt{LogicalMachine} owns code-independent regions, resource streams,
communication channels, capacities, and capabilities. Listing
\ref{lst:device-construction} demonstrates the idiomatic construction workflow.
The compute and memory handles count logical capacity and imply neither an
encoding nor a physical carrier count. Their bidirectional channel permits one
logical transfer at a time without prescribing its QEC protocol or physical
route. The external resource stream exposes $T$-state supply while
deliberately leaving its producer outside this device. A closed architecture
would instead back the stream with a typed factory protocol such as the one in
Listing~\ref{lst:connected-protocols}.

The P2 \texttt{QECMachine} owns the encoded block pools, channel ports,
encodings, packing, and decoder bindings. Because \texttt{Steane} exposes one
protected logical port per block, each \texttt{qec.bind} call derives one block
of capacity. The returned handles name the exact P2 regions that refine
\texttt{compute} and \texttt{memory}.

Calling \texttt{build()} checks refinement and ownership, then freezes the
description graph. Listing~\ref{lst:device-construction} therefore produces a
code-aware P2 device; a P1-only study could omit the QEC bindings.

\begin{wrapfigure}{r}{0.57\textwidth}
  \begin{minipage}{\linewidth}
\begin{lstlisting}[
  style=qlx-python,
  aboveskip=0pt,
  belowskip=0pt,
  caption={Defining a code, encoding, patch type, and QEC region.},
  label={lst:explicit-code-encoding}]
@ql.code
class ResearchCSS:
    block = ql.codes.CSSBlock(data=7, sx=3, sz=3)
    d = 3
    hx = ((0, 1, 2, 3),
        (0, 1, 4, 5),
        (0, 2, 4, 6))
    hz = hx
    lx = (tuple(range(7)),)
    lz = lx
ResearchEncoding = ResearchCSS.encoding(
    name="research_css_primary", 
    presentation=ResearchCSS.default_presentation,
    block="primary", logical_ports={"data": 0},
)
ResearchPatch = ql.patch[ResearchEncoding]
research_pool = ql.devices.QECRegion("research_compute", 
    encoding=ResearchEncoding, block_capacity=8, 
    packing="dense", role="compute"
)
\end{lstlisting}
  \end{minipage}
\end{wrapfigure}
The P3 \texttt{PhysicalMachine} independently owns resource classes,
topologies, native actions and instruments, and an operating point containing
timing, noise, calibration, and cost assumptions. Before freezing, a
schedule-aware study can continue the sequence in
Figure~\ref{fig:architecture-construction} by binding \texttt{compute\_qec} and
\texttt{memory\_qec} to separate carrier pools, or by supplying one pool whose
exact grouping compilation must derive. This physical refinement adds no new
meaning to the P1 regions themselves.

\subsection{From Code Algebra to Executable QEC}
\begin{figure}[b]
  \centering
  % TO DO: figure text contains "realization"; keep only if it labels the gadget body, otherwise rename to "binding"
  \includegraphics[width=0.96\textwidth]{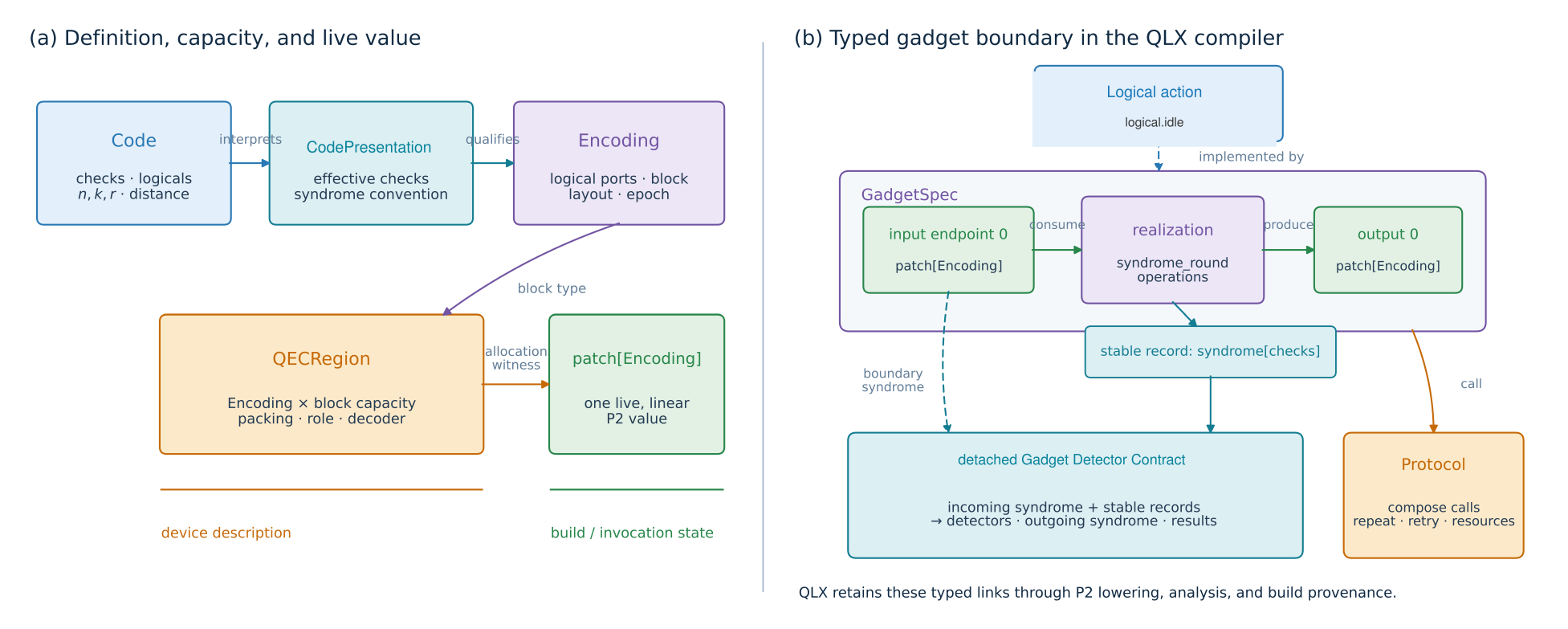}
  \caption{The connected P2 object model in CUDA-Q Logical. (a) A code supplies
    verified algebra, a code presentation supplies a syndrome-boundary
    interpretation, and an encoding exposes the reusable typed interface. A
    device's \texttt{QECRegion} provides capacity for that encoding;
    compilation allocates linear patch values from the pool. (b) A gadget
    specification relates an ideal logical action to an executable realization.
    Typed patch endpoints and stable record families expose its boundary, a
    detached detector contract supplies the syndrome and detector interpretation used by
    analysis, and a protocol composes invocations and owns control policy.}
  \Description{A two-panel diagram. The left panel shows Code leading to
    CodePresentation and Encoding, with Encoding qualifying a QECRegion containing
    a finite number of block slots. An allocation witness connects the region
    to a linear patch value. The right panel shows an ideal logical action linked to
    a gadget specification. A typed input patch enters the gadget realization
    and a typed output patch leaves it; a stable syndrome record exits below.
    A separate detector contract consumes the virtual input syndrome and stable
    record and produces virtual detectors and the virtual output syndrome. A
    protocol box composes calls and contains repeat, retry, and resource-flow
    policy.}
  \label{fig:p2-semantic-chain}
\end{figure}

The principle P2 objects are best understood as a semantic chain of types rather than independent 
descriptions. A \emph{code} states which quantum information is protected. An \emph{encoding} 
states how that code is presented at a typed compiler boundary. A \emph{QEC region} states how 
much capacity a device makes available for blocks with that encoding. A \emph{patch} is one live, linearly
owned value that inhabits such a boundary during compilation; a patch is the compiler's handle on what
Ref.~\cite{kliuchnikov2026composing} calls a code block at a gadget port. \emph{Gadgets} transform patches 
and produce \emph{records}, and gadget detector contracts interpret those records as virtual detectors and virtual output syndromes.
\emph{Protocols} compose gadget invocations into larger behaviors. Figure \ref{fig:p2-semantic-chain} places these 
roles in the CUDA-Q Logical compilation model. 

Listing \ref{lst:explicit-code-encoding} makes the first four roles explicit. The class body defines a seven-data-qubit 
CSS code. The separate \texttt{encoding(...)} call then names one reusable compiler-facing view of that 
algebra -- it selects a code presentation, names the primary block boundary, and maps the code's single
protected logical degree of freedom to the port data. Finally, the patch alias names the corresponding 
dynamic value type, while the region declares that a device can host eight blocks 
of that type. 

The code is the invariant algebraic object. It owns the qubit carrier partitions and the stabilizer, 
protected-logical, and gauge structure needed to distinguish the codespace. 
Construction verifies commutation, rank, and canonical logical pairing and derives $n = 7$, 
$k = 1$, and $r = 0$. The bare $d = 3$ is normalized as a distance claim. The \texttt{CSSBlock} 
partitions the formal frame into data and syndrome-ancilla sets used by realizations -- its indices are 
not physical qubit indices, this gets formalized at the P3 level. 

A code presentation is a boundary interpretation of that algebra. It fixes the,
possibly over-complete, set of stabilizer generators whose signs form the
virtual syndrome at a block boundary, and hence the syndrome convention used
by analysis. The ordinary case is unambiguous, so CUDA-Q Logical derives
\texttt{ResearchCSS.default\_presentation}. A different measured basis,
metacheck construction, or dynamic-code interpretation can instead select a
different presentation without redefining the code. This separation between code
algebra and its syndrome-boundary interpretation follows the compositional
model developed by Kliuchnikov et al. \cite{kliuchnikov2026composing}.

An encoding is not an encoding circuit. It is an immutable interface that
binds a code and presentation to an ordered set of logical ports, a primary block
boundary, an optional hierarchy or layout, and an epoch schema. The explicit
encoding in Listing~\ref{lst:explicit-code-encoding} changes only the reusable
interface---in particular, it gives logical port 0 the semantic name
\texttt{data}. For the canonical view, \texttt{ql.patch[ResearchCSS]} remains
concise syntax for the compiler-generated default encoding. Preparation,
conversion, growth, and code switching are executable gadgets; they are never
hidden inside an \texttt{Encoding} value.

A \texttt{QECRegion} belongs to the device, not to the dynamic program. It is
a finite pool described by an encoding, encoded-block capacity, packing rule,
role, and optional decoder binding. It does not identify which workload value
occupies a block or which physical carriers host it. In normal device
authoring, \texttt{builder.qec.bind(...)} constructs the primary region while
refining a P1 logical region; the explicit constructor shown here is also used
for architecture-owned region templates such as scratch or factory capacity.
Compilation later records the allocation witness that maps P1 owners to region
slots and, after P3 lowering, to concrete carriers.

By contrast, \texttt{ResearchPatch} is a type for one invocation-time value.
A patch carries the selected code, presentation, encoding, and epoch in its typed
boundary. It is linear: an operation consumes the current owner and returns the
updated owner, while partition views such as \texttt{block.data} and
\texttt{block.sx} borrow from that same value. The distinction between region
capacity and patch ownership lets one device support many workloads
without embedding any workload-specific allocation in the architecture.

\subsection{Gadgets, Detector Contracts, and Protocol Composition}
\label{sec:gadget-protocol-context}

\begin{wrapfigure}{l}{0.55\textwidth}
  \begin{minipage}{\linewidth}
\begin{lstlisting}[
  style=qlx-python,
  aboveskip=0pt,
  belowskip=0pt,
  caption={Defining a typed syndrome-extraction gadget.},
  label={lst:typed-syndrome-gadget}]
@ql.gadget(implements=ql.logical.idle)
def syndrome_round(
    block: ResearchPatch,
) -> ResearchPatch:
    block = ql.reset(block.sx)
    block = ql.h(block.sx)
    block = ql.cx(block.sx, block.data,
        schedule=ql.codes.Schedule.HX)
    block = ql.h(block.sx)
    block = ql.reset(block.sz)
    block = ql.cx(block.data, block.sz,
        schedule=ql.codes.Schedule.HZ)
    block, syndrome = (ql.ops.read_syndrome_ancillas(
            block, record="syndrome"))
    return block
\end{lstlisting}
  \end{minipage}
\end{wrapfigure}
A gadget turns the preceding static vocabulary into a typed transformation.
Its logical action states the ideal logical behavior to preserve; its realization
consumes and produces patches and classical results. In the common case,
CUDA-Q Logical infers the gadget interface from the annotated parameters, return type,
and record-producing operations. Listing~\ref{lst:typed-syndrome-gadget}
therefore defines a one-input/one-output block flow without restating a port
schema by hand.

The annotated argument and return value become distinct typed endpoints even
though both use the diagnostic name \texttt{block}. Their encoding, presentation,
epoch, state, and ownership qualifications must agree with the inferred
one-to-one transform flow. The outgoing patch is the new linear owner; the
incoming patch cannot remain live. Preparation, terminal measurement,
code-changing transforms, merge, and split use the same interface model with
zero, one, or several endpoints on either side. Those less regular boundaries
may require an explicit interface assertion, but they do not introduce a
second notion of a patch.

The readout also establishes a stable classical record family. Records are the
realization outcomes of Ref.~\cite{kliuchnikov2026composing}, given compiler
identities so that detector contracts and protocols can name them. A record's identity is
attached to the producing operation and result path, not to the local Python
variable \texttt{syndrome} or a later MLIR SSA name. This stable identity lets
analysis and protocol composition refer to the result after tracing, inlining,
or lowering. Classical records may be copied when their schema permits;
encoded patch owners may not.

\begin{wrapfigure}{r}{0.54\textwidth}
  \begin{minipage}{\linewidth}
\begin{lstlisting}[
  style=qlx-python,
  aboveskip=0pt,
  belowskip=0pt,
  caption={Attaching detector and observable interpretations to a gadget.},
  label={lst:gadget-detector-contract}]
@ql.gadget(implements=ql.logical.measure_z)
def logical_z_probe(block: ResearchPatch,
) -> tuple[ResearchPatch, bool]:
    block, _ = ql.extract_syndrome(block)
    block, outcome = ql.mpp(ql.types.Z(block[0]))
    return block, outcome

records = logical_z_probe.records
checks = records.syndrome(0).checks
logical_z_analysis = ql.gadgets.GadgetDetectorContract(
    logical_z_probe,
    detectors=(ql.gadgets.Detector(
        logical_z_probe.inputs.only().syndrome
        ^ checks),),
    observables=(ql.gadgets.LogicalObservable(
        records.product(0), index=0),),
    boundary={
        logical_z_probe.outputs.only().syndrome:
            checks},
    name="logical_z_analysis",
)

noise = ql.noise.NoiseModel((ql.noise.Depolarizing2(rate=1e-3),),
    name="entangling_noise",
)
dem_build = ql.detectors.build_dem(ql.compile(logical_z_probe),
    analysis=logical_z_analysis,
    noise=noise,
)
dem_text = ql.targets.dem_of(dem_build,
    condition_zero_input_seams=True,
    project_boundary_outputs=True,
    categorical_faults="approximate_independent",
)
\end{lstlisting}
\end{minipage}
\end{wrapfigure}
The virtual syndrome at a gadget boundary is related to, but distinct from, the
measurements produced inside the realization. An input endpoint carries the
\emph{virtual input syndrome}: the signs of the code presentation's stabilizers as
established by the classical record of earlier gadgets. A detached gadget detector
contract interprets that syndrome together with stable record families to
define \emph{virtual detectors}, observable or success results, and the
\emph{virtual output syndrome map} that fixes the outgoing virtual
syndrome~\cite{kliuchnikov2026composing}.

Selecting a different contract can therefore change the analysis
interpretation without copying or modifying the executable realization.
Conversely, it cannot change the gadget's operations, ownership flow, or
schedule.

An idle syndrome round has no logical result to label as an observable.
Listing~\ref{lst:gadget-detector-contract} therefore uses a closely related
logical-$Z$ probe: it extracts one syndrome and then performs a nondestructive
logical measurement. This produces both kinds of stable evidence needed to
show the detector-contract boundary honestly---check records for virtual detectors and a
product record for the logical result.

The detector contract is a separate immutable object, not code embedded in the gadget.
The gadget's typed input endpoint supplies the virtual input syndrome, and
the check-record accessor denotes the stable family produced by the extraction
operation. Their vector parity creates one virtual detector per check. The
product record produced by \texttt{mpp} is attached as logical observable 0;
this is the gadget's outcome map. Finally, the \texttt{boundary} entry is the
virtual output syndrome map: it carries the extracted checks to the outgoing
patch endpoint, where a later invocation consumes them as its virtual input
syndrome.

These attachments are structural: they refer to typed endpoints and record
families rather than Python variable names or textual IR. The detector-contract verifier
checks their dimensions and, for supported Clifford realizations, checks that
the virtual detectors are deterministic in the fault-free branch and that the
outcome map and virtual output syndrome map agree with the realization.
\texttt{verify\_gadget} runs this detector-contract check together with the
logical-action--realization equivalence check, so a passing receipt certifies both
conditions of the gadget definition (Section~\ref{sec:gadget-study}). A protocol may select this
detached contract through its \texttt{analysis=} argument. Choosing a different
valid contract changes the interpretation used by analysis, not the operations
executed by the probe.

The latter half of Listing~\ref{lst:gadget-detector-contract} shows one immediate use of
that separation. \texttt{build\_dem} propagates a selected noise model through
the compiled realization and interprets the resulting record effects through
\texttt{logical\_z\_analysis}. The result is an immutable P2 EDEM
build containing the fault schema, probabilities, response matrix, selected
detector contract, and provenance. \texttt{dem\_of} then emits Stim-compatible DEM text:
detector indices refer to the contract's virtual detectors, while \texttt{L0}
refers to its logical observable 0. The same artifact can be inspected in
CUDA-Q Logical, sampled through the DEM target, or passed to a decoder such as Stim or
PyMatching.

The two boundary options are explicit because this standalone probe has live
virtual input and output syndromes. They condition the former to zero and
project out the latter for this isolated study, which turns the EDEM into a
closed DEM; a closed preparation--memory--
readout protocol needs neither assumption. Likewise, two-qubit depolarizing
noise contains mutually exclusive Pauli alternatives that flat Stim DEM text
cannot represent exactly. CUDA-Q Logical rejects that projection by default; the
\texttt{categorical\_faults} argument requests and records the displayed
independent-prior approximation.
\begin{wrapfigure}{r}{0.55\textwidth}
  \begin{minipage}{\linewidth}
\begin{lstlisting}[
  style=qlx-python,
  aboveskip=0pt,
  belowskip=0pt,
  caption={Composing gadgets into memory and production protocols.},
  label={lst:connected-protocols}]
@ql.protocol(implements=ql.logical.idle)
def memory_cycle(
    block: ResearchPatch) -> ResearchPatch:
    return ql.ops.repeat(4,
        carries=(block,),
        body=lambda value: syndrome_round(value),
    )

prepare_y = ql.gadgets.stabilizer_preparation(
    ResearchEncoding,
    logical_stabilizers=(ql.types.Y(0),))

@ql.protocol(implements=ql.logical.produce(
        ql.standard.Y_STATE))
def y_state_factory(
) -> ql.types.resource[ql.standard.Y_STATE]:
    payload = ql.ops.allocate_patch(
        ResearchEncoding)
    payload = prepare_y(payload)
    return ql.ops.pack_resource(
        payload, kind=ql.standard.Y_STATE)
\end{lstlisting}
  \end{minipage}
\end{wrapfigure}

Figure~\ref{fig:p2-semantic-chain}(b) deliberately stops at this interface. The
detailed algebra for detector-contract construction and EDEM composition
is a specialized QEC contribution in its own right~\cite{kliuchnikov2026composing}. CUDA-Q Logical's role is to give
that model a stable place in the compiler: typed endpoints and records at P2,
explicit detector-contract selection for analysis, provenance through lowering, and
well-typed virtual-syndrome ports between invocations.

A protocol composes these callable boundaries into a folded graph. In
Listing~\ref{lst:connected-protocols}, \texttt{memory\_cycle} carries one patch
through four linked gadget invocations, while \texttt{y\_state\_factory}
allocates a patch in the same encoding and consumes it into a typed resource.
The protocol owns iteration, resource flow, retry, timeout, and fallback
policy; none of that policy is stored on the gadget realization or detector contract.

The explicit carry preserves patch ownership across folded iterations, while
invocation-qualified record paths keep repeated syndrome results distinct. The
factory's output type is the protocol boundary; its packed patch retains the
encoding and preparation provenance needed for estimation or later expansion.
A device supplies finite production and scratch capacity through QEC regions,
and compilation witnesses which regions and protocols satisfy each request.

Together, these distinctions expose the choices a co-design study needs without
turning every concern into one monolithic object. Codes and encodings remain
reusable across devices; regions can change capacity or packing without
rewriting gadgets; detector contracts can change detector interpretation without
rewriting realizations; and protocols can change composition or control policy
without changing the meaning of their component gadgets. The selected objects
and the links among them are retained in the immutable build, so later lowering,
resource estimation, simulation, and execution all refer to the same P2
commitments.

\subsection{QEC Architectures}
An \emph{architecture} is the coherent assembly that connects these reusable
objects. It combines logical capabilities, code and encoding choices, supported
logical actions, gadget and protocol roots, resource-production models, packing and
scratch rules, lowering strategies, and the physical or control capabilities
required below them. The source-level \texttt{QECArchitecture} closes the P2
portion of that composition; a device binding connects it to logical spaces and,
when available, to physical and control machines. An architecture is thus
more than an estimator configuration: for every supported logical action it must
either participate in lowering and expose the resulting evidence, or report
that the path is unsupported.

\begin{wrapfigure}{l}{0.53\textwidth}
  \begin{minipage}{\linewidth}
\begin{lstlisting}[
  style=qlx-python,
  aboveskip=0pt,
  belowskip=0pt,
  caption={Composing a QEC architecture.},
  label={lst:qec-architecture}]
scratch = ql.devices.QECRegion(
    "syndrome_scratch",
    encoding,
    block_capacity=1,
    packing="dense",
    role="scratch")

research_arch = ql.devices.QECArchitecture(
    name="research_css",
    encoding=encoding,
    link_roots=(memory_cycle, y_state_factory),
    packing="dense",
    auxiliary_regions=(scratch,))

study = ql.devices.DeviceBuilder("ArchStudy")
compute = study.logical.add_compute(capacity=8)
primary = study.qec.bind(
    compute, architecture=research_arch)
(scratch_qec,) = primary.auxiliary_regions
\end{lstlisting}
  \end{minipage}
\end{wrapfigure}

Listing~\ref{lst:qec-architecture} turns the running memory and
resource-production stack into one complete P2 architecture. Its auxiliary
region is a template for attempt-local QEC workspace, not a second P1 logical
owner.

Beyond its name, this \texttt{QECArchitecture} has five semantic fields.
\texttt{encoding} fixes the primary code boundary. \texttt{packing} states how
P1 owners occupy its logical ports. \texttt{link\_roots} names only typed QEC
definitions---codes, presentations, encodings, gadgets, detector contracts, protocols, and QEC
lowerings---from which construction seals the reachable definition graph.
The auxiliary-region tuple contains finite \texttt{QECRegion} templates whose
role must be \texttt{scratch}; each specifies an encoding, block capacity,
packing, and optional decoder. Finally, immutable metadata may identify a
family or source without changing these semantics.

Binding the architecture is intentionally stronger than passing
\texttt{encoding=}. The builder selects the architecture's encoding and packing
together, derives the primary pool capacity from the P1 capacity and code rate,
and expands every scratch template under the primary region's name. In the
example, \texttt{scratch\_qec} is the returned handle for the template prefixed
with \texttt{compute}. It appears in the P2 machine but not in the P1 logical
machine; the architecture owns its purpose, and a P3 device author must bind
the handle rather than reconstructing its generated name. Packing cannot be
overridden at the bind site, and
\texttt{encoding=} and \texttt{architecture=} are mutually exclusive.

The linked roots bound provider visibility and provenance. Their transitive
closure includes definitions referenced through typed gadget and protocol
boundaries and declared dependencies. A provider selected for one workload may
derive a site-specialized gadget within that closed transaction, but it
cannot introduce an unrelated late definition into the architecture. Adding a
new root, packing rule, or scratch template therefore constructs a new
\texttt{QECArchitecture}; it does not mutate one already retained by a device
or build.

The Pinnacle and ZSZ architectures elaborated on in Section \ref{sec:evaluation} 
illustrate the breadth of this contract. Pinnacle combines
generalized-bicycle compute and memory blocks, dense logical-port packing,
preparation and storage operations, transversal operations, magic-resource
engines, and joint-measurement and rotation protocols. Published block and
engine footprints remain named analytical or physical inputs until a device
author supplies the mapping needed for P3~\cite{webster2026pinnacle}. The ZSZ
architecture uses the same object roles for finite lifted-product codes,
graph-gauging ancillas and checks, merge/extract/split protocols, frame recovery,
and reconfigurable-atom materialization. Keeping those definitions separate makes
the source code family, graph-surgery construction, extraction schedule, and
physical assumptions independently inspectable~\cite{hong2026zszlp}. An external
architecture enters through the same contracts rather than modifying the
semantic core.

\subsection{Providers and Capability Negotiation}
\begin{wrapfigure}{r}{0.54\textwidth}
  \begin{minipage}{\linewidth}
\begin{lstlisting}[
  style=qlx-python,
  aboveskip=0pt,
  belowskip=0pt,
  caption={Adding a typed MPP lowering provider.},
  label={lst:mpp-provider}]
from my_qec.codes import ResearchCSS
from my_qec.surgery import specialize_mpp

encoding = ResearchCSS.default_encoding

@ql.qec.qec_lowering(
    logical_action=ql.logical.mpp, codes=(encoding,),
    requires=(ql.architecture.capability.lattice_surgery,),
    dependencies=(syndrome_round,))
def research_mpp(site, context):
    return specialize_mpp(site, context)

research_arch_with_mpp = (ql.devices.QECArchitecture(
        name="research_css_with_mpp",
        encoding=research_arch.encoding,
        link_roots=(*research_arch.link_roots,
                    research_mpp),
        packing=research_arch.packing,
        auxiliary_regions=research_arch.auxiliary_regions))
\end{lstlisting}
  \end{minipage}
\end{wrapfigure}
CUDA-Q Logical separates an open extension contract from the disclosure policy of an
implementation. A collaborator may publish a complete architecture package so
others can inspect, modify, and compile toward it; distribute an out-of-tree
provider that emits standard CUDA-Q Logical artifacts; or expose an opaque service that
returns only the artifact, report, and evidence named by its contract. All
three participate through typed capabilities rather than private compiler
forks.

The distinction remains visible. Every provider supplies a stable identity and
version, accepted types and capabilities, output schema, assumptions, and the
validation evidence it makes available. CUDA-Q Logical can verify the shared boundary and
the returned artifact, but it does not infer properties hidden behind an opaque
provider. A shared architecture can therefore support independent compilation
and detailed audit, while a proprietary component supports interoperability
with a deliberately narrower evidence claim.

\emph{Providers} connect external, independently authored definitions to compiler action in the 
CUDA-Q Logical model. A provider declares the logical-action family it implements, accepted code and
encoding boundaries, and required machine capabilities. Listing \ref{lst:mpp-provider}
shows the fully inspectable source-package form and completes the running example. The same
contract may be supplied by an independently distributed implementation or a service adapter
without changing how the compiler checks its boundary.
Specifically, we demonstrate how one might contribute custom lattice surgery 
compilation for Pauli-product measurements. The provider returns an ordinary gadget or 
protocol and CUDA-Q Logical performs compatibility and whole-result verification before selection. The derived 
value reuses the immutable encoding, packaging, and existing roots, while adding 
\texttt{research\_mpp} through the constructor that reseals and revalidates the closure. 
\texttt{syndrome\_round} remains in that closure even though it is reached through \texttt{memory\_cycle}. The
architecture still does not select a multi-Pauli-product (MPP) gadget for a particular
workload. That decision occurs only when a placed P1 action site is matched
during compilation.

Negotiation is performed over the definitions linked into the experiment. The
compiler first filters candidates by logical action, code and encoding boundary,
placement, device and target capabilities, dependencies, and evidence policy.
It then applies the requested selection objective and a deterministic tie-break,
records the chosen provider and specialization in a witness, and seals the
selected transitive closure into the build. Import order and process-local
object identity do not participate in this decision.

\subsection{Authoring and Co-Design Loop}

\begin{wrapfigure}{l}{0.58\textwidth}
  \begin{minipage}{\linewidth}
\begin{lstlisting}[
  style=qlx-python,
  aboveskip=0pt,
  belowskip=0pt,
  caption={Retargeting an architecture study.},
  label={lst:co-design-loop}]
def device_for(architecture):
    builder = ql.devices.DeviceBuilder("Study")
    compute = builder.logical.add_compute(capacity=8)
    builder.qec.bind(
        compute, architecture=architecture)
    return builder.build()

device = device_for(pinnacle.gb30)
build = ql.compile(
    pbc_workload,
    pipeline=ql.compiler.pipelines.qec(),
    device=device)
estimate = ql.analysis.estimate(
    build, tier=ql.analysis.Tier.STATIC)

candidate_device = device_for(
    research_arch_with_mpp)
\end{lstlisting}
  \end{minipage}
\end{wrapfigure}

Listing \ref{lst:co-design-loop} follows the same lifecycle for built-in and
external components. Only the builder and its namespaced handles are mutable.
Calling \texttt{build()} validates cross-field and adjacent-level bindings,
detaches caller-owned containers, and freezes the device. Compilation then
derives workload-specific assignments and selected gadgets and protocols, and analysis
observes the resulting build. Authors supply the facts that cannot be derived:
code algebra, noncanonical boundaries, provider specialization, architectural
capabilities, physical assumptions, and evidence. CUDA-Q Logical derives canonical
views, unique mappings, assignments, and witnesses when their inputs determine
one answer.

Here \texttt{compute} is a typed handle into the builder, \texttt{device} is a 
description, and \texttt{build} is the workload-specific compiler
product. Replacing \texttt{pinnacle.gb30} with
\texttt{research\_arch\_with\_mpp} creates a new device and, on recompilation,
a new build without rewriting
\texttt{pbc\_workload}. The two builds can be inspected at the same semantic
level because each retains its selected definitions, witnesses, assumptions,
and analysis inputs. This is the concrete construction loop required by R2:
target, analyze, revise the architecture, retarget, and reanalyze while holding
logical meaning fixed. Section~\ref{sec:levels} next defines how these authored
objects are represented and verified as P0--P4 compiler contracts.

%% file: appendices/semantic-profiles.tex
\section{Progressive IR Contracts from Logical Intent to Control}
\label{sec:levels}

The construction model in Section~\ref{sec:construction} supplies the objects
that a CUDA-Q Logical experiment may use. This section follows those objects through the
compiler. The organizing principle is the \emph{semantic stage}: P0--P4 state
which facts have become commitments, which facts remain deliberately unknown,
and which evidence justifies the transition to the next stage. A dialect is a
vocabulary used to express those commitments. The two notions are related but
not interchangeable. For example, \texttt{cflow} may preserve a loop from P0
through P3, and a P1 \texttt{lvm.kernel} may retain \texttt{qlx} action
attributes. 

Table~\ref{tab:stage-dialect-map} gives the semantic compiler layering. Each row answers the same five
questions: what enters, which new facts become admissible, which earlier meaning
must be preserved, what the verifier establishes, and which artifact and witness
may proceed. Facets such as detector interpretation, noise, scheduling, and
evidence attach to the stage that owns their facts. 
A compiler path may also stop early when its endpoint does
not require later commitments. Detector-model analysis can terminate at a
verified P2 artifact, offline physical simulation at P3, and realtime execution
continues through P4.

We use the nondestructive logical measurement of $X\otimes Y\otimes Z$ from
Section~\ref{sec:overview} as a running example. P0 fixes the observable and its
logical state transition. P1 decides where its three logical owners reside. P2
selects a code-compatible measurement gadget and detector interpretation.
P3 maps that gadget's realization to physical carriers, events, routes, and a schedule.
P4 binds the scheduled dependencies to controller and decoder tasks. The
fragments below emphasize the fact added at each boundary rather than presenting
five unrelated programs.

\begin{table*}[b]
  \caption{The P0--P4 semantic spine. Dialects in the third column are the
  principal vocabularies, not exclusive namespaces: retained \texttt{qlx},
  \texttt{cflow}, and \texttt{event} operations may cross several stages when
  their semantics remain valid.}
  \label{tab:stage-dialect-map}
  \small
  \setlength{\tabcolsep}{2.5pt}
  \begin{tabular}{@{}p{0.06\textwidth}p{0.16\textwidth}p{0.17\textwidth}p{0.27\textwidth}p{0.27\textwidth}@{}}
    \toprule
    \raggedright Stage & \raggedright Canonical root
      & \raggedright Principal vocabulary
      & \raggedright New commitment
      & \raggedright Verified output and evidence \tabularnewline
    \midrule
    \profile{P0}
      & \raggedright \texttt{ql.program}
      & \raggedright \texttt{qlx}, \texttt{cflow}, \texttt{event}
      & \raggedright Ideal outcome-free actions and instruments, typed outcomes,
        linear logical ownership, and structured control
      & \raggedright Target-independent logical build. Frontend and verifier
        evidence for logical-action classification, ownership, and admissible P0
        operations \tabularnewline
    \midrule
    \profile{P1}
      & \raggedright \texttt{lvm.kernel} on an \texttt{lvm.domain}
      & \raggedright \texttt{lvm} plus retained logical and control vocabulary
      & \raggedright Logical residency, capacity, capability, communication
        intent, and stable action-site identity
      & \raggedright Placed logical build plus a replayable placement witness.
        no code, physical route, or timing claim \tabularnewline
    \midrule
    \profile{P2}
      & \raggedright Selected \texttt{fabric.protocol} closure
      & \raggedright \texttt{fabric} plus retained structured control and events
      & \raggedright Codes, presentations, encodings, gadgets, detector contracts, protocols, record
        interpretation, and detector semantics
      & \raggedright Verifier-accepted QEC network plus selection and proof
        evidence. No physical coordinate or schedule claim \tabularnewline
    \midrule
    \profile{P3}
      & \raggedright \texttt{phys.graph}, optionally with \texttt{phys.schedule}
      & \raggedright \texttt{phys} plus retained \texttt{cflow} and \texttt{event}
      & \raggedright Physical allocation occurrences, native events, records,
        routes, timing, noise bindings, and resource exclusion
      & \raggedright Physical build plus mapping, routing, record-projection,
        provenance, and, when scheduled, timing witnesses \tabularnewline
    \midrule
    \profile{P4}
      & \raggedright Portable control plan bound to the P3 build
      & \raggedright Runtime/controller contracts such as \texttt{rt\_sched},
        \texttt{rt\_decode}, and \texttt{rt\_abi}
      & \raggedright Destinations, clocks, queues, transports, decoder sessions,
        deadlines, feedback, and result ABI
      & \raggedright Deployable control plan and run-provenance contract.
        hardware latency claims require a target that verifies them \tabularnewline
    \bottomrule
  \end{tabular}
\end{table*}

\subsection{P0: State the Logical Action}

P0 accepts logical intent without a machine. Its root, \texttt{ql.program},
contains ideal actions and instruments over values of type
\texttt{!ql.logical\_qubit}. Each value represents exclusive access to one
logical degree of freedom at one program point. \texttt{ql.prepare} begins an
SSA ownership chain. A state-preserving operation consumes the current access
value and returns its successor, while destructive measurement and discard end
the chain. This convention makes no-cloning and lifetime errors ordinary
dataflow errors rather than properties reconstructed after lowering.

The \texttt{qlx} dialect owns logical values, ideal logical-action declarations, and
their application. An outcome-free \emph{action} is deterministic, outcome-independent, and
ownership-preserving. An \emph{instrument} may create or end logical degrees of
freedom and may return classical outcomes. Built-in Clifford actions can be
carried as attributes on \texttt{ql.apply}. Custom logical actions are symbols whose
signatures and bodies state ideal behavior independently of any realization.
The \texttt{cflow} dialect carries structured repetition, conditionals, and
loops, while \texttt{event} carries asynchronous readiness handles. Neither
vocabulary commits the program to a code or target.

A custom logical action is represented by \texttt{ql.decl.action} or
\texttt{ql.decl.instrument}. The symbol may be interface-only, or it may carry
a P0 body that defines the ideal behavior. For Clifford bodies, the
\texttt{clifford\_action} attribute records the composed logical action as a
binary symplectic matrix and phase vector. The attribute verifier establishes
symplecticity; the frontend separately establishes that the tableau agrees with
the body. This division is representative of the stage contract: well-formed
metadata is not, by itself, evidence that the declared logical action is correct.

For the running example, the logical instrument consumes current access to
three owners, returns their successor states, and produces one Boolean result.
Its observable is the Hermitian Pauli product $X\otimes Y\otimes Z$.
Listing~\ref{lst:p0-xyz} makes those facts explicit.
The listing intentionally says nothing about whether the measurement is
transversal, surgery-based, ancilla-mediated, or implemented by a native
multi-Pauli operation. Those are realization choices, so admitting them at P0
would prevent the same workload from being compared across architectures.

\begin{wrapfigure}{r}{0.50\textwidth}
  \begin{minipage}{\linewidth}
\begin{lstlisting}[
  style=qlx-mlir,
  % basicstyle=\scriptsize\ttfamily,
  aboveskip=0pt,
  belowskip=0pt,
  caption={The P0 product-measurement logical action and its target-independent
    ownership transition.},
  label={lst:p0-xyz}]
ql.decl.instrument @measure_xyz :
    (...) -> (..., i1)

ql.program @trial : (...) -> i1
    attributes {ql.stage = "p0"} {
  %a1, %b1, %c1, %m = ql.instrument @measure_xyz(
          %a0, %b0, %c0)
      : (...) -> (..., i1)
  ...
}
\end{lstlisting}
  \end{minipage}
\end{wrapfigure}

Authors can reach this representation through \texttt{@ql.program} and
\texttt{@ql.logical\_action}, or by importing supported CUDA-Q/Quake input. The
decorators retain lazy Python definitions; \texttt{ql.compile} materializes one
by tracing it with proxies for MLIR SSA values. With
\texttt{@ql.logical\_action(kind="auto")}, the signature supplies an initial family:
an \texttt{i1} result or unequal logical input and result arities selects an
instrument, and other signatures initially select an action. Tracing may only
strengthen that classification. Preparation, measurement, discard, selection,
or a reachable instrument call requires an instrument; if the signature
initially selected an action, the author must request
\texttt{kind="instrument"}. An explicit false \texttt{kind="action"} claim is
rejected, as are runtime and resource-orchestration operations inside an
logical-action body.

The two import paths make equally explicit promises. \texttt{import\_quake}
accepts pre-normalized, value-semantic Quake. \texttt{import\_cudaq} specializes
a CUDA-Q kernel's arguments, collects reachable helpers, and runs CUDA-Q
normalization and linear-value passes before using the same Quake-to-P0
converter. That converter deliberately accepts a strict subset: supported
gates and measurements, constant rotations, constant-trip loops, and
single-level adaptive conditionals. Controlled rotations, negated controls,
dynamic angles, operations without a P0 conversion, and logical values left
live at the boundary fail closed.
\begin{wrapfigure}{r}{0.55\textwidth}
  \begin{minipage}{\linewidth}
\begin{lstlisting}[
  style=qlx-mlir,
  % basicstyle=\scriptsize\ttfamily,
  aboveskip=0pt,
  belowskip=0pt,
  caption={An abridged P1 logical machine and its placed product measurement.
    Capabilities, capacity, resource supply, residency, and action-site identity
    are explicit; code and physical materialization remain absent.},
  label={lst:p1-xyz-lvm}]
lvm.domain @machine {
  lvm.space @compute {
    capabilities = [#lvm.capability<"logical_compute">,
      #lvm.capability<"/logical_measurement">
    ], capacity = 3 : i64
  }
  lvm.space @factory {
    capabilities = [#lvm.capability<"logical_factory">]
  }

  lvm.stream @T {
    backing_region = @factory,
    produces = @t_state, capacity = 1 : i64
  }
  lvm.channel @T_supply {
    from = @factory, to = @T,
    capabilities = [#lvm.capability<"resource_transfer">]
  }
  lvm.channel @T_delivery { from = @T, to = @compute,
    capabilities = [#lvm.capability<"resource_transfer">]
  }
}
lvm.kernel @trial_placed on @machine {
^bb0(%a0: !lvm.logical_qubit<
          @machine::@compute>, ...):
  %a1, %b1, %c1, %m = lvm.instrument @measure_xyz(
        %a0, %b0, %c0)
      at [@machine::@compute, @machine::@compute,
          @machine::@compute] {site = 0 : i64}
      : (...) -> (..., i1)
  ...
}
\end{lstlisting}
  \end{minipage}
\end{wrapfigure}

Every frontend must ultimately materialize a P0 root, normalize supported
control flow to \texttt{cflow}, and establish path-sensitive linear use. It must
classify each logical action using both its interface and reachable behavior. Local
declaration verification alone cannot justify an action claim when a callee
performs measurement, preparation, or another instrument.
The P0 verifier therefore rejects later-stage dialect operations, invalid
logical-action bodies, and broken ownership chains. Its folded representation is
also the input to two logical estimators: the C++
\texttt{qlx-estimate-logical} pass and a Python profile walker. Both visit a
\texttt{cflow.repeat} body once and multiply by its trip count; the C++ pass also
follows \texttt{ql.call} without inlining. With checked arithmetic and a
sum-of-arms upper bound for dynamic conditionals, it reports action and
instrument histograms, idle and discard sites, resource requests and
consumption, \texttt{synthesis\_demand}, peak live owners, and a serial-depth
upper bound. Recursion, overflow, unknown logical operations, unsupported
region operations, and dynamic \texttt{while} result in an error in the C++ pass; the
Python walker is deliberately more permissive about dynamic and unknown
regions. These remain P0 observations rather than machine costs, as formalized
in Appendix~\ref{app:resource-analysis}. The output is a verified, target-independent
build whose observable, typed result, and successor-state meaning must survive
every later refinement.
P0 has now answered \emph{what} the program requests. It has not answered where
the three owners can reside or whether the logical machine can support their
joint use. That is the sole new question at the P0-to-P1 boundary.

\subsection{P1: Bind the Logical Action to a Logical Machine}
P1 combines a verified P0 build with an immutable logical virtual machine
(LVM), placement constraints, and placement preferences. It preserves the
logical instrument in Listing~\ref{lst:p0-xyz} but qualifies each owner with
an \texttt{lvm.space}. The result remains code-independent: a P1 slot represents
logical residency and capacity, not a physical qubit, ion, resonator, or encoded
block. P2 may later pack several compatible owners into one encoded block
without changing the meaning of the P1 capacity witness.

An \texttt{lvm.domain} declares the available logical structure. An
\texttt{lvm.space} is a typed pool of capacity whose capabilities state which
classes of logical use it supports. Roles such as \emph{compute} and
\emph{memory} are descriptive shorthand. Capability declarations, rather than
the role name, determine admissibility. An \texttt{lvm.stream} exposes a typed
logical resource supply, such as a magic-state stream, without fixing how that
resource is prepared. An \texttt{lvm.channel} records code-independent
communication capability between components. Ordinary local owners reference a
space directly; \texttt{lvm.placement} is reserved for nonlocal distributed,
trajectory, or topological bindings. The placed executable operation is the
action site. Its deterministic \texttt{site} ordinal gives P2 selection evidence
a durable handle; a materialized \texttt{lvm.action\_site} declaration can retain
the same logical action and placement facts when a symbolic form is required.

Listing~\ref{lst:p1-xyz-lvm} previews the resulting P1 artifact. The compute
space exposes the capabilities needed by the logical action and capacity for its
three simultaneously live owners. The factory-backed stream and its two
channels show how production and delivery remain logical supply facts rather
than QEC or physical mechanisms. In the kernel, placement-refined types and the
\texttt{at} list record residency, while \texttt{site = 0} names the
product-measurement action site. The \texttt{on} clause statically binds the
kernel to one domain. Operations such as \texttt{lvm.prepare},
\texttt{lvm.apply}, and \texttt{lvm.measure} preserve their P0 logical
semantics---including inline \texttt{\#ql.action} attributes---and add only the
residency commitment.

In the running example, the three qualified operands reside in
\texttt{@compute}. Their exact local slots belong to the immutable Build
placement witness rather than semantic IR; the domain states only that three
slots exist and support the requested logical use. Simultaneously live owners
must occupy distinct slots; a slot may be reused only after its previous
owner's lifetime has ended on every control-flow path. The \texttt{@T} path is
part of the reusable logical machine, but the product measurement does not
consume it. P1 still cannot say which lattice-surgery seam, transversal
circuit, cat state, or physical path will realize that measurement. This
boundary is important: failure to find a legal P1 placement is an
architectural-capability result, whereas failure to find a QEC gadget is a P2
result.

The public entry points converge on that contract. A logical machine may be
declared with \texttt{@ql.machine} or obtained from
\texttt{DeviceBuilder.logical}, and \texttt{ql.compiler.place} binds a P0 build
to it. The native \texttt{qlx-to-lvm} pass is a deterministic C++ baseline: it
infers required capabilities, chooses the first compatible space and lowest
available slot, and preserves supported folded control flow. Direct bindings
authored through \texttt{PlacedBuilder} are constraints, not an escape hatch;
they pass through the same materialization and verification path.

Placement policy is separated from P1 validity through immutable
\texttt{PlacementProblem} and \texttt{PlacementPlan} artifacts exposed by
\texttt{placement.problem}, \texttt{placement.solve}, and
\texttt{placement.apply}. The problem records P0 lifetimes, LVM structure, hard
constraints, soft preferences, and the placement objective. On the current
implementation, \texttt{solve} supports the compiler-owned
\texttt{strategy="first\_fit"}, and \texttt{apply} reconstructs that result and
checks its input and machine witness before accepting the plan. The problem and
plan nevertheless define the research seam: a future strategy may propose a
complete assignment, but that assignment is authoritative only as a request,
not trusted as proof. Materialization must preserve the requested bindings and
independently check capabilities, capacity, lifetime overlap, slot exclusivity,
and supported hard constraints; compiler-owned metrics are recomputed from the
emitted \texttt{lvm.kernel}. Placement policy can therefore evolve without
changing what counts as a legal P1 artifact.

\begin{wrapfigure}{r}{0.52\textwidth}
  \begin{minipage}{\linewidth}
\begin{lstlisting}[
  style=qlx-mlir,
  % basicstyle=\scriptsize\ttfamily,
  aboveskip=0pt,
  belowskip=0pt,
  caption={Abridged P2 binding of the placed product-measurement site. The
    protocol selects a gadget and one detached detector interpretation.},
  label={lst:fabric-gadget-detector-contract-protocol-ir}]
fabric.gadget @xyz_mpp for @xyz_mpp_spec
    realization @xyz_body(
      %a: !fabric.patch<...>,
      %b: !fabric.patch<...>,
      %c: !fabric.patch<...>)
    -> (..., i1)

fabric.gadget_detector_contract @xyz_analysis for @xyz_mpp {
  fabric.detector() {
    records = ["xyz.check.0", "xyz.check.1"],
    label = "xyz_consistency"
  }
  fabric.observable() {
    records = ["xyz.result"], index = 0 : i64
  }
}

fabric.protocol @measure_xyz : (...) -> (..., i1) {
  %a1, %b1, %c1, %raw = fabric.call @xyz_mpp(
        %a0, %b0, %c0) {detector_contract = @xyz_analysis}
      : (...) -> (..., i1)
  fabric.protocol_return
      %a1, %b1, %c1, %raw : ...
}
\end{lstlisting}
  \end{minipage}
\end{wrapfigure}
The P1 verifier establishes four properties needed downstream. Every placed
owner still has one linear SSA lifetime. Simultaneously live owners occupy
distinct slots unless the LVM explicitly models another admissible relation.
Every operation is supported by the capabilities of its bound space and
channels. Finally, references to the domain, spaces, streams, channels, and
action sites resolve to one immutable machine description. The placement plan,
recomputed metrics, and source identities form the replayable witness.

P1 has answered \emph{where} the logical action is supported. The stable
site ordinal now gives selection a precise input, but the artifact still
contains no code, encoding, check-extraction method, or detector interpretation.
P1-to-P2 lowering must supply those facts without altering the
$X\otimes Y\otimes Z$ logical action or its placed owner flow.

\subsection{P2: Select and Verify a QEC Binding}

P2 consumes the placed action site, the QEC architecture bound to its LVM
spaces, and the linked candidate definitions. The \texttt{fabric} dialect owns
the QEC vocabulary. A \texttt{fabric.code} records the stabilizer and protected
logical Pauli operators of a code, together with gauge pairs when present. A
\texttt{fabric.code\_presentation} states how checks are extracted and interpreted,
including redundant checks, metachecks, dynamic epochs, and optional recovery
contracts. A \texttt{fabric.encoding} gives that algebra a block interface with
named logical ports. Keeping these symbols distinct allows the same code to be
studied under multiple code presentations or encodings.

A \texttt{fabric.machine} binds those definitions to finite, code-qualified QEC
regions and selected interconnects. Within the executable, a
\texttt{!fabric.patch<@code,@encoding,@epoch>} value denotes exclusive access to
one encoded block at one epoch. State-preserving operations consume it and
return a successor; destructive measurement or disposal ends the lineage.
Syndrome and raw gauge records carry the same code, encoding, and epoch
qualification, so an incompatible handoff is a type error rather than a
downstream inference from names.

The executable side has a parallel separation. A \texttt{fabric.logical\_action}
names ideal behavior, a \texttt{fabric.gadget\_spec} declares its typed QEC
boundary, and a \texttt{fabric.gadget} supplies one code-local realization. A
specification distinguishes in--out, input-only, and output-only patch ports and
requires every endpoint to participate in exactly one block flow. Optional
record schemas and affine outcome maps connect named measurement records to the
Boolean results of the ideal logical action. The gadget body may then use
code-qualified preparation, carrier views, gauge or stabilizer extraction,
product measurements, code transformations, frame updates, and typed resource
consumption. A \texttt{fabric.protocol} composes gadgets and protocols without
copying their bodies. It owns bounded retry, selection policy, resource routing,
and temporary block lifetimes around those code-local realizations.

For $X\otimes Y\otimes Z$, selection matches the placed \texttt{site = 0}
action to a gadget or
generated protocol whose port table accepts the three chosen encodings and whose
declared logical action is the same Pauli-product instrument. The selected closure
contains the code, presentation, encoding, logical action, specification, realization,
detector contract, and every transitive dependency. Listing~\ref{lst:fabric-gadget-detector-contract-protocol-ir}
shows the central relationship. The protocol calls one selected gadget.
The detached gadget detector contract gives the resulting records their virtual-detector
and virtual-output-syndrome meaning. Selecting a different contract changes interpretation
evidence without silently rewriting the realization.

The Python surface preserves the same decomposition with immutable code,
presentation, encoding, gadget, protocol, and gadget-detector-contract objects. Authors use
\texttt{@ql.gadget} for a realization and \texttt{@ql.protocol} for a network.
The gadget's \texttt{implements=} binding records the claimed logical action;
verification, rather than that spelling alone, establishes signature, port, and
outcome-map compatibility. Compilation materializes the code, code presentation, and
encoding before the logical action, gadget specification, realization, and detached
detector contract that refer to them. High-level operations retain a typed expansion contract:
\texttt{ql.extract\_syndrome}, for example, expands to code-derived ancilla
reset, Hadamard, CX, and measurement operations, and each validated metacheck
becomes a declarative \texttt{fabric.detector} row over the stable records.

Detector semantics are a facet of this P2 artifact rather than another stage.
A \texttt{fabric.gadget\_detector\_contract} names stable measurement records and declares
virtual-detector, logical-observable, success, and virtual-output-syndrome parities. Runtime
data remain separate values: operations such as
\texttt{fabric.submit\_syndrome}, \\ \texttt{fabric.await\_correction}, and
\texttt{fabric.decode\_bit} mark where records are produced or consumed. The
contract states how to interpret those records, while P4 later decides where and
when a decoder task runs.

P2 verification combines structural checks with bounded semantic proofs. The
structural layer checks code and presentation matrices, encoding qualification, port
lifecycles, complete endpoint flow, call signatures, unique record producers,
and total outcome maps. For supported permutation realizations, CUDA-Q Logical also proves
the induced logical action. It maps each transformed logical or gauge generator
into a basis containing stabilizers, logicals, and gauges, discards stabilizer
coordinates, and compares the remaining action with the declared logical action. For
the three-qubit repetition code, for example, reversing carrier order sends the
declared $Z_L=Z_0$ to $Z_2$. Because
\(Z_2=Z_0(Z_0Z_1)(Z_1Z_2)\), both representatives act identically on the code
space. The GF(2) reduction makes that equivalence explicit rather than relying
on symbol names.

Supported Clifford gadget detector contracts receive a separate proof, the
encoder--unencoder sandwich of Ref.~\cite{kliuchnikov2026composing}. A symbolic
fault-free run represents every input logical state by entangling each input
with an untouched reference and represents each virtual input syndrome bit by a
free bit. Measurement randomness contributes additional free bits. Every
record and virtual output syndrome is then an exact affine function over GF(2).
Declared virtual detectors and accepted-success parities must be constant,
virtual output syndromes must be determined by the realization outcomes and
virtual input syndromes, and the declared rows must span the complete
deterministic subspace, the gadget's outcome code. This establishes the
detector contract's correctness independently of logical-action equivalence.

Those proofs are intentionally scoped. Signature and parity checks do not prove
fault tolerance, distance preservation, or general non-Clifford equivalence.
Such claims require attached algebraic proof, fault propagation, simulation, or
decoding evidence. P2 lowering fails if selection leaves unresolved intent, if
encodings or ports do not match, or if required evidence is absent. Its output
is therefore a selected, verifier-accepted QEC network and its evidence closure,
not merely a pointer to a gadget with a suggestive name.

P2 has answered \emph{how} the placed logical action is encoded and interpreted.
The patches, records, and protocol still have no physical coordinates, native
routes, or global timestamps. P2-to-P3 lowering must project this exact selected
closure onto a compatible physical machine.

\subsection{P3: Project QEC Semantics to Physical Events}

A P3 root is a \texttt{phys.graph} that names one \texttt{phys.machine} and,
when needed, a \texttt{phys.operating\_point}. It expresses a backend-neutral
physical materialization as allocation occurrences, native actions and instruments,
typed records, resource transfers, and dependencies. A value of type
\texttt{!phys.state<@resource>} represents exclusive access to the state held by
one declared physical allocation at one event. \texttt{phys.acquire} begins the
graph-local lifetime, state-preserving events return successors, and
\texttt{phys.release} ends it. The resulting SSA chains provide physical
dataflow for verification and scheduling.

The machine vocabulary makes the physical meaning behind that dataflow
explicit. A \texttt{phys.resource\_class} declares a finite carrier class and
its linked native \texttt{phys.action} and \texttt{phys.instrument} processes;
a \texttt{phys.resource} is one allocation occurrence whose class and index
identify the concrete carrier. An action is outcome-free. An instrument
produces a stable \texttt{!phys.record<@schema>}; a nondestructive measurement
also returns the successor state, whereas a destructive one returns only the
record. Classical conditions, decoder inputs, and Boolean combinations consume
those typed records rather than rediscovering measurements from event names.

P3 also represents asynchronous supply and delivery with linear
\texttt{!phys.event} and \texttt{!phys.resource\_payload} values. Typed
dependencies distinguish state, record, control, resource, order, timing,
readiness, and folded-region causality. Consequently, \texttt{phys.delay},
\texttt{phys.fence}, and \texttt{phys.barrier} remain distinct operations for
occupation, effect ordering, and synchronization. Detector, observable,
selection, allocation, routing, hierarchy, and noise sidecars bind stable graph
identities to their semantic authorities without masquerading as timed events.

\begin{figure*}[t]
  \centering
  \includegraphics[width=\textwidth]{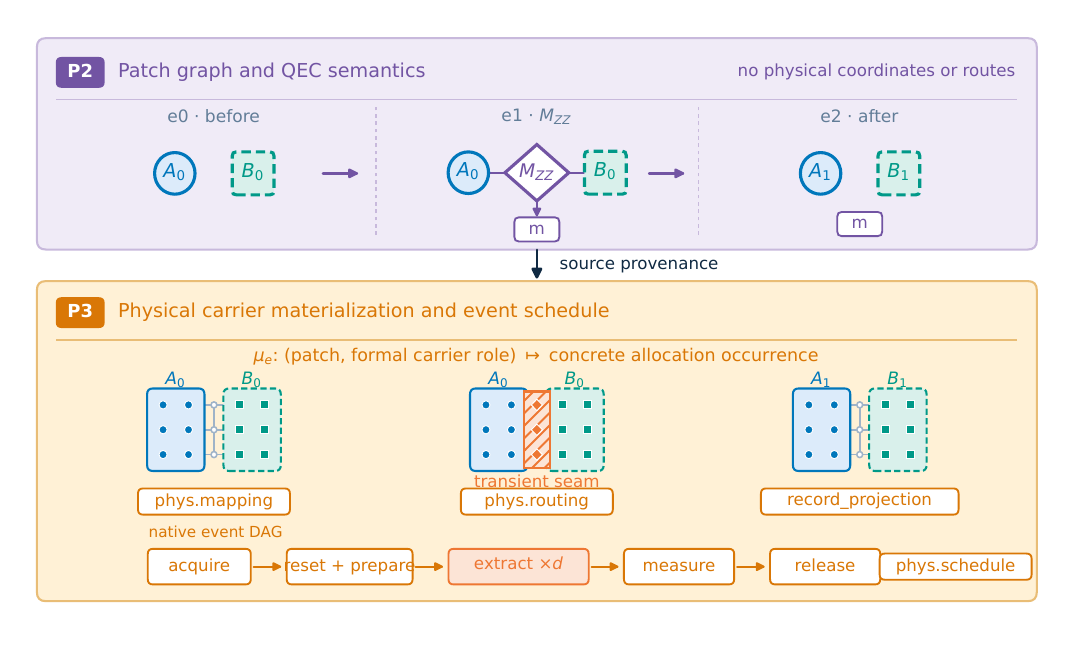}
  \caption{A P2-to-P3 lowering witness for a nondestructive joint
    \(M_{ZZ}\) interaction. The P2 row records typed patch owners, the selected
    QEC interaction, epochs, and record semantics without physical coordinates.
    P3 maps formal carrier roles to concrete allocation occurrences, implements
    the interaction with topology-legal transient workspace and native events,
    and retains mapping, routing, record-projection, provenance, and schedule
    artifacts. This is the local witness shape used when projecting a larger
    product-measurement protocol.}
  \Description{Two aligned rows follow a joint ZZ measurement through three
    epochs. The upper purple P2 row shows separate patches A and B, a joint
    measurement hyperedge, successor patches, and the logical result record.
    The lower orange P3 row shows the same carrier topology at each epoch:
    separate carrier groups before the operation, an active hatched seam with
    native events during the operation, and restored ownership afterward.
    Sidecars name the mapping, routing, record projection, and schedule
    artifacts retained by lowering.}
  \label{fig:patch-carrier-lowering}
\end{figure*}

The transition from the running P2 protocol to P3 is a projection, not a second
QEC selection. Each formal carrier role is mapped to a concrete
\texttt{phys.resource} allocation occurrence. The projector expands the chosen
realization into topology-legal preparation, movement, interaction,
measurement, reset, and release events, and it turns P2 record roles into typed
\texttt{!phys.record} values. Mapping, routing, record-projection, and provenance
sidecars explain which physical subgraph implements each source operation.
Figure~\ref{fig:patch-carrier-lowering} zooms in on an $M_{ZZ}$ interaction of
this kind. The same witness shape applies to the constituent interactions of
the three-patch product-measurement protocol.

Static architecture authoring remains separate from that workload-specific
projection. \texttt{DeviceBuilder} assembles resource classes and topology,
native actions and instruments, operating points, and typed P2-to-P3 bindings.
It neither emits workload SSA nor chooses allocation occurrences; calling
\texttt{build()} freezes the declarations as an immutable \texttt{Device} for a
later projection to select.

P3 supports carrier- and patch-granular physical models. A carrier-granular
resource denotes one physical carrier. A patch-granular resource denotes a
schedulable encoded patch with a stated footprint, avoiding the need to
materialize every carrier in a large estimate. Granularity does not move code,
encoding, or logical meaning out of P2. At either resolution, allocation,
event, record, and folded-call identities must be deterministic, and the
projector must preserve every linear state and resource owner.

Hierarchy also remains explicit. A region-bearing \texttt{phys.call} receives
the current physical states as block arguments and returns their successors
through \texttt{phys.yield}.
Its \texttt{callee} retains the selected P2 definition, whereas
\texttt{instance} and \texttt{event\_id} identify this physical occurrence.
Each invocation therefore has a unique instance path even when it shares a
canonical body. A \texttt{cflow.repeat} can carry one typed body template and
its state across many dynamic iterations, while \texttt{phys.call\_template}
uses \texttt{template\_event} to name a previously verified body without
reusing that invocation's records or runtime identity. Bounded
\texttt{cflow.if} and \texttt{cflow.while} regions preserve the same ownership
discipline through explicit arguments and \texttt{cflow.yield}. Body sharing is
not analytical multiplicity. Even when
\texttt{state\_boundary\_elided} omits an all-state identity boundary, the
scheduler reconstructs readiness, internal exclusion, and final availability
from the canonical body. Folded structure therefore reduces IR size without
making repeated work disappear from causality, occupancy, or estimation.

An unscheduled \texttt{phys.graph} is a partial order. SSA dependencies,
resource exclusion, explicit ordering, timing constraints, and structured
ancestry constrain execution but do not assign every event a global timestamp.
Scheduling is a separate P3 facet represented by \texttt{phys.schedule}. Its
rows bind stable event identities to start times, durations, resources,
predecessors, and hierarchy. The deterministic baseline scheduler constructs a
canonical \texttt{greedy\_asap} schedule; it is a reproducible construction, not a
makespan optimizer. Its verifier reconstructs dependencies, resource keys,
earliest starts, and structured envelopes from the graph rather than trusting
submitted rows. A schedule that is otherwise legal but shifts, pads, or
reclassifies an event is therefore rejected as a different schedule rather than
accepted under the canonical witness.

Physical lowering and scheduling consequently have distinct failure modes.
Projection rejects unsupported native operations, missing routes, invalid
topology, incomplete record bindings, duplicated owners, and unresolved
sidecars. Scheduling rejects conflicting reservations, unavailable inputs,
invalid durations, and rows that do not match the claimed construction. A
scheduled P3 build can support occupancy, movement, critical-path,
native-event-rate, detector-bandwidth, and utilization analyses, as developed in
Appendix~\ref{app:resource-analysis}. An unscheduled graph may still support mapping,
routing, noise, or target analysis, but it cannot claim schedule-derived
quantities.

The generic graph is also the physical extension contract. A hardware dialect
or provider must declare which generic event semantics it refines and supply
serialization, verification, and legalization for every added invariant.
Targets derive behavior from linked \texttt{phys.action} and
\texttt{phys.instrument} process declarations, never from suggestive symbol
spelling. Nonlocal interactions, out-of-capacity resources, missing causal
producers, duplicated owners, unresolved sidecars, and unauthenticated compact
plans consequently remain verifier errors even for provider-specific lowering.

P3 has answered \emph{on which physical resources, with which dependencies, and
under which schedule} the QEC protocol can run. It has not yet assigned
controller destinations, decoder workers, transport endpoints, bounded queues,
or response deadlines. Those facts belong to P4.

\subsection{P4: Bind Realtime Control}

P4 consumes the verified P3 graph and schedule together with the detector and
observable semantics retained from P2. It adds the facts needed by an endpoint
whose classical work participates in the same shot: instruction destinations,
clock domains, controller fragments, decoder sessions, typed input windows,
queues, transports, waits, branches, correction latches, deadlines, and result
ABI. Portable contracts such as \texttt{rt\_sched}, \texttt{rt\_decode}, and
\texttt{rt\_abi} describe these obligations. A target later lowers them to its
native controller and transport mechanisms.

For the product measurement, P2 already determines which physical records form
detectors and which parity represents the $X\otimes Y\otimes Z$ result. P3
determines when those records are produced and which later events depend on
them. P4 binds a detector window to a decoder session, assigns the task to a
worker and queue, states when its correction becomes visible, and connects that
typed correction to a later frame update or branch. It may not reorder detector
rows, reinterpret the observable, replace the selected decoder model, or hide a
schedule dependency.

The P4 verifier checks the resulting join across stages. Every decoder input
must resolve to the ordered P2 record schema and to P3 producers. Every task and
response must have an authenticated destination and a compatible ABI. Queue,
transport, backpressure, and deadline claims must be supported by the selected
control machine and target. Feedback edges must preserve the schedule's causal
order, and result provenance must identify the build, controller, decoder,
model, and deployment binding that produced it. Unsupported dynamics, missing
latency capabilities, or unbounded backpressure fail closed rather than being
silently converted to offline postprocessing.

CUDA-Q Logical currently distinguishes functional realtime semantics from a hardware
timing claim. The local execution path described in
Section~\ref{sec:execution-model} opens decoder tasks inside a shot and applies
their typed corrections before dependent computation continues. That path
exercises the P3-to-P4 causal contract, but it does not establish controller
deadlines, transport jitter, or sustained decoder throughput on hardware. A
hardware target must verify those additional capabilities before its P4 output
can support such claims.

The P4 artifact is a deployable control plan with a run-provenance contract.
Execution is not P5: a launch policy selects the requested observation and a
target selects a verified lowering recipe for an already-defined P2, P3, or P4
artifact. Section~\ref{sec:execution-model} develops that boundary and explains
why offline emission, sampling, detector-model analysis, and realtime execution
can share one target mechanism without pretending that every endpoint consumes
the same semantic stage.

\subsection{Operations and Types That Persist Throughout the Pipeline}

Several mechanisms recur because they preserve meaning across transitions
rather than belonging to only one stage. First, linear SSA ownership provides
the common dataflow discipline. P0 threads logical owners, P1 threads placed
logical owners, P2 threads encoded patches and syndromes, and P3 threads
physical states and records. The type becomes more committed, but a lowering
must still account for the predecessor, successor, creation, and termination of
every owner.

Second, structured control and hierarchy remain folded. \texttt{cflow} is not a
P0-only dialect. It carries explicit arguments and yields wherever repeated or
conditional work must preserve typed owners. Calls, protocols, and physical
templates similarly preserve body sharing while assigning each dynamic
invocation its own identities and effects. Analyses may multiply a verified
summary by a repeat count, but neither lowering nor estimation may confuse a
shared definition with one execution.

Third, stable identity and provenance connect the evidence chain. Structural
P0 sites acquire placement identities at P1, selection witnesses cite those
sites at P2, mapping and record-projection sidecars cite selected P2 symbols at
P3, and control bindings cite P3 event and record identities at P4. Every pass
derives a new immutable build from its verified input. Consequently, changing a
placement heuristic, gadget provider, mapper, scheduler, decoder, or target
creates a new inspectable branch of the experiment rather than mutating the
meaning of an earlier artifact.

Finally, verification follows ownership of facts. Each stage verifies the
commitments it introduces and preserves the verified interface it consumes. It
does not retrospectively redefine earlier meaning or claim properties owned by
a later stage. This yields one continuous argument through the compiler:
P0 states the logical action, P1 establishes logical support, P2 supplies a
proof-carrying QEC binding, P3 constructs a physical event plan, and P4
binds realtime control. Resource estimates and executions observe selected
points on that spine while retaining the witnesses needed to explain how one
result follows from the same original logical program.

%% file: appendices/execution-model.tex
\section{Execution Model: Targets, Simulation, and Realtime Feedback}
\label{sec:execution-model}

\begin{figure*}[b]
  \centering
  \includegraphics[width=\textwidth]{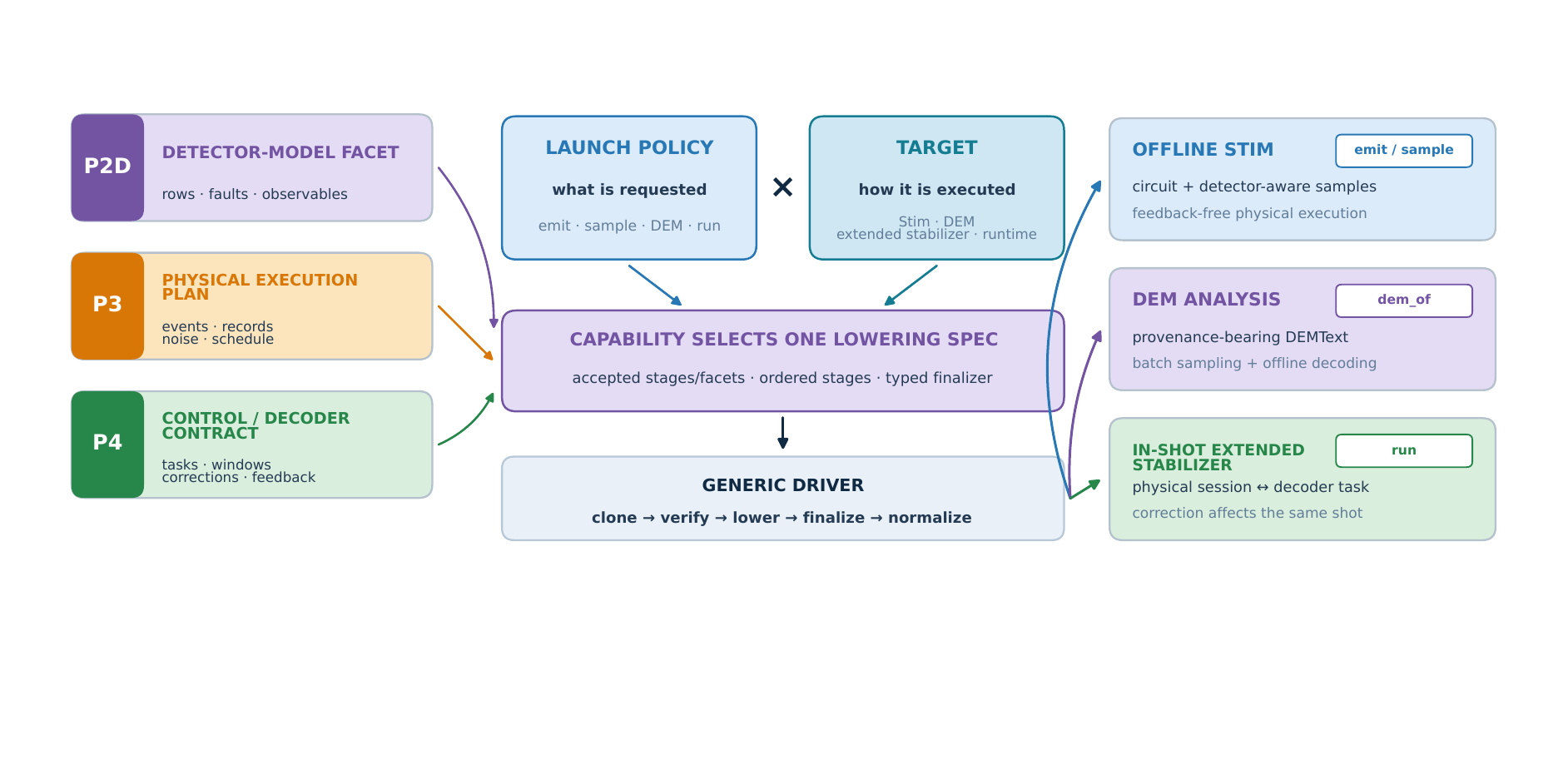}
  \caption{The CUDA-Q Logical policy--target execution boundary. Verified P2D, P3, and P4
  artifacts provide detector, physical, and control facts. A launch policy
  states the requested observation; an immutable target supplies one lowering
  recipe for each capability it actually implements. The common driver clones
  and verifies the source build before producing offline Stim circuits and
  samples, detector-model analyses, or in-shot extended stabilizer execution with decoder
  feedback.}
  \Description{A wide rounded-card diagram shows verified P2D detector-model,
  P3 physical-plan, and P4 control-contract artifacts entering the product of
  a launch policy and a target. Capability negotiation selects one lowering
  specification, and a generic driver clones, verifies, lowers, finalizes, and
  normalizes the request. Three output cards show offline Stim circuit and
  sampling, detector error model analysis, and local XStab execution in which a
  decoder correction affects the same shot. A bottom rail states the preserved
  invariants: immutable build, no target-side QEC selection, explicit policy,
  and provenance-bearing results.}
  \label{fig:execution-model}
\end{figure*}

% \begin{wrapfigure}{r}{0.55\textwidth}
%   \begin{minipage}{\linewidth}
\begin{lstlisting}[
  style=qlx-python,
  % float=tp,
  % basicstyle=\scriptsize\ttfamily,
  aboveskip=0pt,
  belowskip=0pt,
  caption={One physical build projected to Stim circuit, samples, and an offline detector-model decode.},
  label={lst:stim-offline-execution}
]
import cudaq.logical as ql 

p3_unscheduled = ql.compile(memory, pipeline=ql.compiler.pipelines.physical(), device=device)
p3 = ql.compiler.schedule(p3_unscheduled).build

stim_target = ql.targets.stim
stim_text = ql.emit(p3, target=stim_target)
emission = ql.targets.emit_artifact(p3, target=stim_target)
samples = ql.sample(p3, target=stim_target, shots=10_000, seed=7)

dem_text = ql.targets.dem_of(emission, categorical_faults="reject")
matching = pymatching.Matching.from_detector_error_model(stim.DetectorErrorModel(dem_text))
predicted = matching.decode_batch(samples.detectors)
\end{lstlisting}
%   \end{minipage}
% \end{wrapfigure}
Execution is an observation of a verified compiler artifact. 
A P3 build fixes the physical resources, events,
records, noise bindings, and schedule that a simulator or device must execute.
P4 fixes any controller-visible queues, decoder sessions, task
dependencies, and feedback edges needed during that execution. The
execution layer selects an implementation for those already-defined facts; it
may not change the selected code, reinterpret a detector, or repair an
infeasible schedule. CUDA-Q Logical expresses this boundary through a uniform
policy--target launch model (Figure~\ref{fig:execution-model}).

A CUDA-Q Logical \texttt{Target} is an immutable, capability-indexed collection of lowering
recipes.  It is distinct from a device: a device describes the logical, QEC,
physical, and control architecture used during refinement, whereas a target
describes how a verified result of that refinement is emitted, sampled, or
executed.  It is also distinct from a launch policy.  

\begin{wrapfigure}{l}{0.53\textwidth}
  \begin{minipage}{\linewidth}
\begin{lstlisting}[
  style=qlx-python,
  % float=tp,
  % basicstyle=\scriptsize\ttfamily,
  aboveskip=0pt,
  belowskip=0pt,
  caption={Executing a device-owned decoder workflow in the loop with the local extended stabilizer simulation provider.},
  label={lst:xstab-realtime-execution}
]
import cudaq.logical as ql 

@ql.gadget(device=device)
def decoded_round() -> bool:
    block = ql.prepare_zero(ql.ops.allocate_patch())
    block, syndrome = ql.extract_syndrome(block)
    ql.ops.submit_syndrome(syndrome)
    block, raw = ql.mpp(ql.types.Z(block[0]))
    corrected = ql.ops.decode_bit(raw)
    ql.discard(block)
    return corrected

decoding = ql.qec.decoder.workflow(
    analysis=analysis)
p3 = ql.compile(decoded_round,
    pipeline=ql.compiler.pipelines.physical(),
    decoder=decoding)

result = ql.run(
    p3, target=ql.xstab, shots=1_000,
    decoding=decoding)
\end{lstlisting}
  \end{minipage}
\end{wrapfigure}
The policy states what
the caller requests and owns invocation-local choices such as shot count,
random seed, or an explicitly authorized detector-model approximation; the
target states which requests it can satisfy and how.
Each real target capability contributes one \texttt{LoweringSpec} which 
declares the accepted semantic stages and facets, an ordered sequence of 
transformations, a terminal finalizer, its result schema, and its
effect class. The common driver replays the immutable \texttt{Build} into a
fresh module and lowering context, verifies every declared boundary, runs the
finalizer only after all stages succeed, and lets the active policy validate
and normalize the typed result.  Thus \texttt{ql.sample},
\texttt{ql.run}, \texttt{ql.emit}, and
\texttt{ql.targets.dem\_of} are concise launch-policy spellings, not
separate backend pipelines.  A target manifest makes the capabilities,
accepted stages, recipes, finalizers, result schemas, and implementation
identity inspectable and replayable.

% \subsection{Targets as the Execution Boundary}

\subsection{Offline Sampling and Detector-Model Analysis}

Feedback-free execution can stop at \profile{P3}. In
Listing~\ref{lst:stim-offline-execution}, \texttt{memory} is a
detector-profiled workload and the selected device supplies the physical
processes, noise, and timing assumptions. After scheduling fixes event order
and resource exclusion, the Stim target maps the physical graph---not the
retained logical circuit---to an emission and samples the same
projection. Each request privately replays the build,
leaving its source snapshot unchanged.

The typed emission preserves the correspondence between physical records and
detector, observable, success, and erasure roles. Thus
\texttt{SampleResult.detectors} contains only decoder rows; postselection
mismatch and heralded loss stay separately inspectable, and the seed and
circuit/model provenance travel with the samples.

\texttt{dem\_of} provides the adjacent offline-analysis path. It can
serialize a verified \profile{P2} detector-model facet directly, or, as in
the listing, derive the model from the exact typed Stim emission. Virtual-input-syndrome
conditioning and categorical-fault approximation are explicit
\texttt{DEMPolicy} choices, recorded with the source digest in
\texttt{DEMText}. Offline batches can therefore be decoded without severing
the channel-check matrix from the compiled virtual detectors.

Offline sampling applies when decoding does not affect the physical instruction
stream. If a decoded value controls a later operation, CUDA-Q Logical rejects native
sampling and requires a target with a \texttt{run} recipe.

\subsection{Local Simulated Realtime Decoding}

Realtime decoding changes the causal position of decoding. It means that a
decoder task is submitted and consumed within a shot, so its typed correction
can affect a later value or branch. Listing~\ref{lst:xstab-realtime-execution} exposes that
distinction. The selected device owns the decoder implementation, while the
gadget owns the production and consumption sites. The detached analysis owns
the ordered detector and observable rows. From these objects CUDA-Q Logical derives the
decoder context---channel-check matrix, mechanism priors, and
mechanism-to-observable map~\cite{kliuchnikov2026composing}---and freezes the resulting workflow contract into
the build.

An extended stabilizer simulation target (here abbreviated \emph{ql.xstab}) 
is supplied by an independently installable local simulator
provider.  Its \texttt{run} recipe accepts only a P3 
\texttt{SimulationPlan}.  At launch, CUDA-Q Logical checks that \texttt{decoding} is
content-identical to the build's decoder contract and compiles it to an
immutable tasking configuration. For each shot, the core run coordinator
opens a physical extended stabilizer simulation session and walks the physical event graph.  On
\texttt{submit\_syndrome} it sends the canonical detector window to the
prepared decoder; on \texttt{decode\_bit} it applies the returned observable
correction before subsequent computation proceeds. The provider executes
physical commands, but it does not parse detector conventions or choose a
decoder. Each result retains a task trace alongside the shot result, making
the in-the-loop dataflow directly inspectable.

This local path is the executable bridge between P3 and
\profile{P4}. The event graph and physical results remain P3 facts;
decoder sessions, task identities, typed windows and corrections, and their
feedback dependencies are P4 facts. Local simulation demonstrates
functional realtime semantics---the correction is produced at the right
causal point and changes the same shot---but does not establish controller
deadlines, transport jitter, or decoder throughput on hardware.

\subsubsection{Toward a Realtime Runtime}
\begin{table*}[t]
  \caption{Principal execution requests.  Horizontal rules emphasize that
  each row is a distinct policy and evidence contract, even when several rows
  are served by the same target.}
  \label{tab:execution-targets}
  \small
  \begin{tabular}{@{}p{0.18\textwidth}p{0.21\textwidth}p{0.25\textwidth}p{0.29\textwidth}@{}}
    \toprule
    \raggedright Request & \raggedright Public spelling
      & \raggedright Required compiler evidence & \raggedright Result and use \tabularnewline
    \midrule
    \raggedright Circuit emission
      & \raggedright \texttt{ql.emit(..., target=stim)}
      & \raggedright Verified feedback-free physical graph; typed process,
        record, detector, and observable bindings
      & \raggedright Stim circuit text, or a typed Stim emission retaining row
        roles and provenance. \tabularnewline
    \midrule
    \raggedright Offline sampling
      & \raggedright \texttt{ql.sample(..., target=stim)}
      & \raggedright The same \profile{P3} graph plus a complete noise binding
        when noisy behavior is claimed
      & \raggedright \texttt{SampleResult}: logical bits, decoder detectors,
        observables, success bits, erasure rows, seed, and source identities. \tabularnewline
    \midrule
    \raggedright Detector-model analysis
      & \raggedright \texttt{dem\_of(...)}
      & \raggedright A verified detector-model facet at \profile{P2}, or a
        typed Stim projection of a \profile{P3} graph
      & \raggedright Provenance-bearing \texttt{DEMText} for offline sampling,
        decoding, composition, or threshold studies. \tabularnewline
    \midrule
    \raggedright In-shot execution
      & \raggedright \texttt{ql.run(..., target=xstab,
        decoding=workflow)}
      & \raggedright A \profile{P3} simulation plan and a decoder workflow
        identical to the contract frozen into the build
      & \raggedright Per-shot \texttt{RunResult}, including physical results and
        decoder-task traces whose corrections may affect later computation. \tabularnewline
    \bottomrule
  \end{tabular}
\end{table*}
The target boundary makes a future realtime runtime a substitution of
implementations rather than a new program model. A controller target can add
a \texttt{run} recipe that accepts the same verified P3/P4 facts,
lowers the \texttt{rt\_sched}, \texttt{rt\_decode}, and \texttt{rt\_abi} contracts to its
native queues and clocks, and returns the declared result schema.  A hardware
physical session can then replace extended stabilizer target, and a persistent CPU, GPU, or service
decoder can replace the local decoder binding, without moving physical meaning
into the target or decoder meaning into the device.

Such a target must additionally verify bounded queues, supported event
operations, deadline and backpressure semantics, transport and worker
identity, and complete result provenance. These requirements are deliberately
stronger than the local functional demonstration. The important architectural
result is that the compiler/runtime seam is already exercised: the same
immutable physical plan, typed decoder context, tasking contract, and
policy--target negotiation used by local extended stabilizer execution are the artifacts a
realtime runtime must consume.

%% file: appendices/resource-estimation.tex
% \newpage %for now for wrapfigure to look nice
\section{Resource Estimation as Progressive Compiler Analysis}
\label{app:resource-analysis}

% \begin{wrapfigure}{r}{0.50\textwidth}
% \begin{minipage}{\linewidth}
\begin{lstlisting}[
style=qlx-python,
aboveskip=0pt,
belowskip=0pt,
caption={Resource estimates for one experiment at successive compiler profiles.},
label={lst:profile-estimates}
]
import cudaq.logical as ql 

budget = ql.estimate.FailureBudget(0.01)

p0 = ql.compile(program, pipeline=ql.compiler.pipelines.logical())
p2 = ql.compile(p0, pipeline=ql.compiler.pipelines.qec(),device =device)
p3 = ql.compile(p2, pipeline=ql.compiler.pipelines.physical(), device=device)
schedule = ql.compiler.schedule(p3)

logical = ql.estimate(p0)
static = ql.estimate(p2)
analytical = ql.estimate(p2, p_phys=1e-3, failure_budget=budget)
scheduled = ql.estimate(schedule, p_phys=1e-3, failure_budget=budget)

samples = ql.sample(schedule.build, target=ql.targets.stim, shots=100_000)
observed = ql.estimate(samples, failure_budget=budget)
\end{lstlisting}
%   \end{minipage}
% \end{wrapfigure}
Resource estimation in CUDA-Q Logical is a family of analyses indexed by the semantic
profile of the verified compiler artifact they observe: logical demand at P0,
architectural constraints at P1, selected-QEC accounting and analytical
modeling at P2, physical and schedule-derived accounting at P3, and realtime
control at P4. Each typed result retains its source roots, assumptions, and
evidence, preventing an estimate from silently substituting a different code,
protocol, device, operating point, or schedule. Samples, decoded observations,
and run measurements remain provenance-linked evidence rather than additional
profiles, making the customary layers of fault-tolerant resource estimation
\cite{beverland2022assessing} inspectable compiler boundaries rather than
detached cost models, spreadsheets, or scripts.

\subsection{Analysis Across Semantic Profiles}

A resource estimate becomes more specific as implementation decisions are
fixed. Listing~\ref{lst:profile-estimates} applies
\texttt{ql.estimate} to the same experiment after logical compilation, QEC
selection, physical scheduling, and sampling. The input artifact determines
the calculation that can be performed. The API does not require one result
object to be passed into the next call. Instead, provenance links each artifact
to the earlier decisions and evidence on which it depends; a sample, for
example, is analyzed against its source build rather than a previously
computed schedule summary.

At P0, \texttt{LogicalProfile} describes the logical workload before a code or
device has been selected. It records logical operations, liveness and depth
bounds, synthesis requirements, and abstract resource demand. P1 placement and
capacity checks attach code-independent feasibility and communication evidence
to the build. After QEC selection, \texttt{FabricCounts} traverses the
executable P2 call graph and counts the selected gadgets, protocols, detectors,
and observables. Folded calls and static repetitions remain symbolic, so these
counts can be compared without expanding the program. P0 therefore supports
comparisons between logical formulations, whereas P2 supports comparisons
between QEC bindings.

The P2 build can also be evaluated under an analytical hardware model.
Supplying a physical-error probability, code-scaling law, cycle time, and
failure budget produces a \texttt{FabricEstimate} of code distance, logical
error, physical qubits, execution time, and budget compliance. These quantities
are meaningful only with their assumptions, which are stored in the result.
CUDA-Q Logical rejects the calculation when distance evidence is absent, the operating
point is incompatible, required resources are unbound, or the producer model
is unsupported. \texttt{FailureBudget} specifies the total acceptable failure
probability; its allocation among memory, factories, communication, and
decoding belongs to the evaluation model.

\begin{wrapfigure}{r}{0.50\textwidth}
  \begin{minipage}{\linewidth}
\begin{lstlisting}[
style=qlx-python,
aboveskip=0pt,
belowskip=0pt,
caption={Compiler-driven P3 estimation from a selected QEC build.},
label={lst:compiler-driven-estimate}
]
p3 = ql.compile(p2,
  pipeline=ql.compiler.pipelines.physical(),
  device=device)
schedule = ql.compiler.schedule(p3)
estimate = ql.estimate(schedule, p_phys=1e-3,
  failure_budget=ql.estimate.FailureBudget(0.8),
  scaling=ql.estimate.Scaling(prefactor=0.03, threshold=0.01),
  cycle_time=1e-6,
  evidence_policy=ql.estimate.EvidencePolicy())
\end{lstlisting}
  \end{minipage}
\end{wrapfigure}
Scheduling introduces information that an analytical model does not contain:
the dependencies, overlap, and resource contention of a particular physical
execution. A \texttt{ScheduleEstimate} is therefore computed from a verified
P3 \texttt{phys.schedule} and retains the matching P2 result, device, physical
model, and operating point. It summarizes makespan, resource occupancy,
utilization, retry and exhaustion behavior, and the limiting resource. The
summary does not replace the schedule; the schedule remains the evidence for
how work was ordered and resources were allocated in space and time.

Sampling addresses whether the compiled experiment exhibits the modeled
behavior. A P3 sampling backend may classify failures directly, and a P4 run
may additionally record in-shot decoding and control. When failure is evaluated
outside the backend, \texttt{q.estimate.decoded\_observation} binds one
failure bit per shot to the exact sample and decoder. \texttt{TwinEstimate}
then reports the conditional logical-error estimate, exact Clopper--Pearson
intervals, and classifies the budget test as \texttt{met},
\texttt{not\_met}, or \texttt{inconclusive}. The exact intervals remain valid
for the limiting cases of zero failures and all failures.

Table~\ref{tab:estimation-by-profile} (Section~\ref{sec:overview-resource-analysis})
summarizes the analyses available at each profile, together with their
authoritative inputs and reported evidence.

Static counting rejects unresolved or recursive calls. The analytical engine
supports a closed set of scaling laws and deterministic producer models;
unsupported retry, probabilistic, and shared-resource cases are reported as
errors. These restrictions prevent unsupported stochastic behavior from being
hidden in a point estimate and ensure that every P3 result remains tied to the
P2 result and device configuration it refines.

Physical analysis can represent a selected P2 operation either as detailed P3
events or through a characterized compact model for a factory, synchronous
protocol, or transport service. Both representations enter the same scheduler
and undergo the same dependency and capacity checks. They differ in the
resolution and provenance of the physical model, not in the meaning of the
resulting schedule estimate.

\subsection{Compiler-Driven Schedule Estimation}

In the compiler-driven path, the schedule estimate is derived from the program
and device description. Placement and QEC compilation first select a
\profile{P2} binding. Physical lowering then replaces its selected
operations with timed events, resource claims, and dependencies, from which the
scheduler constructs a verified \profile{P3} schedule. The estimate is a
summary of that schedule, so its qubit and runtime values remain traceable to
the physical events that produced them. No operation count or synthetic
schedule is supplied separately by the application.

Listing~\ref{lst:compiler-driven-estimate} shows the remaining steps when a
selected \profile{P2} build is already available. The call to
\nolinkurl{ql.compiler.schedule} exposes the immutable schedule before it is
passed to \texttt{ql.estimate}. The device, scheduling policy, and objective
are fixed by this artifact; changing any of them requires a new physical build
and schedule. Calls, static repetitions, and bounded control remain folded in
the IR. Their multiplicities scale the work represented by each event, while a
conditional contributes the cost of its longest executable branch. For a
bounded retry, the result reports first-attempt, expected, and maximum costs
separately.

\subsection{Detailed and Compact Physical Models}
Expanding every physical event is unnecessary when the same factory, protocol,
or transport service is invoked many times. A service implementation may also
be provider-specific and unavailable for inlining. A compact model records the
timing, dependencies, and resource use required by the P3 scheduler while
retaining the identity and provenance of the physical materialization. CUDA-Q Logical
supports three such models:

\begin{enumerate}
  \item \texttt{FactoryModel}, an asynchronous producer with startup and
        output cadence;
  \item \texttt{SpacetimePlanModel}, a synchronous physical model with latency,
        initiation interval, phase DAG, and typed resource and factory claims;
        and
  \item \texttt{TransportModel}, an exact selected channel with latency,
        initiation interval, exact route reservations, and per-transfer
        endpoint occupancy.
\end{enumerate}

These models describe physical materializations of selected \profile{P2} concepts.
They do not redefine the logical operation or estimate an arbitrary workload.
For example, the semantics of an MPP or lattice-surgery protocol remain in its
P2 definition; the compact P3 model describes how that selected protocol uses
time and physical resources.

CUDA-Q Logical accepts three sources of physical evidence. A compiler-driven materialization
retains the full event graph. Compiler characterization begins with a
representative detailed schedule, verifies it, and extracts a compact model.
For early studies, a researcher may instead assert the model parameters
directly together with their provenance and validity domain. In each case the
model lowers to generic P3 operations and enters the same scheduler. The
difference lies in the evidence supporting the model parameters, not in the
subsequent dependency or capacity analysis.

A compiler-characterized model is bound to the build and schedule from which it
was obtained, as well as to the selected protocol or channel, provider,
architecture, operating point, timing, and footprint. The compiler credential
is opaque: copying the visible parameter values does not reproduce the evidence
that established them. Synchronous and transport models also contain a hash of
their canonical description. On replay, CUDA-Q Logical recomputes this hash and checks the
timing, dependencies, policies, and resource claims against the typed IR. A
changed binding or model parameter therefore invalidates the characterization.

\begin{figure}[t]
  \begin{minipage}{\columnwidth}
\begin{lstlisting}[
style=qlx-python,
aboveskip=0pt,
belowskip=0pt,
caption={Characterizing and reusing a synchronous physical model.},
label={lst:characterized-model}
]
detailed = ql.compile(selected_protocol, pipeline=ql.compiler.pipelines.physical(),
  device=detailed_device)
detailed_schedule = ql.compiler.schedule(detailed)
plan = ql.compiler.spacetime_plan_model(detailed_schedule, protocol=selected_protocol)

compact_device = (detailed_device.with_spacetime_plan(plan))
compact = ql.compile(repeated_matching_calls, pipeline=ql.compiler.pipelines.physical(),
  device=compact_device)
compact_schedule = ql.compiler.schedule(compact)
estimate = ql.estimate(compact_schedule, p_phys=1e-3, failure_budget=ql.estimate.FailureBudget(0.01))
\end{lstlisting}
  \end{minipage}
  \Description{Python code that derives a synchronous spacetime plan model
  from a detailed physical schedule and reuses it for repeated protocol calls.}
\end{figure}

Listing~\ref{lst:characterized-model} applies this procedure to a synchronous
protocol. It derives a \texttt{SpacetimePlanModel} from one detailed schedule
and then uses the model for a program containing repeated calls to that
protocol. The detailed event body is not copied at each occurrence. Factory
and transport characterization use the corresponding
\nolinkurl{ql.compiler.factory_model} and
\nolinkurl{ql.compiler.transport_model} entry points.

\begin{figure}[t]
  \begin{minipage}{\columnwidth}
\begin{lstlisting}[
style=qlx-python,
aboveskip=0pt,
belowskip=0pt,
caption={Characterizing transport from a scheduled physical transfer.},
label={lst:transport-characterization}
]
builder.physical.bind_channel(delivery_qec, to=links, 
  transport_claims=(ql.devices.PhysicalResourceClaim(links.resource, units=1)), endpoint_occupancy=(1, 1))
builder.physical.set_operating_point(timing={
  "cycle_ns": 1.0,
  "transport_resource_ns": 5.0,
  "transport_resource_initiation_interval_ns": 2.0})
detailed_device = builder.build()

detailed = ql.compile(one_delivery, pipeline=ql.compiler.pipelines.physical(), device=detailed_device)
transport = ql.compiler.transport_model(ql.compiler.schedule(detailed),channel=delivery_qec.channel)
compact_device = detailed_device.with_transport_model(delivery_qec.channel, transport)
\end{lstlisting}
  \end{minipage}
  \Description{Python code that binds calibrated transport timing and route
  occupancy, schedules a representative transfer, and derives a transport
  model for reuse.}
\end{figure}

Transport adds route and endpoint constraints to the same characterization
procedure. A transport model requires calibrated latency and initiation
interval, an exact route reservation, and the occupancy imposed at both
endpoints. In Listing~\ref{lst:transport-characterization}, these facts are
attached to the physical channel before a representative transfer is compiled
and scheduled. Characterization verifies that schedule and records the
resulting model. It neither selects a route from an available pool nor infers
service cadence from the spacing of transfers in the example workload.

Compact operations use the same P3 resource model as detailed events. Factory
availability, synchronous occupancy, route reservations, and endpoint
occupancy are checked as separate constraints. The resources themselves belong
to the selected device and are counted once; a compact model states how they
are occupied but does not add a second copy of them. For transport, the
scheduler assigns a complete acquisition lane to each occurrence, records the
selected physical members and endpoint units, and exposes those assignments to
schedule analysis.

A directly asserted model follows the same scheduling rules but carries weaker
evidence. CUDA-Q Logical marks the model as asserted in Python, IR, replay, and the final
estimate, and checks its protocol or channel, architecture, operating point,
resource claims, timing, and code distance when it is attached. Moving the
model to a different device requires a new assertion. The present contract
supports deterministic \texttt{guaranteed} behavior and explicitly conditioned
\texttt{single\_shot} behavior. Retry averaging, stochastic queues, dynamic
routing, packetization, boundary-changing surgery, and realtime control are
rejected because their execution semantics are not represented by these
models.

%% file: appendices/zsz-construction.tex
\section{ZSZ-LP-100 Construction Details}
\label{app:zsz-construction}
% TO DO (Yifan): this appendix holds the construction text moved out of Section~\ref{sec:ZSZ demos}, lightly compressed. Please check it, and decide whether it should stay or be replaced by a citation to your paper. Delete this comment to sign off.

This appendix records the construction behind the code and surgery gadgets of Section~\ref{sec:ZSZ demos}, which were built and certified outside the compiler and supplied to it as data.

\paragraph{Regular representations and logical seeds.}
ZSZ is short for ``$\mathbb Z$ semidirect $\mathbb Z$'': the family of metacyclic groups $\mathbb Z_m\rtimes\mathbb Z_n$. The block form \eqref{eq:zsz100-five-block} arises from the lifted or balanced product of two classical codes whose parity-check matrices have a $1\times2$ block form; codes with this structure are also called ``mitten'' codes~\cite{bhardwaj2026mitten}. The regular representation assigns one data qubit to each element of $G$, so the code is five copies of $G$ with checks linking qubits by the left and right group actions of $A,B,C,D$. Let $S_\ell$ be the $\ell\times\ell$ cyclic shift $S_\ell\ket{i}=\ket{i+1}$, $T_5$ the affine stretch $T_5\ket{i}=\ket{4i}$, and $\Pi_j=\ket{j}\!\bra{j}$ for $j\in\mathbb Z_4$, indices modulo the cyclic order. The left- and right-regular representations of $x^\alpha y^\beta\in\mathbb Z_5\rtimes_4\mathbb Z_4$ are
\begin{align}
    L[x^\alpha y^\beta] = S^\alpha_5 T^\beta_5 \otimes S^\beta_4 \;,\qquad R[x^\alpha y^\beta] = \sum_{j=0}^{3} S^{4^j\alpha}_5 \otimes S^\beta_4 \Pi_j \, ,
\end{align}
and $A=L[a]$, $B=L[b]$, $C=R[c]$, $D=R[d]$ with the trinomials of \eqref{eq:zsz100-trinomials}. The logical seeds in \eqref{eq:zsz100-ogs-basis} are
\begin{align}
 v_L &= x+x^2+x^3+x^4 + xy +(x^3+x^4)y^2 + (x+x^3)y^3 \, , \nonumber\\
 v_R &= 1+x+x^2+x^4 + x^2y + (1+x^2)y^2 + (x+x^2)y^3 \, .
 \label{eq:zsz100-ogs-seeds}
\end{align}
Both bases restrict to the identity on the first data block and have disjoint supports elsewhere; each seed has 9 monomials, so each logical operator has weight 10, and row $i$ of $L_Z$ anticommutes only with row $i$ of $L_X$.

\paragraph{Surgery ancillas.}
Code surgery generalizes lattice surgery~\cite{horsman2012lattice, landahl2014surgery} to general LDPC codes~\cite{Cohen_2022, Williamson_2026_gauging, cross2025improved, Swaroop_2026_adapters, he2025extractors, zheng2025high, cowtan2025fast, Cowtan_2026_parallel, yuan2026parsimonious}; when the ancilla is built by weight reduction~\cite{hastings2023weight, hsieh2025weight}, both the data and merged codes stay LDPC. The stored artifacts use an opposite-check basis $\widehat H_Z$ with $\operatorname{row}(\widehat H_Z)=\operatorname{row}(H_Z)$ in \eqref{eq:X-meas merged code}, which admits smaller ancillas; the weights in Table~\ref{tab:zsz100-merged-resources} refer to that presentation. Two templates are used. A \emph{graph} for a single logical measurement~\cite{Williamson_2026_gauging, cross2025improved, Swaroop_2026_adapters, he2025extractors, yuan2026parsimonious} has one vertex per qubit of the logical support and one ancilla per edge, with vertex incidence and cycles defining the ancilla checks (for $\bar X$: vertices to $X$-checks, cycles to $Z$-checks; for $\bar Z$ the reverse). A \emph{hypergraph} for parallel measurement of many commuting products~\cite{Williamson_2026_gauging, zheng2025high} measures them in one deformation. Expansion~\cite{Williamson_2026_gauging} or soundness~\cite{zheng2025high} conditions prevent deformed logical operators from losing weight on the ancilla qubits. The edge expansion of each graph ancilla was computed exactly by enumerating every nontrivial vertex cut; the parallel-$XX$ hypergraph's soundness was certified exactly by scanning all equivalence classes of ancilla-check subsets modulo products of the intended measurement modes.

\paragraph{Versatile graphs.}
One graph is built on the support of the seed $\bar X_0$ and one on $\bar Z_0$; the group action permutes logical indices and data-qubit labels together, so the two seeds serve all $k=20$ representatives of each type. For pair products $\bar X_i\bar X_j$ or $\bar Z_i\bar Z_j$, a separate monolithic template joins two identical 10-vertex seed sectors, translated independently onto supports $i$ and $j$, by eight disjoint bridge edges; the committed $XX$ ($ZZ$) template uses 15 (16) internal edges per sector for 38 (40) ancillas. Keeping the sectors distinct handles overlapping representatives, since duplicate attachments cancel modulo two, and global connectivity ensures the pair product but neither factor is measured. Thus one $X$ and one $Z$ seed cover all 40 single-logical measurements and one $XX$ and one $ZZ$ template cover all $\binom{k}{2}=190$ pairs, which is what makes a targeted CNOT between any two logical qubits available from four artifacts. The parallel hypergraph, by contrast, is specific to its pairing.

\paragraph{Protocol and artifacts.}
For an $X$-type ($Z$-type) measurement the compiler prepares the ancillas in $\ket0$ ($\ket+$), runs merged-code rounds to attach, measures the ancillas in $Z$ ($X$) to detach, and applies the Pauli frame update implied by the $\Gamma_Z$ block of \eqref{eq:X-meas merged code}. The machine-readable bundle, to be released with the code, stores the parity checks and equivariant logical bases and, per graph gadget, the edge list, attachment and incidence matrices, check deformation, cycle-space basis, logical target, measurement rows, and complete merged matrices; the hypergraph artifact additionally stores the canonical matching, ordered targets, target kernel and readout map, and its merged matrices.